\documentclass[12pt]{iopart}

\usepackage[colorlinks,allcolors=blue]{hyperref}
\usepackage{graphicx}
\usepackage{braket}
\usepackage{amssymb}
\usepackage[numbers,sort&compress]{natbib}
\usepackage{booktabs}

\expandafter\let\csname equation*\endcsname\relax
\expandafter\let\csname endequation*\endcsname\relax

\usepackage{amsmath}

\begin{document}

\title{Bio-inspired natural sunlight-pumped lasers}

\author{Francesco Mattiotti$^{1,2,3,4}$, William M. Brown$^5$, Nicola Piovella$^{6,7}$, Stefano Olivares$^{6,7}$, Erik M. Gauger$^5$ and G. Luca Celardo$^{8,9}$}

\address{$^1$ ISIS (UMR 7006) and icFRC, University of Strasbourg and CNRS, 67000 Strasbourg, France}
\address{$^2$ Department of Physics, University of Notre Dame, Notre Dame, IN 46556, USA}
\address{$^3$ Dipartimento di Matematica e Fisica and Interdisciplinary Laboratories for Advanced Materials Physics, Universit\`a Cattolica, via Musei 41, Brescia I-25121, Italy}
\address{$^4$ Istituto Nazionale di Fisica Nucleare, Sezione di Pavia, via Bassi 6, Pavia I-27100, Italy}
\address{$^5$ SUPA, Institute of Photonics and Quantum Sciences, Heriot-Watt University, Edinburgh, EH14 4AS, United Kingdom}
\address{$^6$ Dipartimento di Fisica ``Aldo Pontremoli'', Università degli Studi di Milano, via~Celoria 16, Milano I-20133, Italy}
\address{$^7$ Istituto Nazionale di Fisica Nucleare, Sezione di Milano, via Celoria 16, Milano I-20133, Italy}
\address{$^8$ Benem\'erita Universidad Aut\'onoma de Puebla, Apartado Postal J-48, Instituto de F\'isica,  72570, Mexico}
\address{$^9$ Dipartimento di Fisica e Astronomia, Università degli Studi di Firenze,
50019 Sesto Fiorentino, Firenze, Italy}

\ead{mattiotti@unistra.fr}

\begin{abstract}
Even though sunlight is by far the most abundant renewable energy source available to humanity, its dilute and variable nature has kept efficient ways to collect, store, and distribute this energy tantalisingly out of reach.
Turning the incoherent energy supply provided by the Sun into a coherent laser beam would overcome several of the practical limitations inherent in using sunlight as a source of clean energy: laser beams travel nearly losslessly over large distances, and they are effective at driving chemical reactions which convert sunlight into chemical energy.
Here we propose a bio-inspired blueprint for a novel type of laser with the aim of upgrading unconcentrated natural sunlight into a coherent laser beam. Our proposed design constitutes a novel and different path towards sunlight-pumped lasers. 
In order to achieve lasing with the extremely dilute power provided by natural sunlight, we here propose a laser medium comprised of molecular aggregates inspired by the architecture of natural photosynthetic complexes. Such complexes 
exhibit a very large internal efficiency in harvesting photons from a power source as dilute as natural sunlight. Specifically, we consider a hybrid structure, where  photosynthetic complexes in purple bacteria ({\it Rhodobacter Sphaeroides}) surround a  suitably engineered molecular dimer composed of two strongly coupled chromophores. We show that if pumped by the surrounding photosynthetic complex, which efficiently collects and concentrates solar energy, the core dimer structure can reach population inversion, and reach the lasing threshold under natural sunlight. The design principles proposed here will also pave the way for developing other bio-inspired quantum devices.
\end{abstract}

\noindent{\it Keywords\/}: Sunlight-pumped lasing $|$ Bio-inspired photonic devices $|$ Organic nanophotonics $|$ Novel lasing approaches $|$ Photosynthetic quantum technologies

\section{Introduction}
One of the most remarkable aspects of many natural molecular aggregates is their ability to efficiently process extremely weak sources of energy or signals for biological purposes. Examples of this include the ability of avian magneto-receptors to sense the extremely weak geomagnetic field~\cite{shuRP, rodgersRP, gaugerRP, hiscockRP}, or the ability of aquatic bacterial photosynthetic systems to harvest sunlight in deep murky waters, where incident light levels are much reduced beyond the already dilute level on land~\cite{schulten:1,schulten:2}. For instance, purple bacteria have the ability to exploit extremely weak light sources~\cite{schulten:1,schulten:2} (less than 10 photons per molecule per second) and some species of green sulfur bacteria even perform photosynthesis with geothermal radiation from deep-sea hydrothermal vents at about 400$^\circ$C~\cite{greenPNAS}. 
A possible origin of this incredible ability of bacterial photosynthetic systems to utilise weak sources of incoherent light and funnel the collected energy to specific molecular aggregates, could be based on the high level of symmetry and hierarchical organization characterizing the antenna complexes of bacterial photosynthetic organisms. Nevertheless the role of symmetry in the efficiency of such systems is still under debate~\cite{kassal,kassal2,schulten:1}
and it is not essential to the proposal discussed in this manuscript, which is based only on the well-known efficiency of natural bacterial photosynthetic complexes.  
Photosynthetic antenna complexes~\cite{schulten:1,schulten:2,schulten1:1,schulten1:2,photo,photoT,mukameldeph,mukamel,srfmo,srrc} are comprised of a network of chlorophyll molecules which are typically modelled as two-level systems (2LS) capable of absorbing radiation and transporting the resulting electronic excitation to the reaction center where charge separation occurs, a process which precedes and drives all other photosynthetic steps. Each 2LS has an associated transition dipole moment (TDM) which determines its coupling with both the electromagnetic field and also with other proximal chlorophyll molecules.
Owing to the low solar photon density, photosynthetic aggregates operate in the single-excitation regime, meaning at most one excitation is present in the system at any one time.
Many molecular aggregates, both naturally occurring as well as artificially synthesized, display bright and dark states in their single-excitation manifold~\cite{Hdimer:1,Hdimer:2,Hdimer:3,Hdimer:4}: J-aggregates are characterized by a bright state below the energy of the monomer absorption peak, while H-aggregates are characterized by a bright state above the energy of the monomer absorption peak. 
Cooperative properties, such as those seen in photosynthetic aggregates, have inspired many proposals for engineering artificial light-harvesting devices~\cite{scullyPRL,scully1,scully2,creatore,erik2017,superabsorb,zhang, guideslide}. Moreover, the lasing properties of molecular aggregates, such as organic crystals (3D  molecular aggregates) which display strong cooperative effects in the form of H- or J-aggregates, have been widely investigated~\cite{Hlasers,Mlasing:1,Mlasing:2}. 

\begin{figure}
    \centering
    \includegraphics[width=\columnwidth]{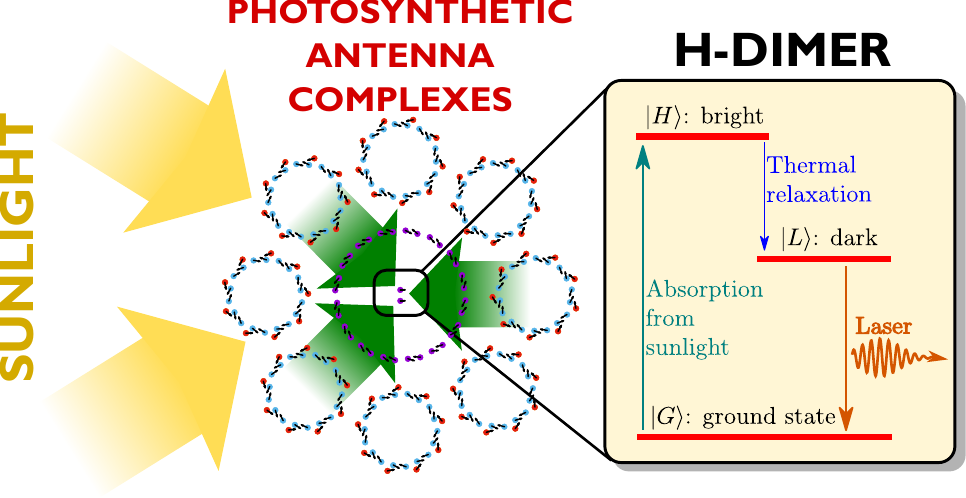}
    \caption{Pictorial representation of the design principle presented in the text. A photosynthetic antenna complex collects energy from sunlight, which is converted to an electronic excitation and efficiently funneled to an H-dimer placed in the middle. Here, the excitation is absorbed to a bright, high-energy state and it relaxes quickly to a dark, low-energy state. This mechanism prevents re-emission and allows population inversion between the $\ket{L}$ and $\ket{G}$ states and, therefore, lasing.}
    \label{fig:intro}
\end{figure}

The design principles of natural photosynthetic complexes  rely on several levels of organization. On the lower level, single molecular aggregates feature a high degree of symmetry which favours the formation of bright (superradiant) or dark (subradiant) states with, respectively, large or small dipole strength~\cite{schulten:1,schulten:2,mukamel,gulli}. On a higher level, photosynthetic systems assemble many of these symmetric aggregates into larger structures, characterized by a hierarchy of energy levels, to maximize light harvesting and energy transport. For instance, in purple bacteria, symmetric rings of chlorophyll molecules (LHI and LHII) surround the reaction center. These rings are J-aggregates with superradiant bright states which favor the absorption of light and the transfer of the excitation between each other~\cite{schulten:1,schulten:2}. 

In this Article, inspired by the design of the photosynthetic apparatus of purple bacteria~\cite{schulten:1,schulten:2}, we show that bio-mimetic molecular aggregates hold the promise of significantly lowering the threshold requirements for sunlight-pumped lasers. 

Sunlight is by far the most abundant renewable energy source on Earth
(a single hour of sunlight provides all the energy humanity uses in a whole year). Despite this, there remain significant limitations in utilising sunlight as it is both dilute and variable. Therefore, efficient storage and distribution of energy harvested from sunlight is paramount. In this respect, sunlight-pumped lasing is an extremely promising technology for energy harvesting, distribution and storage of solar energy~\cite{sunlaserbook}. Sunlight-pumped lasers transform natural incoherent sunlight into intense beams of coherent light, which can be used to efficiently distribute the collected energy and drive chemical reactions as a way to efficiently store solar energy. Indeed, sunlight-pumped lasers have been proposed as essential elements in several renewable energy technologies such as the magnesium cycle~\cite{yabe:1,yabe:2,Natphotonic}. 

Since the power density in natural sunlight is very dilute, typically concentrated sunlight is needed to cross the lasing threshold. Experimentally, a concentration of $10^5$ `suns' (1~sun = 0.14~W/cm$^2$) is reachable,  but clearly the smaller the `number of suns' required for reaching the lasing threshold the better, since this lowers cost, technical demand, and increases efficiency. The first sunlight-pumped laser was realized in 1963~\cite{sunlaserbook} and most sunlight-pumped lasers operate above 1000 suns~\cite{sunlaser1}.
Typically, the concentration of sunlight relies on imaging or non-imaging concentrators. One of the most efficient way to concentrate sunlight is through black-body cavity pumping, where concentrated sunlight collected by a mirror heats a black-body cavity to temperatures which range from 1000~K to 3000~K~\cite{sunlaserbook}.  Recently, sunlight-pumped lasing has been demonstrated under natural sunlight intensity~\cite{masuda2020}. In this groundbreaking experimental work, a dye solution and a dichroic mirror are used to boost the pumping of the solid-state gain medium: the dye solution absorbs sunlight but its emitted fluorescent light is trapped in the cavity due to the dichroic mirror; this effectively increases the pumping rate of the lasing medium.
By contrast, we propose a fully organic architecture based on natural photosynthetic antenna complexes. Besides bio-compatibility and non-toxicity, this may potentially offer advantages for realising nanolasers, as well as broad scope for molecular engineering and tailoring of different antenna complexes. 

Our proposed bio-inspired molecular architecture has at its core a suitably engineered molecular H-dimer. In an H-dimer the interaction between the excited states of each molecules creates a bright state at high energy and a dark state at low energy. Under illumination, energy is absorbed mainly by the bright high-energy state and quickly transferred by thermal relaxation to the lower-energy state, which, being dark, loses energy by re-emission very slowly. Thus, an H-dimer is an ideal candidate to achieve population inversion (which is a main requirement for lasing) and its lower dark excitonic state can be exploited for the lasing transition. Nevertheless, natural sunlight is so weak that the required level of darkness of the lower excitonic state to achieve lasing would be unrealistically high. Indeed, the very long required excitonic lifetime of the lowest-excited state in the single-excitation manifold might be difficult to achieve in practice, due to disorder and competing nonradiative decay processes.
In order to increase the pumping on the bright dimer state, so that the requirement on the darkness of the lower excitonic state can be relaxed, we consider
surrounding a suitably engineered H-dimer
by photosynthetic antenna complexes~\cite{schulten:1,schulten:2}], see Fig.~\ref{fig:intro}. 
Indeed, natural antenna systems are extremely efficient precisely at collecting and funneling natural sunlight energy to specific locations. 
We show that a randomly positioned ensemble of such hybrid molecular aggregates inside a double-mirror cavity can lase under natural (unconcentrated) sunlight for realistic dimer parameters.

\section{Lasing equations for molecular aggregates} We derived lasing equations for a generic ensemble of molecular aggregates, see details in the Supplementary Material (SM),  each made of $N$ identical molecules, that are placed in an optical lasing cavity with suitably chosen frequency. 
The Hamiltonian of the molecular aggregate is written with the usual Pauli operators as
\begin{subequations}
\label{ham}
\begin{eqnarray}
  \hat{H}_S = \sum_{j=1}^N \frac{\hbar\omega_A}{2} \hat{\sigma}_j^z + \sum_{i,j} \Omega_{i,j} \left( \hat{\sigma}_i^+\hat{\sigma}_j^- + \hat{\sigma}_j^+\hat{\sigma}_i^- \right) ~,
\end{eqnarray}
where
\begin{eqnarray}
  \Omega_{i,j}=\frac{\vec{\mu}_i\cdot\vec{\mu}_j}{r_{ij}^3}-3\frac{(\vec{\mu}_i\cdot\vec{r}_{ij})(\vec{\mu}_j\cdot\vec{r}_{ij})}{r_{ij}^5}
\end{eqnarray}
\end{subequations}
is the dipolar inter-molecular coupling~\cite{gulli,phil,mukamel} with $\vec{\mu}_j$ being the TDM of the $j$-th molecule in the aggregate and $\vec{r}_{ij}$ the vector between the $i$-th and the $j$-th molecule~\footnote{Closely spaced molecules require a modification of the dipole interaction term which is then not solely determined by the TDM~\cite{schulten1:1,schulten1:2}.}.
Equation~(\ref{ham}) represents a molecular aggregate where each molecule is approximated as a 2LS with splitting $\omega_A$. 
Under the relatively weak pumping conditions considered here, rather than retaining the full Hilbert space of dimension $2^{N}$ it suffices to limit our analysis to the overall aggregate ground state $\ket{G}$ and the single-excitation manifold comprised of $N$ states $\ket{j}$ where the $j$-th molecule is excited while all the other ones are in their respective ground states. 

We capture thermal relaxation by coupling each molecule to an independent bath of harmonic oscillators~\cite{guideslide,erik2017}, and for simplicity we here neglect vibronic effects~\cite{Hdimer:4}. The interaction of molecular aggregates with solar radiation has been widely discussed in the literature~\cite{whaley2018,brumer2017,plenio2020,kassal3,scholes2015,AndreasSun,brumer2012,caosunlight,scully1,scully2}. Here, we consider the well-established Bloch-Redfield (B-R) formalism for the interaction with a black-body radiation at the temperature of the Sun, which also underpins the recent work in Refs.~\cite{brumer2017,scully1,scully2}. As the B-R master equation has not been secularised, coherence and population dynamics are not decoupled, however, as we will later only be interested in the effective optical pumping rate, we do not expect subtleties regarding the absorption process and transient coherences~\cite{brumer2017} to be important. 
Within this framework, the interaction of the molecular aggregates with the phonon and photon bath is then governed by the master equation
\begin{eqnarray}
    \frac{d\hat{\rho}(t)}{dt} = -\frac{i}{\hbar} \left[ \hat{H}_S, \hat{\rho}(t) \right] + {\cal D}_{BB} [\hat{\rho}(t)] + {\cal D}_{T} [\hat{\rho}(t)] \, , \label{mas}
\end{eqnarray}
where ${\cal D}_{BB}$ and ${\cal D}_T$ are the Bloch-Redfield dissipators for the coupling to the black-body cavity and phonon environments, respectively (see SM).
In our simulations phonon bath parameters have been chosen in order to effect thermal relaxation within a few picoseconds, typical of molecular aggregates~\cite{schulten:1,schulten:2}.
Equation~(\ref{mas}) can be largely simplified under well-motivated assumptions: first, as we check and validate numerically in the SM, for the parameter regimes of interest we may safely secularise and reduce our master equation to Lindblad form~\cite{petru}; this is consistent with the approach of the recent works~\cite{whaley2018,AndreasSun}. 
In deriving the laser equation, we shall make a further assumption, which is analyzed in detail and validated by numerical simulations in the SM:
since thermal relaxation  is typically the fastest time scale for molecular aggregates at room temperature (RT), we can assume that the populations in the single-excitation manifold are always at thermal equilibrium. 

Using the above mentioned assumptions, the coupling with the black-body photon bath is well-approximated by rate equations for the populations (see SM and Ref.~\cite{AndreasSun}), with absorption rates between the $\ket{G}$ and the single-excitation states $\ket{k}$ given by
\begin{eqnarray}
\label{gk}
R_k=n^k_T \gamma_k \, , \quad \mbox{with}\quad \gamma_k=\frac{\mu_k^2 \omega_k^3}{3 \pi \epsilon_0 \hbar c^3} \quad \mbox{and} \quad n^k_T=\frac{1}{e^{ E_k/k_BT_{BB}}-1} ~,
\end{eqnarray}
where $\gamma_k$ is the spontaneous decay rate of the $k$-th state, $\omega_k$ and $\mu_k$ its transition frequency and TDM, respectively, $E_k=\hbar\omega_k$ the transition energy and $n^k_T$ the photon occupancy at the black-body temperature $T_{BB}$.

As our laser gain medium, we consider an ensemble of molecular aggregates randomly distributed with density $n_A$ inside a lasing cavity of frequency $\omega_c$ and containing a classical oscillating field $\vec{E}=E_0\hat{\epsilon}\cos(\omega_ct)$. The aggregate's single-excitation states $\ket{k}$ couple coherently to the cavity mode with Rabi frequencies $\Omega_k=(\vec{\mu}_k\cdot \hat{\epsilon})E_0/\hbar$ that depend on the cavity polarization $\hat{\epsilon}$ and field amplitude $E_0$.
In molecular aggregates under weak pumping, the Rabi frequency is typically smaller than the interband excitonic dephasing rate, $\Gamma_{\phi} \gg \Omega_k$. Therefore, instead of coherent Rabi oscillations we obtain incoherent transition rates proportional to $\Omega _k^2$, 
as derived in the SM and Ref.~\cite{lev}.
The field intensity $I= \epsilon_0 |E_0|^2 c /2$  can also be written as $I= \hbar \omega_c n c/V$, where $n$ is the number of photons in the cavity, $V$ the cavity volume, and $c$ the speed of light. This allows us to express the cavity-induced transition rate between  $\ket{G}$ and $\ket{k}$ state in terms of the number of cavity photons $n$ as
\begin{eqnarray}
\label{Bk}
    n B_k = n \, \frac{1}{3} \frac{|\mu_k|^2 \omega_c}{V \hbar \epsilon_0}
\frac{\Gamma_{\phi}}{\Gamma_{\phi}^2+(\Delta_k/\hbar)^2} = 
    \frac{\Omega_k^2}{2}\frac{\Gamma_\phi}{\Gamma_{\phi}^2+(\Delta_k/\hbar)^2} ~,
\end{eqnarray}
where $\Delta_k=(E_k-\hbar\omega_c)$ is the energy detuning between the single-excitation state $k$ and the cavity mode. 
The factor $1/3$ derives from averaging over the random aggregate orientations.

Under the above assumption we can write lasing rate equations that couple the populations of the molecular aggregates with the number of photons in the cavity. For this purpose, let us define the density of aggregates in the excited states as $N_e=n_A P_e$, the density of aggregates in the ground state as $N_G=n_A P_G$, and the population difference per unit volume  as $D=N_e-N_G$. This gives the lasing equations 
\begin{eqnarray}
    \frac{d D}{dt} =  & -D [R_d+R_u+ (B_{tot}+\langle B \rangle)n] + n_A  [R_u-R_d+n(B_{tot}-\langle B \rangle)] \nonumber \\
    \frac{d n}{dt} = ~& V(B_{tot}+\langle B \rangle)\frac{n D}{2} - V(B_{tot}-\langle B \rangle)\frac{n n_A}{2} - \kappa n ~, 
\label{Leq2}
\end{eqnarray}
where $R_u = \sum_k R_k $ is the total absorption rate and $R_d = \sum_k (R_k+\gamma_k) p_k$ is the spontaneous and stimulated emission rate from the single-excitation manifold.
Further, $B_{tot}= \sum_k B_k$ and $\langle B \rangle = \sum_k B_k p_k$  are, respectively, the total upwards and downwards transition rates between $\ket{G}$ and the single-excitation manifold that are induced by the coupling to the cavity mode. 

From Eq.~\eqref{Leq2} we obtain the stationary values of the population difference per unit volume $D_0$ and the stationary number of photons $n_0$ in the cavity
\begin{eqnarray}
    D_0=  & \frac{2 \kappa}{V (B_{tot}+\langle B \rangle)} + n_A \bar{B}~, \nonumber \\
    n_0 = & \frac{V(n_A D_{eq}- D_0)}{2 \kappa} (R_u+R_d) ~,
\label{laser5}
\end{eqnarray}
where $\bar{B}=\left(B_{tot}-\langle B \rangle\right)/\left( B_{tot}+\langle B \rangle\right)$ and $D_{eq}=(R_u-R_d)/(R_u+R_d)$
is the equilibrium population difference in absence of driving from the cavity.
Above the lasing threshold, i.e.~having $n_0>0$ stationary photons in the cavity, the laser intensity and output power will be, respectively,
    \begin{eqnarray}
        I &= \frac{\hbar \omega_c c}{V} n_0 ~, \nonumber \\
        P_{out} &= \frac{\kappa V}{c} I = \kappa \hbar \omega_c n_0 ~.
    \label{powerint}
    \end{eqnarray}
We turn to the question under which conditions we achieve lasing. Imposing $n_0>0$ in Eq.~\eqref{laser5} we require $n_A D_{eq}- D_0>0 $, which can be written as
\begin{eqnarray}
\label{nA}
n_A (D_{eq}-\bar{B})> \frac{2 \kappa}{V(B_{tot}+\langle B\rangle)} ~.
\end{eqnarray}
Using the definitions of $R_u$ and $R_d$ and for $n^k_T \ll 1$,
\begin{eqnarray}
D_{eq}\approx \frac{\sum_k \chi_k n^k_{T} - \langle \chi \rangle}{\sum_k \chi_k n^k_{T} + \langle \chi \rangle} ~,\quad \mbox{with} \quad \langle \chi \rangle =\sum \chi_k p_k~,
\label{Deq}    
\end{eqnarray}
where $\chi_k=\gamma_k/\gamma_0$ indicates the relative brightness of the state $\ket{k}$ and
$\langle \chi \rangle$ is the thermal average of the relative decay rates of all the single-excitation states. Moreover, $\gamma_0= (\mu^2 \omega_A^3)/(3 \pi \epsilon_0 \hbar c^3)$
is the spontaneous decay rate of a single molecule.  
We reiterate that Eq.~(\ref{Deq}) is generically valid subject to fast thermal relaxation and with negligible occupation of states containing more than one excitation. Both assumptions are realistic for molecular aggregates under black-body radiation pumping.

Equation~(\ref{nA}) determines the critical density of molecular aggregates to achieve lasing, implying
\begin{eqnarray}
    \label{Lcond}
D_{eq}>\bar{B} \, .
\end{eqnarray} 
Since $\bar{B} \ge 0$ by definition,  unsurprisingly we require population inversion, $D_{eq}>0$, to achieve lasing. Considering Eq.~\eqref{Lcond} with Eq.~\eqref{Deq} and  recalling that $R_u
 = \gamma_0 \sum_k \chi_k n_T^k$, implies  $\langle \chi \rangle \leq \frac{R_u}{\gamma_0} \frac{1-\bar{B}}{1+\bar{B}}$, which can be recast as: 
\begin{eqnarray}
    \langle \chi \rangle \leq \frac{R_u}{\gamma_0} \frac{\langle B \rangle}{B_{tot}} \, .
    \label{chicr}
\end{eqnarray}
Equation~(\ref{chicr}) clearly shows that given a non-zero (but  realistically small) value for $\langle \chi \rangle$, two conditions need to be met for lasing: $(i)$ the ratio $\langle B \rangle / B_{tot}$ should be as large as possible, given $\langle B \rangle \le B_{tot}$ this is maximized for $\langle B \rangle \approx B_{tot}$. This condition  can be realised by a lasing state that is well-gapped  (w.r.t.~$k_BT$ at RT) 
below all other states in its excitation manifold; ($ii$) the absorption rate $R_u$ should be as large as possible.
Lasing under very weak pumping requires  a highly dark aggregate (i.e.~small $\langle \chi \rangle$), even if $\langle B \rangle \approx B_{tot}$. This is not easy to achieve, and the situation is compounded by nonradiative losses typically present in molecular aggregates. 
As we shall show in the following, a bio-inspired molecular architecture can help to mitigate this stringent demand and make lasing achievable.

Throughout this Article, when considering black-body optical pumping, we choose a temperature $T_{BB}=3000$~K which is attainable using sunlight concentrated by a mirror of a few m$^2$~\cite{sunlaserbook} with an input power into the black-body cavity of few kW. 
To compute the laser output power we assume a typical gain medium volume of $V=11.3$~cm$^3$ (radius of 6~mm and length of 10~cm). The densities of the laser medium are chosen to be lower than 1~aggregate/(10 nm)$^3$, corresponding to realistic densities for dye lasers: $n_A^{\rm (max)}=10^{18}$~cm$^{-3}=1.6$~mmol/L~\cite{dyelaser}. This choice ensures that direct interactions between the molecular aggregates can be neglected. Moreover, it also keeps the output power below 1~kW, so that thermal balancing with the black-body cavity under realistic sunlight pumping can be maintained. 
To remove the need for and complexity of sunlight concentration we shall also consider the possibility to achieve lasing under direct natural sunlight illumination. To model natural sunlight we consider pumping under a  black-body at $T_{BB} \approx 5800$~K but with rates in Eq.~\eqref{gk} reduced by a factor $f_S$ representing the solid angle of the Sun as seen on Earth~\cite{solarcellbook},
\begin{eqnarray}
    \label{fac1Sun}
    f_S = \frac{\pi r_S^2}{4\pi R_{ES}^2} = 5.4\times 10^{-6} \, ,
\end{eqnarray}
with $r_S$ being the radius of the Sun and $R_{ES}$ the Sun-to-Earth distance.  In this case we limit the output power to 1~W since the incident power on our chosen lasing cavity is just a few W.

\section{Lasing with dimers}

\begin{figure*}[!ht]
  \centering
  \includegraphics[width=\linewidth]{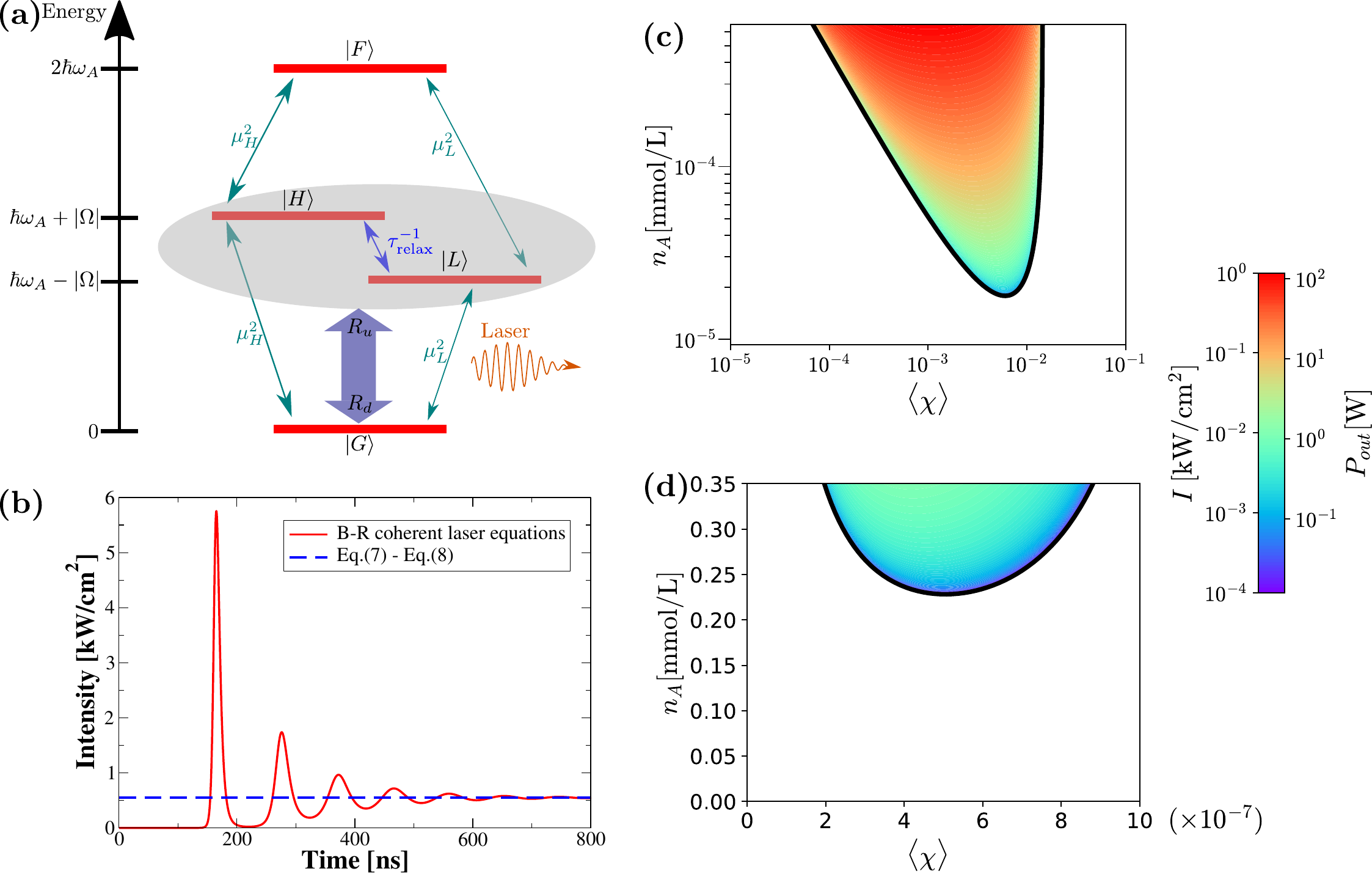}
  \caption{{\bf (a)} Eigenstates of the dimer Hamiltonian $\hat{H}_S$ [Eq.~\eqref{ham}] and transition rates between them. 
  Optical rates (green arrows) are proportional to the squared dipole strengths [see Eq.~\eqref{gk}], while the thermal relaxation rate (blue arrow) is $\tau_{\rm relax}^{-1} \approx 1$~ps$^{-1}$. All these transitions obey detailed balance, i.e.~upwards and downwards rates are proportional to the Bose-Einstein occupation number $n_T (\hbar \omega)$ and $[1+n_T (\hbar \omega)]$, respectively. 
  The thick double arrow represents the resulting effective rates linking the ground state $\ket{G}$ to the single-excitation manifold (grayed area), as utilised in Eq.~(\ref{Leq2}a). The $\ket{L}-\ket{G}$ transition is further coupled to a resonant lasing cavity.
  {\bf (b)}
  Temporal evolution of the laser intensity under black-body pumping: Bloch-Redfield model (solid red) for dimers that are coherently coupled to the lasing cavity (details in the SM), and steady-state solution (dashed blue) of our lasing equations, Eqs.~\eqref{laser5} and \eqref{powerint}.
  Parameters: $\mu=10.157$~D, $\hbar\omega_A=1.17$~eV, $\Omega=2000$~cm$^{-1}$, $T_{BB}=3000$~K, $\Gamma_\phi=1/(10$~ps$)$, $\kappa/(2\pi)=50$~MHz, $\langle \chi \rangle=0.005$, $n_A=5\times10^{-4}$~mmol/L, and a lasing cavity volume $V=11.3$~cm$^3$ (cylindrical shape of radius $R=6$~mm and length $L=10$~cm).
  {\bf (c)-(d)} Laser intensity and output power for the
  same parameters as (b) except for $\langle\chi\rangle$ (thermal average of the relative brightness of single-excitation states with respect to a single molecule) and $n_A$ (dimer density in millimol/L) which are varied along the axes.
  The black line represents the lasing threshold~[Eq.~\eqref{nA}]. In (c) optical pumping occurs via a black body radiation at $T_{BB}=3000$~K, whereas in  (d) the lasing medium is illuminated by  natural sunlight.}
  \label{dip}
\end{figure*}

Let us consider a dimer comprising two identical chromophores. Each molecule (labeled $j=1,2$) has one relevant optical transition, so that we may model it as a 2LS with ground state $\ket{g_j}$ and excited state $\ket{e_j}$. Excitation energy $\hbar\omega_A$ and magnitude $\mu$ of the electric TDM are identical between the molecules, while the direction of the optical dipole $\vec{\mu}_j$ depends on the orientation of its chromophores and may differ.

For this dimer system the Hamiltonian in Eq.~\eqref{ham} is diagonalized by a set of four states (see Fig.~\ref{dip}a): $\ket{G}=|g_1\rangle |g_2\rangle$, where both molecules are in their respective ground state; $\ket{L}$ and $\ket{H}$ are the lowest and highest single-excitation states, where only one excitation is present in the system, delocalised over both molecules; these states  span the single-excitation manifold and they correspond to the symmetric and anti-symmetric states $\left(|g_1\rangle |e_2\rangle \pm |e_1\rangle |g_2\rangle \right)/\sqrt{2}$.
Finally, $\ket{F}=|e_1\rangle |e_2 \rangle$ has both molecules in their respective excited state. The corresponding energies
are shown in Fig.~\ref{dip}a.

Optical transitions between the levels are determined by the relative orientation of the single molecules. We indicate the transition dipoles between the ground state $\ket{G}$ and the states $\ket{L}$ and $\ket{H}$ with $\vec{\mu}_L$ and $\vec{\mu}_H$,
respectively, see  Fig.~\ref{dip}a. The conservation of total oscillator strength demands that
$\mu_L^2 + \mu_H^2 = \mu_1^2 + \mu_2^2 = 2 \mu^2$ and we have an H-dimer if $\mu_L<\mu$ while if $\mu_L>\mu$ we have a J-dimer. 
The coupling $\Omega$ between molecules can either have a dipolar or a different origin, see SM.  
Typical H-dimers feature splittings between their bright and dark states of several $k_B T$, and the lower state may be many hundreds times less bright than the upper state~\cite{Hdimer:1,Hdimer:2,Hdimer:3,Hdimer:4}. 
Here, we consider a dimer with excitation energy in the near-infrared $\hbar \omega_A=1.17$~eV ($\lambda \approx 1060$~nm), transition dipole of $\mu=10.157$~D (as for the bacteriochlorophyll-a molecule) and a coupling $\Omega=2000$~cm$^{-1}$ as in similar H-dimers~\cite{SWNIR}. 

Under black-body illumination, primarily the bright state $\ket{H}$ undergoes excitation, followed by rapid thermal relaxation to the lower dark $\ket{L}$ state. The large energetic separation between $\ket{H}$ and $\ket{L}$ makes this relaxation one-way, preventing environmental re-excitation into $\ket{H}$. In SM we show that it is possible to achieve population inversion provided the absorption rate $\ket{G}\rightarrow \ket{H}$ dominates over the spontaneous emission rate $\ket{L}\rightarrow\ket{G}$.
We proceed to couple the $\ket{L}\rightarrow\ket{G}$ transition to a resonant lasing cavity and evaluate the lasing performance of the system using the equations derived in the previous section. 
Figure~\ref{dip}b shows the resulting laser intensity: 
once the stationary regime has been reached, there is perfect agreement between the intensity predicted by Eq.~\eqref{Leq2} with our numerically obtained results from a coherent Bloch-Redfield model. The latter, derived in Eqs.~(S37, S43) of the SM,   treats both photon and phonon environments in the Bloch-Redfield formalism, includes the doubly excited state, and -- as its main assumption -- treats the laser field semi-classically, but nonetheless coherently coupled to the aggregates similarly to Refs.~\cite{scullybook,scullypaper}. 
In Fig.~\ref{dip}c  we show the dependence of the laser intensity and power output on
$n_A$ and $\langle \chi \rangle$ based on realistic choices for all other parameters (see caption). 
The white area highlights the region below the laser threshold Eq.~\eqref{nA} (black continuous line), where the dimer density $n_A$ is too low to permit lasing.
As one can see, an intensity of up to $1$~kW/cm$^2$ can be reached with a very low dimer concentration.

To assess the possibility of lasing under direct natural sunlight illumination (i.e.~without a  black-body cavity heated by concentrated sunlight), we show the lasing threshold and output power for this scenario in Fig.~\ref{dip}d. Clearly, lasing is still theoretically possible but only for very low values of $\langle \chi \rangle$. 
In practice, this is
challenging due to the competition of nonradiative decay and other sources of noise in realistic situations.  
Nevertheless, as we  show in the next section, the critical value of $\langle \chi \rangle$ increases by orders of magnitude if the dimer is placed inside a purple bacteria molecular aggregate. Finally, note that for lasing we require a small yet finite value of $\langle \chi \rangle$. In the case of a homodimer (where for parallel TDMs the $\ket{L}$ dimer state would be fully dark) this can either arise as a consequence of the relative orientation of the TDMs, or through the presence of structural or energetic disorder. 

\begin{figure*}[!ht]
  \centering
  \includegraphics[width=\linewidth]{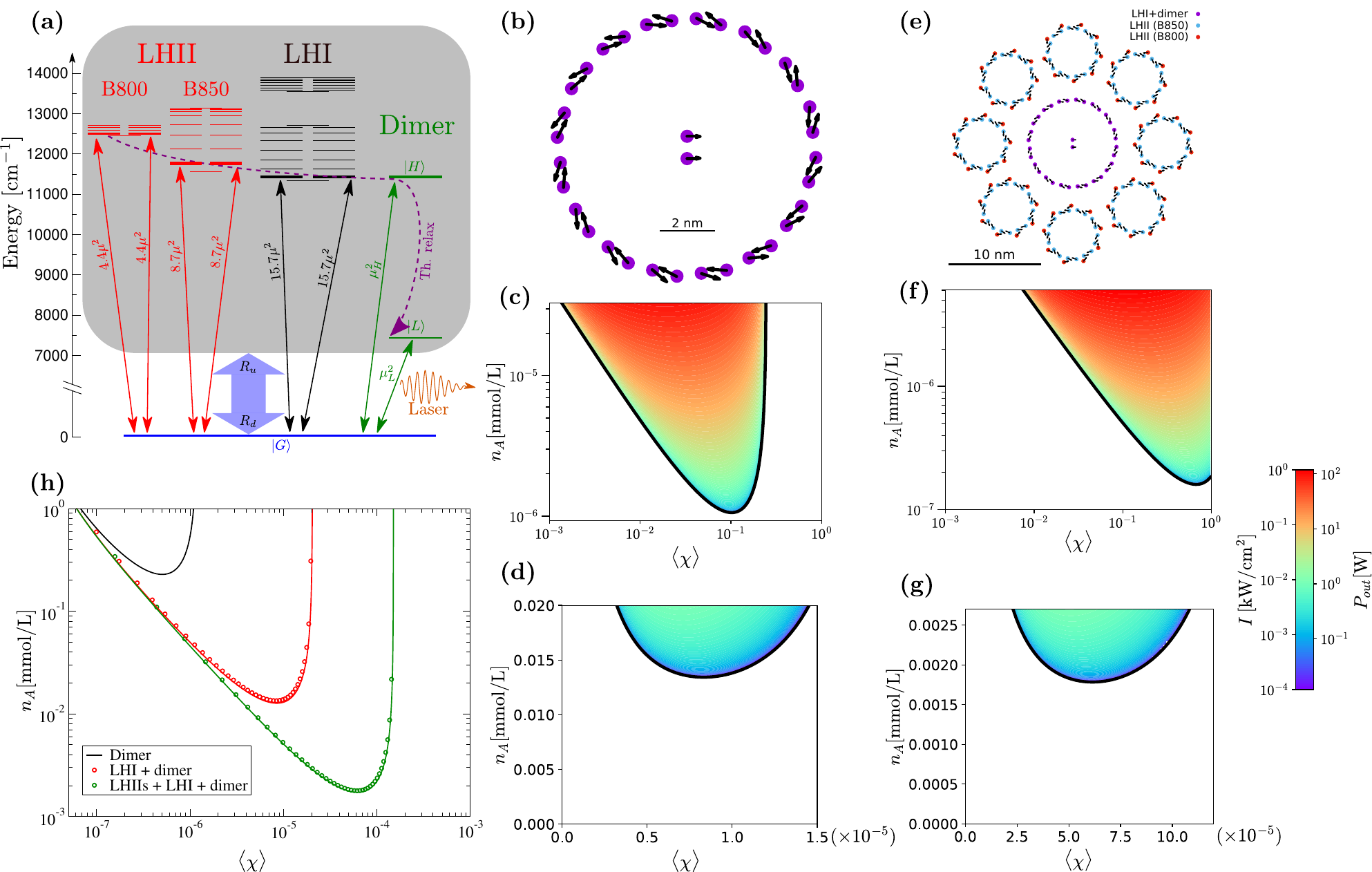}
  \caption{{\bf (a)} Single-excitation eigenstates of the LHII complex (with a distinction between the B800 and the B850 subunits), LHI complex, and H-dimer. 
  The strongest optical rates, proportional to the squared dipole strengths, are indicated alongside the two-headed arrows. 
  As before, the thick double arrow represents the resulting effective rates linking the ground state $\ket{G}$ to the single-excitation manifold (grayed area) and the $\ket{L}-\ket{G}$ dimer transition is also coupled to a lasing cavity. The H-dimer parameters are: $\mu=10.157$~D, $\hbar\omega_A=1.17$~eV, $\Omega=2000$~cm$^{-1}$. Further, $\Gamma_\phi=1/(10$~ps$)$, $\kappa/(2\pi)=50$~MHz, and $V=11.3$~cm$^3$. 
  {\bf (b), (e)} Positions of the chromophores (circles) and transition dipole orientations (arrows) for the bio-mimetic complexes.  Excitation energies and nearest-neighbour couplings for bio-mimetic aggregates are taken from Refs.~\cite{schulten1:1,schulten1:2,schulten2,schulten3,kassal}. 
  {\bf (c), (d), (f), (g)} Laser intensity and output power for an LHI ring surrounding an H-dimer (c,d) and with eight additional B800/850 LHII rings surrounding the LHI ring (f,g).
 On the axes we vary $\langle\chi\rangle$ of the H-dimer and $n_A$ (the aggregate density in mmol/L). In all panels the black line represents the lasing threshold~[Eq.~\eqref{nA}]. In (c,f) we have black-body pumping at temperature $T_{BB}=3000$~K, whereas {(d,g)} are for natural sunlight illumination. {\bf (h)} Lasing threshold density under natural sunlight for a bare dimer, one LHI surrounding the dimer as in (b), and eight LHIIs around an LHI ring containing the dimer as in (e). The red and green continuous curves are for a dimer with a re-scaled $R_u$ pumping rate: by a factor of 17 (red curve) and by a factor of 125 (green curve), corresponding to an enhancement factor $N/2$ with $N$ being the number of chromophores in the aggregate. The circles are obtained from the full set of optical transitions as indicated in (a).}
  \label{dip2}
\end{figure*}

\section{Bio-inspired lasers}

Whilst lasing with a gain medium composed of suitable H-dimers is realistic under black-body cavity pumping, achieving the extremely high optical darkness (small $\langle \chi \rangle$) for direct sunlight-pumped operation is a tall order.  
According to Eq.~\eqref{chicr}
$\langle \chi \rangle$ is upper-bounded by the pumping rate in units of $\gamma_0$ (assuming a favourable ratio $\langle B \rangle / B_{tot} \approx 1$). Thus increasing the effective pumping rate reduces the  stringency of required darkness. A possible way of increasing pumping is to surround the dimer by a molecular aggregate that is capable of efficiently absorbing photons and transferring the resulting energy excitations to the $\ket{H}$ dimer state. The aggregate should absorb at an energy larger than the $\ket{H}$ dimer state energy in order to preserve the gap which separates the dimer lasing state from other states, so that $\langle B \rangle / B_{tot} \approx 1$. Note that under this condition $\langle \chi \rangle$ of the whole aggregate can remain very close to $\langle \chi \rangle$ of the dimer alone. 

In the following, in Sec.~\ref{sec-purple} we consider a dimer surrounded by the photosynthetic antenna complex of purple bacteria  that thrive in very low light intensity~\cite{schulten:1,schulten:2,schulten3,kassal,kassal2}. 
Moreover, in Sec.~\ref{sec-GSB} a green sulfur bacteria antenna complex is used to increase the pumping of the dimer $\ket{H}$ state even further.

\subsection{Purple bacteria}
\label{sec-purple}

Purple bacteria feature a hierarchical structure of symmetric molecular aggregates which absorb light and direct the collected energy to the specific molecular aggregate of the reaction center (RC). The purple bacteria RC contains a bacteriochlorophyll (BChl) dimer called the special pair, and is surrounded by an LHI (Light-Harvesting system I) ring comprising 32 BChl molecules. 
The LHI ring is a J-aggregate with two superradiant states at $875$ nm that are polarized in the ring plane and close to the lowest excitonic state. The LHI ring is surrounded by several LHII rings, each featuring the B850 ring, a J-aggregate composed of 18 BChl molecules with two superradiant states at $850$ nm, and the B800 ring composed of 9 BChl molecules with main absorption peak at $800$~nm. This hierarchical structure is able to absorb photons at different frequencies and guide the collected energy down an energetic funnel to the RC through a process dubbed supertransfer, resulting from the coupling between the LHII-LHI superradiant states and the LHI-RC aggregates \cite{kassal, kassal2, schulten:1,schulten:2}.

We here propose a hybrid structure,  substituting the RC in the purple bacteria  with an H-dimer whose $\ket{H}$ state is resonant with the superradiant states of LHI. Photons absorbed by the LHI ring would then contribute to the pumping of the dimer's $\ket{H}$ state, with the strong coupling between $\ket{H}$ and the bright state of the LHI complex ensuring fast transfer. The geometrical arrangement for this envisioned aggregate is shown in Fig.~\ref{dip2}b~\cite{schulten1:1,schulten1:2,schulten2,schulten3}.
Specifically, at the centre of the LHI we place a homo-dimer with its two optical dipoles separated by 8~\AA\ (similar to BChls in the special pair of purple bacteria reaction centers~\cite{schulten:1,schulten:2}). We consider two transition dipoles that have an equal projection $\cos \theta$ onto the plane defined by the LHI ring, and opposite components $\pm \sin \theta$ orthogonal to the LHI plane. Controlling $\theta$ and keeping the coupling between the dimer molecules fixed, we can change the dimer brightness $\langle \chi \rangle$, so that we go from an H-dimer ($\theta \approx 0$, $\langle \chi \rangle \approx 0$) to a J-dimer ($\theta \approx \pi/2$, $\langle \chi \rangle \approx 1$). 
We chose the excitation frequency of the molecules in the dimer to be $1.17$~eV corresponding to $\approx 1060$~nm, which is in the near infrared wavelength. Since the coupling in the dimer is 2000~cm$^{-1}$ the dimer $\ket{H}$ state has an excitation wavelength of about $875$~nm at resonance with the superradiant states of the LHI aggregate. This choice ensures a large supertransfer coupling  and fast thermal relaxation between the LHI and the dimer. Indeed due to the symmetric arrangement (the dimer is at the center of the LHI ring) the coupling between the $\ket{H}$ state of the dimer and the superradiant states of the LHI ring will be enhanced by a factor $\approx \sqrt{32}$~\cite{schulten:1}.

To model the purple bacteria antenna, we took the structure of LHI ring from Ref.~\cite{schulten:2,schulten3} and the structure of the LHII ring from Ref.~\cite{kassal2}. 
The antenna complex molecular aggregates is  described by the Hamiltonian in Eq.~\eqref{ham}, where the nearest-neighbour couplings are replaced by the values reported in Table~\ref{tab-par}. Other relevant model parameters are also reported in Table~\ref{tab-par}, while the complete list of positions and dipole moments for each molecule can be found in the SM.
We obtain the eigenvalues and the TDMs of all the energy states by direct Hamiltonian diagonalization, which allows evaluating the lasing equations (\ref{Leq2}) under the already discussed assumptions of negligible nonradiative losses and fast thermal relaxation. The latter is valid in this aggregate owing to supertransfer throughout the aggregate
which entails thermal relaxation on the order of tens of picoseconds~\cite{schulten1:1,schulten1:2}: this is much faster than optical pumping, which ranges from a few nanoseconds (large aggregates, high black-body temperature) down to milliseconds (small aggregates, natural sunlight) and spontaneous decay, of the order of a nanosecond for the brightest states. Moreover, other relevant timescales  are the transition rate due to the coupling to the lasing cavity field which we estimate to be larger than hundreds of picoseconds (for the parameters considered here), and the realistic extraction rate $\kappa$ from the cavity of about three nanoseconds which we considered. In summary, thermal relaxation is clearly the fastest process, justifying the use of Eq.~\eqref{Leq2} for analyzing the lasing response of such bio-inspired aggregates. Moreover, in the SM, results of incoherent laser equations, see Eqs.~(S60) in SM, which do not assume quasi-instantaneous thermalization, are shown to be in excellent agreement with Eq.~\eqref{Leq2}.

\begin{table}[!tb]
    \centering
    \begin{tabular}{ccc}
    \toprule
    Subunit & Site energy [cm$^{-1}$]   & Nearest-neighbor coupling [cm$^{-1}$] \\
    \midrule
    LHII (B850) & 12532, 12728 (alternate)    & 363, 320 (alternate)~\cite{kassal2} \\
    LHII (B800) & 12555 & dipole-dipole \\
    LHI & 12911 & 806, 377 (alternate)~\cite{schulten:1,schulten:2} \\
    Dimer   & 9437  & 2000 \\
    \midrule
    \midrule
    Transition dipole & 10.157~D~\cite{schulten:1,schulten:2} & \\
    \bottomrule
    \end{tabular}
    \caption{Parameters for the aggregate Hamiltonian. The site energies are set to match the main fluorescence peaks at 800~nm (B800), 850~nm (B850) and 875~nm (LHI).}
    \label{tab-par}
\end{table}

The calculated lasing intensity and output power for a disordered ensemble of such aggregates (LHI+H-dimer) is shown in Fig.~\ref{dip2}c for black-body cavity pumping, and in Fig.~\ref{dip2}d for natural sunlight illumination.
Comparing Fig.~\ref{dip}c-d with Fig.~\ref{dip2}c-d, the critical aggregate density to cross the lasing threshold is greatly lowered, and more importantly, the required level of the dimer darkness for the natural sunlight case is reduced by more than an order of magnitude. We obtain a further improvement when surrounding our LHI ring with eight LHII rings, see Fig.~\ref{dip2}e (a magnified version of the figure can be seen in the SM, see Fig.~S7).
The B800/850 LHII aggregate level structure next to that of the LHI and H-dimer  is shown in Fig.~\ref{dip2}a. As shown in Fig.~\ref{dip2}f-g this larger aggregate architecture achieves a further lowering of the lasing threshold, i.e.~increase of the critical value of $\langle \chi \rangle$ for the H-dimer below which lasing is possible. Interestingly, in Fig.~\ref{dip2}f, the addition of the rings has increased the effective pumping of the H-dimer to the extent where the dimer no longer has to feature a dark state to lase: indeed in this case lasing is possible even for $\langle \chi \rangle=1$.

\begin{figure}[!tbh]
    \centering
    \includegraphics[width=\linewidth]{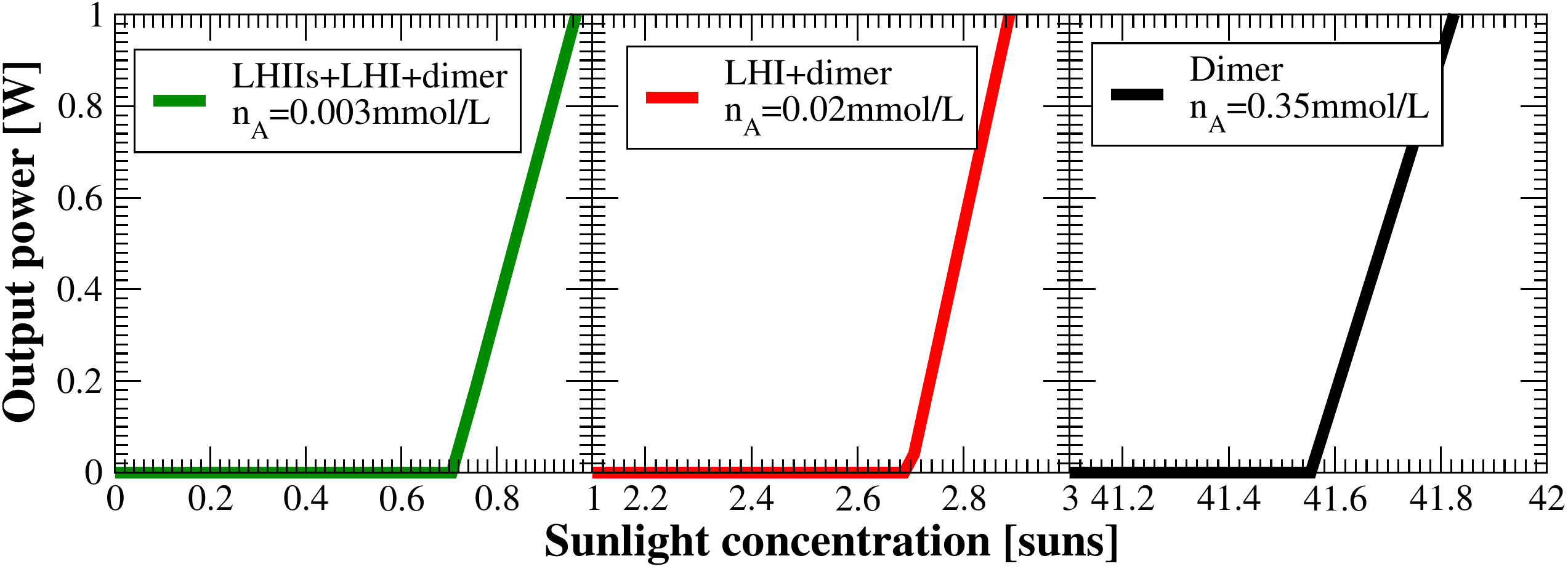}
    \caption{Output laser power, see Eq.~\eqref{powerint}, as a function of the sunlight concentration. Each panel refers to a different configuration, see figure. The sunlight concentration is changed by varying the factor $f_S$, where the value in Eq.~\eqref{fac1Sun} represents `1 sun' in this figure. The dimer brightness is $\braket{\chi}=5\times10^{-5}$ in all panels, while the aggregate density $n_A$ is changed in order to keep the same density of molecules, $Nn_A\approx0.7$~mmol/L (here $N$ is the number of molecules per aggregate: $N=2$ for the dimer, $N=34$ for the LHI+dimer and $N=250$ for the antenna complex + dimer). Other parameters: $\mu=10.157$~D, $\hbar\omega_A=1.17$~eV, $\Omega=2000$~cm$^{-1}$, $\Gamma_\phi=1/(10$~ps$)$, $\kappa/(2\pi)=50$~MHz, and $V=11.3$~cm$^3$.}
    \label{fig:powersuns}
\end{figure}

We have established that surrounding an H-dimer with purple bacteria LHI and LHII rings  not only lowers the necessary threshold density but also enables lasing with much less dark H-dimers.
To better understand and visualise these trends, we map out the lasing transition as a function of threshold  density and average brightness of the H-dimer for natural sunlight pumping in Fig.~\ref{dip2}h~\footnote{The case of a black body pumping at $T_{BB}=3000$~K is discussed in the SM.}.
This uses an H-dimer system described by Eqs.~\eqref{nA} and \eqref{Deq} but with  effective pumping $R_u$ increased by the factor $N/2$, where $N$ is the number of molecules in the bio-inspired aggregate including the H-dimer. This is a simplified approach of approximating enhanced pumping compared to the approach above based on the known TDMs of LHI/LHII states. Interestingly, the resulting threshold lines in  Fig.~\ref{dip2}h perfectly reproduce those obtained in the presence of the whole aggregate (symbols). 
This confirms that the crucial role of adding bio-inspired aggregates is  to increase the effective pumping rate. Moreover, it suggests that our proposed architecture is scalable and an even lower lasing threshold could be achieved with larger J-aggregates surrounding the H-dimer, as we discuss in Sec.~\ref{sec-GSB}, where we consider another bio-inspired architecture where the antenna complex of the green sulfur bacteria pumps the dimer $\ket{H}$ state. 
Nevertheless, caution is necessary when  applying the lasing equations derived here to very large aggregates, where the assumption of thermalization occurring on the fastest relevant timescale can become invalid, in which case an incoherent laser equation might be more appropriate as discussed in the following.

The advantage of our proposed bio-inspired lasing medium over H-dimers can be also seen in terms of the amount of sunlight needed to reach the lasing threshold for a given dimer darkness. To illustrate this point, in Fig.~\ref{fig:powersuns} we show the output laser power as a function of the sunlight concentration, that is determined by the factor $f_S$: the value in Eq.~\eqref{fac1Sun} represents a concentration of `1 sun', while we re-scale $f_S \to xf_S$ to reproduce `$x$ suns'. The density of molecules in the cavity is kept fixed, $Nn_A=0.7$~mmol/L, where $N$ is the number of molecules per aggregate, while $n_A$ is the aggregate density. The latter has to change with the number of molecules if we want to keep $N n_A$ fixed. Indeed, the dimer counts as two molecules, LHI ring contains 32 molecules and each LHII ring contains $18+9=27$ molecules. The dimer brightness is also kept fixed at $\braket{\chi}=5\times 10^{-5}$: at this value, lasing under natural sunlight (`1 sun') is possible only for the full purple bacteria antenna complex (compare Fig.~\ref{dip2}g with Figs.~\ref{dip}d-\ref{dip2}d). Here, Fig.~\ref{fig:powersuns} shows that a minimal sunlight concentration is needed to reach the lasing threshold, confirming the advantage of our proposed lasing medium: in fact, while for the isolated dimer a sunlight concentration of more than 41~suns is needed (see Fig.~\ref{fig:powersuns}, right panel), this requirement drops to $\approx 2.7$~suns if each dimer is surrounded by a single LHI ring (central panel), and to less than 1~sun when the full purple bacteria antenna complex surrounds the dimer (left panel).

\subsection{Green sulfur bacteria}
\label{sec-GSB}

While in the previous subsection we considered a bio-inspired aggregate which mimics the architecture of purple bacteria antennae, here we consider a different bio-inspired aggregate which mimics the architecture of another natural photosynthetic complex: the green sulfur bacteria (GSB) antenna complex~\cite{GSB}. We are motivated in doing so since, as we pointed out above, the lasing efficiency of the bio-inspired aggregate increases with the number of molecules composing the aggregate; the GSB antenna photosynthetic complex is the largest and most efficient antenna complex present in nature~\cite{GSB}. Indeed, GSB antenna complexes can contain up to 250,000 BChl molecules. Even if the precise structure of the GSB antenna complex is not fully known and it can vary a lot in natural samples, one of the most important molecular structures present in GSB antennae are certainly constituted by self-aggregated BChl-c nanotubular structures~\cite{gulli}. Typically in GSB an ensemble of molecular structures (molecular nanotubes and lamellae) absorb sunlight and transfer excitation to a two-dimensional aggregate called the `baseplate', which lies below the nanotubular structures. FMO complexes, which transfer excitations from antennae  to  reaction centres are attached to the baseplate.

\begin{figure}[!htbp]
    \centering
    \begin{tabular}{ll}
    {\large \bf a} & {\large \bf b} \\
    \includegraphics[width=0.47\linewidth]{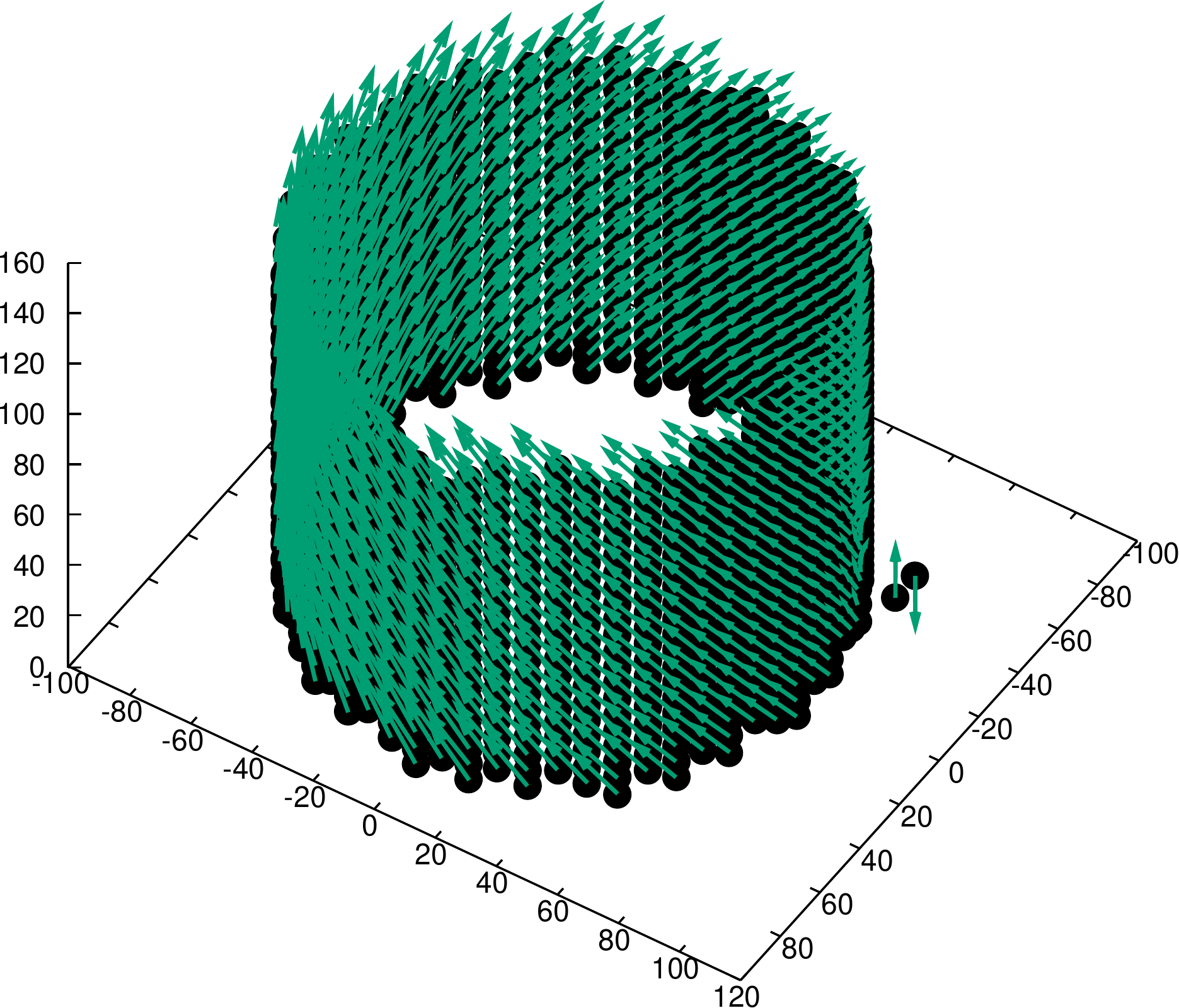} & \includegraphics[width=0.47\linewidth]{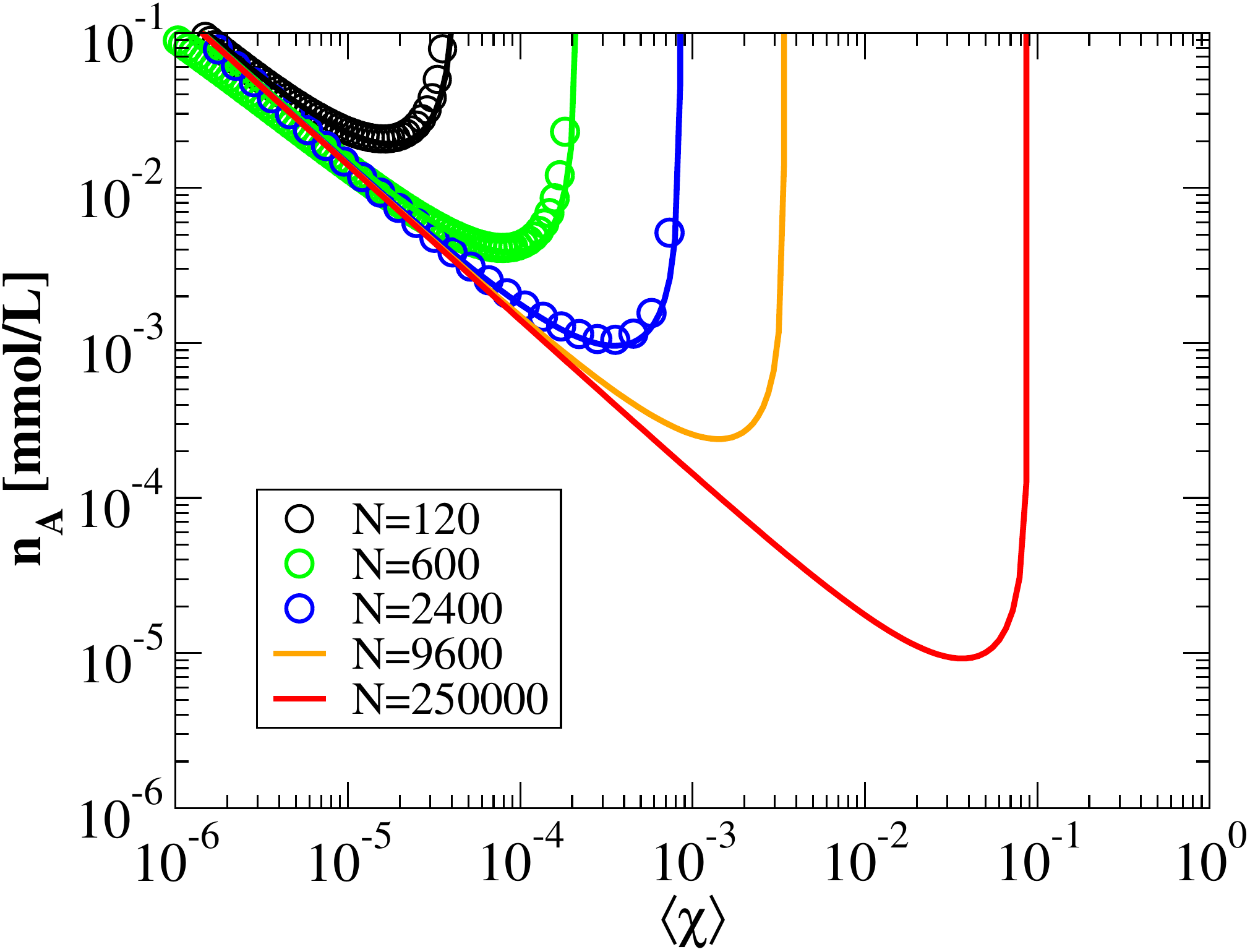}
    \end{tabular}
    \caption{{\bf (a)} Positions of the chromophores (circles) and transition dipole orientations (arrows) for the bio-mimetic complexes inspired to the GSB antenna complex considered in Section~\ref{sec-GSB}. Axes are in \AA\ units. Excitation energies, position, TDM orientation  for this bio-mimetic aggregate are taken from Refs.~\cite{Huh,Linnanto,gulli}. {\bf (b)} Lasing threshold density  under natural sunlight for a  dimer coupled to a green sulfur bacteria nanotube.  Symbols have been obtained by diagonalizing the whole  aggregate Hamiltonian and using  Eq.~\eqref{nA}. Continuous curves have been obtained using the lasing equation, Eq.~\eqref{nA}, for the dimer only  with a modified pumping  $R_u \rightarrow N R_u/2$, where $N$ is the total number of chromophores in the aggregate composed of a nanotube and a dimer.  System sizes up to $N=250,000$, corresponding to a realistic number of molecules in the whole green sulfur bacteria chlorosome~\cite{Huh,Linnanto,gulli}, have been considered. However, as we discuss in the text, aggregate sizes $N \gtrapprox 4000$ stretch the validity of the assumptions of our laser equations, and are thus of indicative rather than literal value.}
    \label{GSB-dip}
\end{figure}

Here we propose to enhance the pumping of the dimer $\ket{H}$ state by placing a natural nanotubular BChl-c aggregate close to the H-dimer as shown in Fig.~\ref{GSB-dip}a, where the positions and TDM orientations of the molecules composing the nanotubular structures and the dimers are shown. The nanotube structure is described in~\cite{gulli} and references therein. The nanotube  is composed by BChl-c molecules which have an excitation energy of 1.9~eV and a TDM of $\sqrt{30}$~D. The coupling between the BChl molecules in the nanotube produces a superradiant excitonic state state around 750~nm~\cite{gulli}. As for the dimer, we place it 3~nm from the wall of the nanotube (the same distance of the baseplate). The excitation frequency of the molecules composing the dimer (1.4~eV) and their coupling $2000$~cm$^{-1}$ are chosen so that the dimer $\ket{H}$ state is close in energy to the superradiant state of the nanotube. This configuration is considered here as an example of how increasing the number of chromophores in the antenna could enhance the lasing efficiency. In the future more sophisticated configurations, such as adding a baseplate and substituting the RC connected to the baseplate with H-dimers, could be analyzed. 

As was done for the purple bacteria, we diagonalize the appropriate Hamiltonian, as reported in Ref.~\cite{gulli}, to obtain the eigenvalues and the TDM of the eigenstates of the whole aggregate  and we then use Eq.~\eqref{nA} to obtain the threshold density of lasing as a function of the brightness $\langle\chi\rangle$ of the dimer under natural sunlight pumping. The results are shown in Fig.~\ref{GSB-dip}b, where  symbols refer to the threshold density  obtained from Eq.~\eqref{nA}. The threshold density is shown for different nanotube lengths which contain different number of molecules $N$, see Fig.~\ref{GSB-dip}. The continuous lines represent the density  threshold  of a dimer alone whose $R_u$ pumping factor is multiplied by $N/2$, where $N$ is the total number of molecules in the whole aggregate (nanotube plus dimer).

As one can see, the effect of this different bio-inspired aggregate is similar to the  Purple Bacteria molecular aggregate, which
is to increase the pumping of the dimer by a factor $N/2$. The curves corresponding to the largest aggregates sizes ($N=9600,250000$) have been obtained only using the dimer equations with enhanced pumping. The results presented are very promising since they indicate the possibility to achieve lasing under natural sunlight for relatively large  dimer brightness $\langle\chi\rangle \sim 10^{-1}$ (which is reachable experimentally~\cite{Hdimer:1}).

Notwithstanding the promising results in Fig.~\ref{GSB-dip}b for larger GSB aggregates, a word of caution is in order: as discussed in the following, applying our lasing equations to very large aggregates stretches the assumptions we have made in deriving Eqs.~\eqref{laser5}-\eqref{powerint}-\eqref{nA}-\eqref{Deq} beyond its strict regime of validity. This renders quantitative conclusions unreliable, however, we believe the extrapolation can nonetheless give an indicative picture of the expected qualitative trend. 

Specifically, we expect the implicit assumption that thermalization in the overall system Hamiltonian eigenbasis is the fastest timescale, to break for very large aggregates. On the one hand, it could be argued that disorder in large structures will lead to more localised states and the picture of completely delocalised eigenstates becomes questionable. On the other hand -- and neglecting disorder for now -- collective enhancements in the radiative rates will at some point change the hierarchy of timescales. 
To discuss this issue we now compare the fluorescence and the thermalization timescales: in large aggregates the dipole strength $|\mu_{SR}/\mu|^2$ of the superradiant state in the nanotube scales with the number of molecules as $|\mu_{SR}/\mu|^2 \approx  0.6 N$~\cite{gulli}. This means that the radiative decay time of the superradiant state is $\tau_{fl}=\tau_0/|\mu_{SR}/\mu|^2$, where $\tau_0\approx 30$~ns is the radiative decay time of a single molecule. The overall thermalization timescale is given by the thermalization on the nanotube followed by the excitation transfer time from the nanotube to the dimer. Since the latter can be enhanced by lowering the distance between the dimer and the nanotube, we will solely focus on the nanotube thermalization time in the following. An upper bound for the thermalization timescale $\tau_{th}$ can be obtained from the diffusion coefficient $D \approx 200~\mathrm{nm^2/ps}$ as estimated for the GSB nanotubes in~\cite{guzik-diff}, through the relationship $\tau_{th}= L^2/D$, where $L$ is the length of the nanotube. For the natural nanotube considered here we have $L \approx N(0.01~\mathrm{nm})$. The given estimates of thermalization and fluorescence times suggest that for a nanotube of $N \approx 4000$ molecules we obtain $\tau_{th} \approx \tau_{fl}$. We would therefore expect the laser equations derived in this manuscript to become quantitatively unreliable for $N \gtrapprox 4000$.

Adequately capturing the dynamics of larger aggregates with an explicit model is beyond the scope of the current work, and would require the derivation of lasing equations without the assumption of quasi-instantaneous vibrational thermalization,  while keeping the large dephasing assumption inducing incoherent cavity driving. This means that one should use incoherent laser equations for these larger aggregates similar to Eqs.~(S60), used in the SM. 

For the above reason, we only show circled data points for nanotubes up to $N=2400$ in Fig.~\ref{GSB-dip}b. Up to this point  the required lasing dimer threshold brightness approaches $10^{-3}$, and  we note that this value is still almost an order of magnitude better than the best result shown in the previous subsection using the purple bacteria bio-inspired aggregate (cf.~Fig.~\ref{dip}h). 
However, we believe the solid extrapolation curves shown in  Fig.~\ref{GSB-dip}b for larger $N$ will nevertheless capture the trend for longer nanotubes, not least since much larger photosynthetic aggregates are known to be highly efficient at channelling energy excitations over long distances to the reaction centers. This extrapolation suggests that larger photosynthetic structures can indeed further significantly lower the lasing threshold requirements in terms of molecular density and dimer brightness.

\section{Effects of nonradiative decay and quantum yield}

Here we investigate the effects on nonradiative decay on the lasing threshold. As for BChla building blocks of the antenna systems, nonradiative decay gives an individual BChla excitonic lifetime of about $1$~ns~\cite{schulten:2,kassal2}, which is much longer than the time needed to transfer the excitation to the central dimer (of order tens of picoseconds~\cite{schulten1:1,schulten1:2}). For this reason, when considering natural antenna complexes nonradiative effects can be safely neglected\footnote{If we were to include them, they would only mildly reduce the effective pumping rate.}. On the other hand, the effect of nonradiative decay on the H-dimer can be relevant, and we analyze it in more detail in the following. 

Nonradiative decay can be included in our model as an additional decay channel in the dimer, with a decay rate $\gamma_{nr}$. In practice, this is equivalent to replacing $R_d \to R_d + \gamma_{nr}$ in all our equations. We quantify the amount of nonradiative decay by means of the widely used emission quantum yield $\Phi= \gamma_r/(\gamma_r+\gamma_{nr})$, where $\gamma_r$ is the radiative decay rate. For an H-dimer, we have  $\gamma_r= \braket{\chi} \gamma_0$, see~Eq.~(\ref{Deq}), so that we have:
\begin{eqnarray}
\label{QY}
    \Phi = \frac{\braket{\chi}\gamma_0}{\braket{\chi}\gamma_0 + \gamma_{nr}}~,
\end{eqnarray}
where $\gamma_0$ is the radiative decay rate for each of the molecules composing the dimer.

A minimal quantum yield is necessary to obtain population inversion and, therefore lasing. In fact [see Eq.~\eqref{laser5} and the following calculations] we reach population inversion only if $R_u > R_d$, but since the total decay rate is $R_d = \braket{\chi}\gamma_0+\gamma_{nr}$, using Eq.~\eqref{QY} we have the condition
\begin{eqnarray}
\label{minQY}
    \Phi > \frac{\braket{\chi}\gamma_0}{R_u}~.
\end{eqnarray}
For natural sunlight and our choice of parameters for the dimer, we can approximate $R_u \approx 6.22 \times 10^{-7} N \gamma_0$, where $N$ is the number of molecules in the aggregate, so that Eq.~\eqref{minQY} becomes $\Phi > \braket{\chi} 10^7 / (6.22 N)$. Note also that $\Phi \leq 1$ by definition, so if the right-hand side of Eq.~\eqref{minQY} is larger than unity, lasing is not possible.

\begin{figure}
    \centering
    \includegraphics[width=0.5\linewidth]{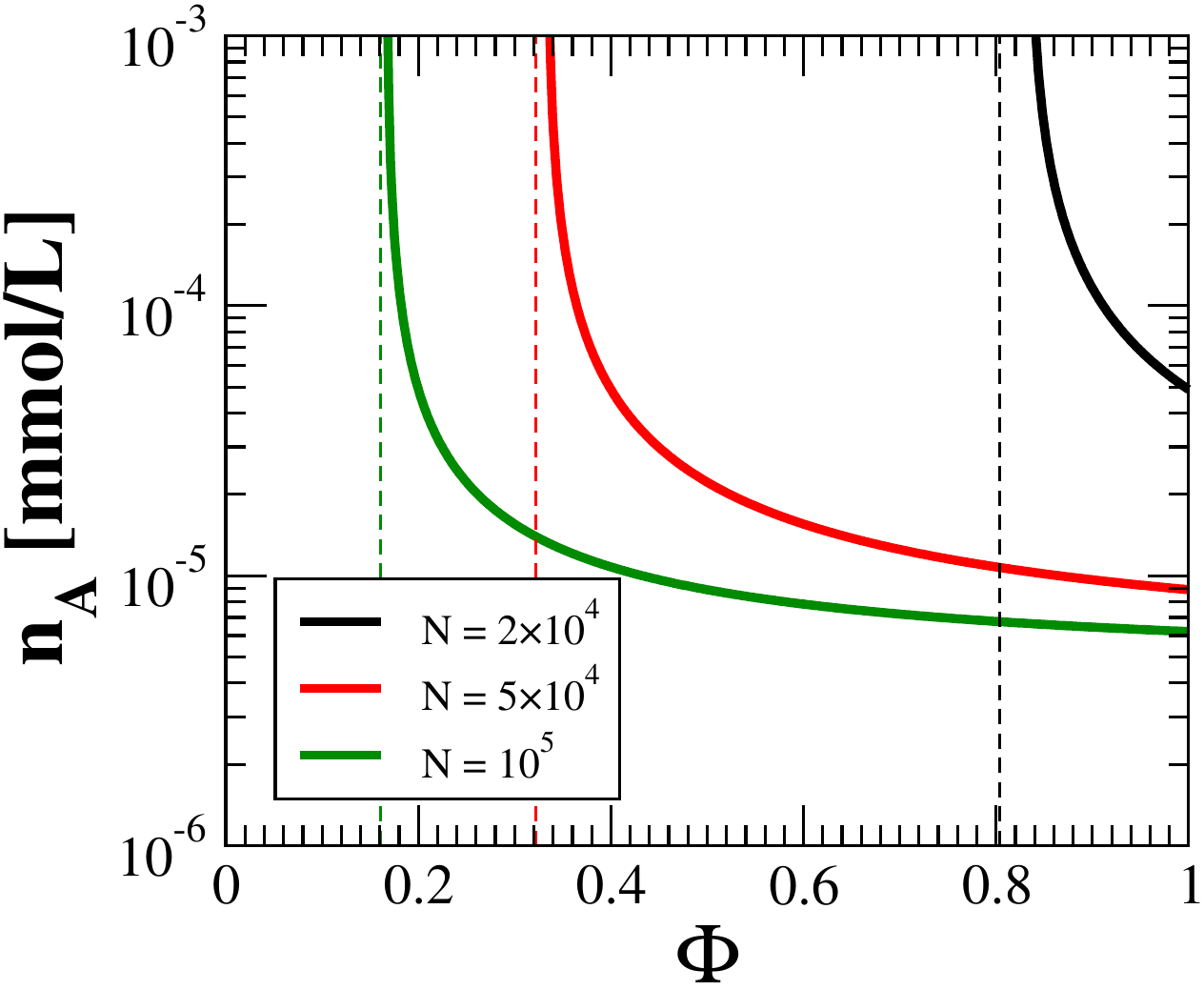}
    \caption{Threshold density, see Eq.~\eqref{nA}, as a function of the emission quantum yield $\Phi$, see Eq.~\eqref{QY}, for a dimer under natural sunlight, with relative brightness $\braket{\chi}=0.01$. Here we re-scale the absorption rate $R_u$ by a factor $N/2$ to reproduce the effect of surrounding the dimer with an aggregate of $N-2$ molecules, as discussed in the text. Different values of $N$ are considered, see figure. The vertical dashed lines represent the minimal quantum yield, see Eq.~\eqref{minQY}. Other parameters: $\hbar\omega_A=1.17$~eV, $\Omega=2000$~cm$^{-1}$, $\Gamma_\phi=1/(10$~ps$)$ and $\kappa/(2\pi)=50$~MHz.}
    \label{fig-QY}
\end{figure}

In Fig.~\ref{fig-QY} the dependence of the threshold density $n_A$ (continuous lines) on the quantum yield $\Phi$ is shown, under natural sunlight. Data in Fig.~\ref{fig-QY} have been obtained  for a dimer brightness $\braket{\chi}=0.01$, which is consistent with  literature~\cite{Hdimer:2}.  We increase the pumping by a factor $N/2$ to account for pumping by the surrounding aggregate, see discussion related to Figs.~\ref{dip2}h,\ref{GSB-dip}b. As one can see from Fig.~\ref{fig-QY}, the threshold density increases and diverges as the quantum yield approaches the minimal value from Eq.~\eqref{minQY}, see the dashed vertical lines. For these parameters, a large aggregate size $N>10^4$ is required to meet the lasing threshold under natural sunlight.
Indeed, as Fig.~\ref{fig-QY} shows, the minimal quantum yield is $\approx0.8$ for $N=2\times 10^4$, but it can be reduced to $\approx 0.3$ by increasing $N$ just by a factor of $2.5$. These values of the quantum yield are realistic for several H-dimers reported in literature~\cite{Hdimer:2}.

\section{Conclusions and perspectives}
Efficient sunlight-pumped lasers could revolutionize renewable energy technologies. Here we have shown  how lasing with natural sunlight pumping can be achieved by mimicking the architecture of photosynthetic antenna complexes.

We first considered an ensemble of molecular H-dimers inside an optical cavity pumped by black-body cavity radiation. When considering realistic values of the H-dimer darkness, lasing is possible for high black-body temperatures which can be achieved by heating up a cavity with concentrated sunlight. 
Nevertheless, lasing with H-dimers under natural sunlight would require a very high  level of darkness, which is very difficult to achieve. The main limitation  is due to the very weak pumping produced by natural sunlight. Larger molecular H-aggregates  might be able to lower the lasing threshold by absorbing more light and thus increasing the pumping. On the other hand, if not properly designed,  large aggregates have an increased density of states which would suppress the thermal population of the lasing state. This effect would compete with the advantage gained by more absorbed light. 
In order to increase the absorbed light without suppressing the population of the lasing state, one possibility is to consider an aggregate which absorbs light on a hierarchy of energy scales and which is able to efficiently funnel the absorbed energy  to a low energy lasing state which is well-gapped below other excitonic states. In this way, its thermal population will not be suppressed. These are precisely the features which characterize many natural antenna photosynthetic systems. 
Our proposed bio-inspired molecular aggregate serving as the lasing medium is composed of an H-dimer operating at low energy, surrounded by LHI and LHII rings of the purple bacteria antenna complex, which absorb at higher energy and efficiently funnel the absorbed energy to the H-dimer.  In this configuration, we show that lasing should be possible even under natural sunlight illumination. 

For a single-mode lasing cavity we can expect an output spectrum with properties similar to that of other organic dye lasers, that is well approximated by a Lorentzian with a width determined by the cavity finesse.
The specific bio-inspired parameters we have analysed would implement a short wavelength infrared laser, which has the advantage of being able to efficiently distribute converted solar energy due to the low dispersion in this wavelength range~\cite{SWNIR}. This is why this spectral regime is used in optical fiber communications. 
As an interesting prospect, our bio-mimetic molecular aggregates should be able to lase in nanocavities with volume of $(\lambda/20)^3$~\cite{nanocavity} and could thus be engineered into sunlight-pumped nanolasers~\cite{nanolaser}.
Indeed, we have shown that the required densities for achieving lasing under natural sunlight are about $n_A \approx 3~\mu$mol/L, meaning about one dimer surrounded by a purple bacteria antenna composed of an LHI and 8 LHII (reaching a length of 15~nm) every 80~nm. This density is thus achievable in a single nanocavity. 
We also note that purple bacteria antenna complexes are naturally arranged on spherical shells, each about 70~nm in diameter and containing $\approx 30$~RCs~\cite{schulten:2}.
Thus in this spherical shell each RC is surrounded by about one/two hundreds of chlorophyll molecules, which is the same order of magnitude considered in this manuscript for the largest purple bacteria antenna complex analyzed (1 LHI and 8 LHII), see Fig.~\ref{dip2}e.
If we replaced all the RCs with H-dimers, the volume of a single spherical shell would contain a dimer density of 0.3~mmol/L, that is 100 times higher than required for lasing, see Fig.~\ref{dip2}g. Therefore, placing such spherical shells in a cavity with a volumetric density of 1 part per 100, would still be enough to achieve lasing under natural sunlight illumination, provided that the H-dimers are dark enough.

Our idea can be generalized to other bio-inspired molecular architectures where molecular (J-)aggregates efficiently pump the bright state of a homo-dimer; this would allow lasing in other spectral regimes, and utilising other photosynthetic systems, e.g.~the chlorosome of GSB \cite{guzikGS} as already discussed here, or photosynthetic membranes such as in Photosystem II \cite{AmarnathPSII} to feed excitations into $\ket{H}$ and make them available for the lasing transition. Biomimetic antennae with controllable spacing between pigments~\cite{lindsey2015} might offer further avenues for improving and tuning pumping efficiency.
Even in its current form, however, 
our proposed biomimetic architecture is sufficiently efficient at collecting light from weak sources that it holds the potential to serve as a spring-board architecture for other bio-mimetic quantum devices, including photon sensing, improved solar cells, or even quantum batteries.

\section*{Author contributions}
G.L.C. and F.M. conceived the study. N.P. and S.O. developed the semi-classical laser theory. G.L.C. and F.M. developed the incoherent laser equations  and derived the  laser equations presented in the main text. E.G. and W.B. developed the Bloch-Redfield master equation and the coherent laser equations. F.M. and W.B. did most of the numerical simulations. E.G. and G.L.C. directed the research. All authors contributed to writing of the manuscript.

\ack 
We thank Alessia Valzelli for her help on the green sulfur bacteria antenna system. G.L.C and F.M.~acknowledge Fausto Borgonovi for useful discussions. W.M.B.~thanks the EPSRC (grant no. EP/L015110/1) for support.
S.O.~acknowledges the support from the MAECI project PGR06314-ENYGMA. E.M.G.~acknowledges  support  from  the  Royal  Society  of  Edinburgh  and  Scottish  Government  and  EPSRC  Grant  No. EP/T007214/1. G.L.C.~acknowledges the funding of ConaCyt Ciencia Basica project A1-S-22706.
N.P. acknowledges the support by the European Union (EU) Horizon 2020 program in the framework of the European Training Network ColOpt, grant agreement 721465.

\bibliography{main}

\end{document}


\title{Supplementary Material --
Bio-inspired natural sunlight-pumped lasers}

\author{Francesco Mattiotti}
\affiliation{ISIS (UMR 7006) and icFRC, University of Strasbourg and CNRS, 67000 Strasbourg, France}
\affiliation{Department of Physics, University of Notre Dame, Notre Dame, IN 46556, USA}
\affiliation{Dipartimento di Matematica e Fisica and Interdisciplinary Laboratories for Advanced Materials Physics, Universit\`a Cattolica, via Musei 41, Brescia I-25121, Italy}
\affiliation{Istituto Nazionale di Fisica Nucleare, Sezione di Pavia, via Bassi 6, Pavia I-27100, Italy}
\author{William M. Brown}
\affiliation{%
 SUPA, Institute of Photonics and Quantum Sciences, Heriot-Watt University, Edinburgh, EH14 4AS, United Kingdom
}
\author{Nicola Piovella}
\affiliation{%
 Dipartimento di Fisica ``Aldo Pontremoli'', Università degli Studi di Milano, via~Celoria 16, Milano I-20133, Italy
}
\affiliation{%
 Istituto Nazionale di Fisica Nucleare, Sezione di Milano, via Celoria 16, Milano I-20133, Italy
}
\author{Stefano Olivares}
\affiliation{%
 Dipartimento di Fisica ``Aldo Pontremoli'', Università degli Studi di Milano, via~Celoria 16, Milano I-20133, Italy
}
\affiliation{%
 Istituto Nazionale di Fisica Nucleare, Sezione di Milano, via Celoria 16, Milano I-20133, Italy
}
\author{Erik M. Gauger}%
\affiliation{%
 SUPA, Institute of Photonics and Quantum Sciences, Heriot-Watt University, Edinburgh, EH14 4AS, United Kingdom
}
\author{G. Luca Celardo}%
\affiliation{%
 Benem\'erita Universidad Aut\'onoma de Puebla, Apartado Postal J-48, Instituto de F\'isica,  72570, Mexico
}

\maketitle

\tableofcontents

\newpage

\section{Introduction}
\label{sec-intro}

\begin{table}[!b]
    \centering
    \begin{tabular}{cccccc}
        \toprule
        \multirow{3.5}{*}{Description} & \multirow{3.5}{*}{Reference} & \multicolumn{4}{c}{Approximation} \\
        \cmidrule{3-6}
        & & Secular & \begin{tabular}{c}
            Single \\
            excitation
        \end{tabular} & \begin{tabular}{c}
            Thermal \\
            equilibrium
        \end{tabular} & \begin{tabular}{c}
            Incoherent \\
            cavity rates
        \end{tabular} \\
        \midrule
        B-R coherently coupled to field & Eqs.~\eqref{mas2}-\eqref{field2} & & & & \\
        Incoherent laser equations (dimer) & Eqs.~\eqref{maslindcav} & \checkmark & & & \checkmark \\
        Incoherent laser equations (aggregate) & Eqs.~\eqref{ILE} & \checkmark & \checkmark & & \checkmark \\
        Laser equations & Eqs.~(5-7) main text & \checkmark & \checkmark & \checkmark & \checkmark \\
        \bottomrule
    \end{tabular}
    \caption{Different laser equations used in the main text and in this SM, and approximations used in each case (see Sec.~\ref{sec-intro} for more details).}
    \label{tab-app}
\end{table}

In this document, we provide more technical detail and complementary supporting calculations for the main text.
To guide the reader we give below a detailed description of each section with its connections to the main text.

Specifically, in Sec.~\ref{sec-geo} we show an explicit example of a dimer system with the required properties for serving as a laser gain medium.
In Sec.~\ref{sec-BR} we briefly derive a Markovian, non-secular Bloch-Redfield (B-R) master equation that describes the coupling to thermal light (black-body radiation) and to the phonon environment, see Eq.~\eqref{mas} and also Eq.~(2) in the main text. Under such B-R master equation, the time evolution of populations and coherences is coupled. Note that our approach has been already used in literature to describe sunlight absorption and phonon relaxation~\cite{brumer2017,scully1,scully2}.
Further, in Sec.~\ref{sec-BRLind} we apply the secular approximation to the B-R master equation, obtaining a Lindblad master equation, see Eqs.~\eqref{maslind}, where the populations are decoupled from the coherences. This approximation is commonly used in the literature to describe the absorption of thermal light~\cite{whaley2018,AndreasSun} in regimes where the absorption-induced coherences are irrelevant.
Then, in Sec.~\ref{sec-popinv} we consider a dimer interacting with a phonon bath at $T=300$~K and black-body radiation at different temperatures by taking into account both the single and double dimer excitation manifolds, using the B-R master equation. The results for the stationary dimer population obtained from the B-R master equation are compared with the expression derived in the main text, see Eq.~(9), which has been obtained using the following assumptions: (i) secular approximation for the photon and phonon bath; (ii) only the single excitation manifold is considered; (iii) thermal equilibrium within the single-excitation subspace. Under these three approximation  we show that Eq.~(9) in the main text  reproduces very well the results of the B-R non-secular master equation~\eqref{mas}, see Fig.~\ref{DimerBrightCoupPopDiffSuppFINv4} for a dimer under natural sunlight and $T_{BB}=3000$~K pumping.

In Sec.~\ref{sec-las} we report additional details on the derivation of the laser equations, Eqs.~(5) in the main text, that have been obtained assuming: (a) the secular approximation, (b) no more than one excitation present in the system, (c) thermal equilibrium within the single-excitation manifold, (d) incoherent transfer rates between the lasing medium and the cavity. In Secs.~\ref{sec-cav}-\ref{sec-BRrate} we discuss the validity of those approximations for a dimer in the lasing regime. To do so, we derive some laser equations under different degrees of approximation, see Table~\ref{tab-app} for an overview, and we compare the results obtained from the different approaches.
Specifically, in Sec.~\ref{sec-cav} we include a coherent coupling to a semi-classical cavity mode into the Hamiltonian, see Eq.~\eqref{eqn:dimWITHf}.
Then, in Sec.~\ref{sec-deph} we obtain the non-secular B-R master equation, Eq.~\eqref{mas2}, including the coherent coupling to the cavity mode in the Hamiltonian and adding the cavity dephasing.
In Sec.~\ref{sec-4levBR} we derive a semi-classical equation for the cavity field amplitude, see Eq.~\eqref{field2}, that, coupled to the B-R master equation Eq.~\eqref{mas2}, forms a closed set of dynamical equations for the dimer and the field. Note that Eqs.~\eqref{mas2} and \eqref{field2} have been derived without applying any of the approximations (a-d) mentioned above, see Table~\ref{tab-app}.
In Sec.~\ref{sec-BRstat} we consider the validity of the single-excitation approximation (b) using just the B-R master equation Eq.~\eqref{mas2}, for a dimer coupled to a static field of different intensities. Fig.~\ref{fig:StaticField} shows that the population of the double-excited state is negligible for the values of intensity and dephasing considered in the main text.
In Sec.~\ref{sec-rates} we derive incoherent transfer rates between the cavity and the lasing medium, that are valid under strong dephasing.
Further, in Sec.~\ref{sec-rateeq} we perform two of the approximations mentioned above: (a) the secular approximation and (d) we use the incoherent transfer rates derived in the previous section. Under those approximations we derive a set of incoherent laser equations for a dimer coupled to a cavity, Eqs.~\eqref{maslindcav}. These equations include the double-excited state and they do not assume thermal equilibrium in the single-excitation subspace, see Table~\ref{tab-app}.
Finally, in Sec.~\ref{sec-BRrate} we show the validity of the approximations (a-d) which led us to Eqs.~(5-7) of the main text, by comparing the three different approaches mention above, namely: (1)  non-secular B-R master equation coupled to the field equation, Eqs.~\eqref{mas2}-\eqref{field2}; (2) incoherent laser equations, Eqs.~\eqref{maslindcav}; (3) laser Eqs.~(5-7) of the main text, see Table~\ref{tab-app}.  The results in Fig.~\ref{dyncav} for a dimer under $T_{BB}=3000$~K photon pumping show that the three approaches give the same results at the steady state, for realistic values of dephasing. Fig.~\ref{dyncav} also shows that the population of the double-excited state remains negligible in the lasing regime and, most importantly, it confirms the validity of the approximations (a-d) that lead to the laser equations Eqs.~(5-7) of the main text for the parameters that we considered.

The last two sections of this SM contain supplemental details and data about the Purple Bacteria antenna complex. Specifically, in Sec.~\ref{sec-purp} we report the detailed positions and dipole orientations of the chromophores, see Fig.~\ref{fig-posdip} and Table~\ref{tab-posdip}. In the same section we also generalize the incoherent laser equations for a dimer, Eqs.~\eqref{maslindcav}, to the case of a larger aggregate, see Eqs.~\eqref{ILE}: in such case, we also employ the single-excitation approximation, see Table~\ref{tab-app}. Note that Eqs.~\eqref{ILE} are not assuming fast thermal relaxation, differently from the laser equations in the main text. Despite this important difference, as we show in Fig.~\ref{LHI-Dimer-I}, Eqs.~\eqref{ILE} agree (at the steady state) with the laser equations Eqs.~(5-7) of the main text, thus justifying the use of the thermal equilibrium approximation for large aggregates in the main text.
Finally, in Sec.~\ref{sec-3000purp} we report the threshold density as a function of the dimer brightness $\braket{\chi}$, for pumping at $T_{BB}=3000$~K, comparing the dimer alone, the dimer surrounded by the LHI ring and the dimer surrounded by the LHI ring and by the 8 LHII rings. Fig.~\ref{agg-comp} shows that the molecular aggregates surrounding the dimer effectively increase the pumping by a factor $N/2$ (where $N$ is the total number of chromophores in the aggregate, including the two dimer molecules), therefore lowering the lasing threshold. This effect is similar to what is shown in see Fig.~3h in the main text, for the case of natural sunlight, with the advantage that at $T_{BB}=3000$~K pumping, the large Purple Bacteria pushes the lasing threshold so much that the system can lase even if the $\ket{L}$ dimer state is not dark ($\braket{\chi}=1$).

\section{A possible geometry for an H-dimer}
\label{sec-geo}

Here we introduce a possible configuration for an H-dimer with dipolar coupling. This serves as a simple example for how to achieve the level configuration presented in Fig.~2a of the main text. 

\begin{figure}[!htbp]
  \centering
  \begin{tabular}{l}
    {\Huge \bf a} \\
    \includegraphics[width=0.5\columnwidth]{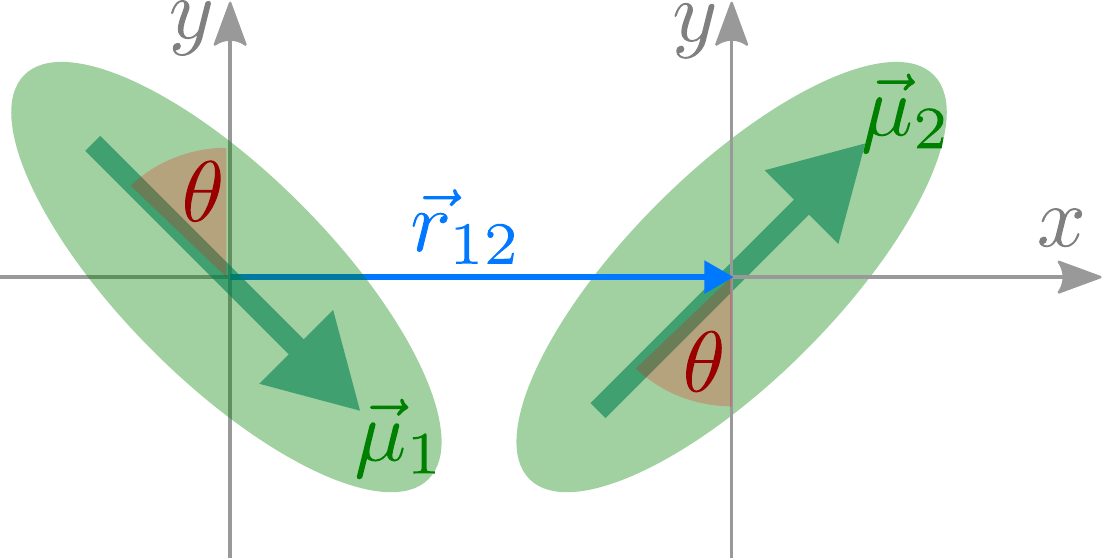} \\
    \vspace{5em} \\
    {\Huge \bf b} \\
    \includegraphics[width=0.5\columnwidth]{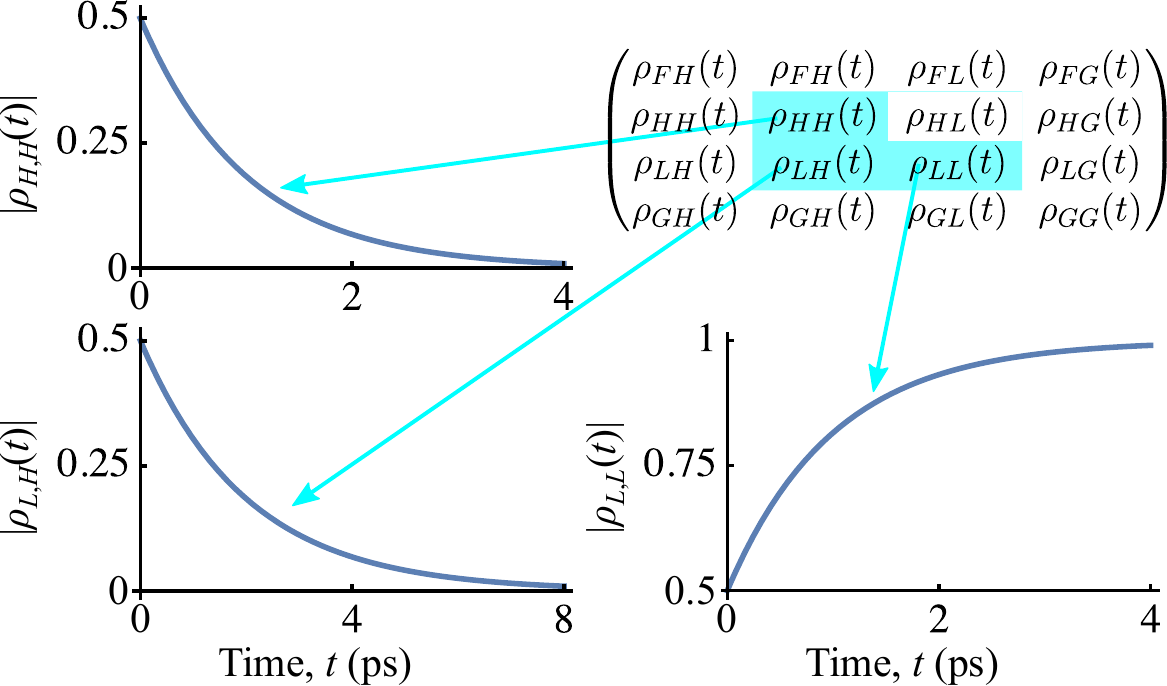}
  \end{tabular}
  \caption{{\bf (a)} Dimer geometry described in~Eq.~\eqref{dipgeo}. Each green ellipse represents a molecule. Orientation of the transition dipole moments (green arrows), vector joining the dipoles (blue arrow) and directions of the $x$, $y$ axes (grey arrows) and $\theta$ angle are indicated. {\bf (b)} Plots of the density matrix component evolution for the populations of, and coherence between $\ket{H}$ and $\ket{L}$. The dimer is only connected to independent phonon baths. The timescale for the relaxation of population in the single excitation manifold and for dephasing between $\ket{H}$ and $\ket{L}$ are both on the order of picoseconds. The dimer parameters are the same  from the main text: ($\hbar\omega_A=1.17$~eV and optical lifetime, $1/\gamma_0=36.8$~ns), with $\theta=0.07$~rad and  coupling $\Omega=$ 2000~cm$^{-1}$ (0.25~eV). The initial density matrix is given by~Eq.~\eqref{eqn:initial}.}
  \label{dip}
\end{figure}

Let us consider the orientation shown pictorially in Fig.~\ref{dip}a: the optical transition dipoles of two identical molecules lie in the $(x,y)$ plane and they have an opposite inclination angle $\theta$ with respect to the $x$ axis. The corresponding transition dipole moment (TDM) for each molecule can be expressed as
\begin{subequations}
\begin{align}
    \label{dipgeo}
    \vec{\mu}_1 &= \mu\sin \theta \hat{x}-\mu\cos \theta \hat{y}~,\\
    \vec{\mu}_2 &= \mu\sin \theta \hat{x} + \mu\cos \theta \hat{y}~.
\end{align}
\end{subequations}
The dipole-dipole coupling between the two molecules at distance $r_{12}$ is
\begin{equation}
  \Omega = -\frac{\mu^2}{r_{12}^3} \left( 1 + \cos^2\theta \right)~,
\end{equation}
and is negative for any value of $\theta$.
Note that the dipole coupling is valid only when the distance between the molecules is larger than the charge displacement producing the TDM. For smaller distances, the whole charge distribution should be taken into account to compute the coupling~\cite{dft-dipole}. For this reason in the main text and in the following sections of this supplementary material the coupling will be considered as an input parameter or a realistic value will be chosen.

Since the two molecules have the same excitation energy $\hbar\omega_A$ and the coupling $\Omega$ is negative, the eigenstates  and the corresponding eigenvalues (relative to the dimer ground state) are
\begin{subequations}
  \label{eig}
  \begin{align}
    \ket{G} &:= \ket{g_1}\ket{g_2}~, & E_G &= 0 ~,\\
    \ket{L} &:= \frac{\ket{g_1}\ket{e_2} + \ket{e_1}\ket{g_2}}{\sqrt{2}}~, & E_L &= \hbar\omega_A-|\Omega| ~, \\
    \ket{H} &:= \frac{\ket{g_1}\ket{e_2} - \ket{e_1}\ket{g_2}}{\sqrt{2}}~, & E_H &= \hbar\omega_A+|\Omega| ~, \\
    \ket{F} &:= \ket{e_1}\ket{e_2}~, & E_F &= 2\hbar\omega_A~.
  \end{align}
\end{subequations}

Optical transitions between the eigenstates are proportional to the TDM, defined as the matrix element of the dipole operator
\begin{equation}
  \label{mu}
  \hat{\vec{\mu}} = \sum_{j=1}^2 \vec{\mu}_j \left( \hat{\sigma}_j^+ + \hat{\sigma}_j^- \right) \, .
\end{equation}
The only non-vanishing elements of the TDM $\bra{k} \hat{\vec{\mu}} \ket{k'}$ between two generic dimer eigenstates $\ket{k}$ and $\ket{k'}$ are:
\begin{subequations}
  \label{mulh}
  \begin{align}
    \vec{\mu}_L &:= \bra{G}\hat{\vec{\mu}}\ket{L} = \bra{L}\hat{\vec{\mu}}\ket{F} = \frac{\vec{\mu}_2 + \vec{\mu}_1}{\sqrt{2}} = \sqrt{2} \mu \sin\theta \, \hat{x} ~, \\
    \vec{\mu}_H &:=\bra{G}\hat{\vec{\mu}}\ket{H} = \bra{H}\hat{\vec{\mu}}\ket{F} = \frac{\vec{\mu}_2 - \vec{\mu}_1}{\sqrt{2}} = \sqrt{2} \mu \cos\theta \, \hat{y}~.
  \end{align}
\end{subequations}
Note that TDMs obey the conservation law
$
  \mu_L^2 + \mu_H^2 = \mu_1^2 + \mu_2^2 = 2 \mu^2 ~
$.

\section{The Bloch-Redfield master equation }
\label{sec-BR}

Here we sketch the derivation of the master equation introduced in Eq.~(2)  of the main text. For convenience we reprint this master equation, which describes the interaction of a molecular aggregate with a collective photon  and individual phonon baths:
\begin{equation}
    \frac{d\hat{\rho}(t)}{dt} = -\frac{i}{\hbar} \left[ \hat{H}_S, \hat{\rho}(t) \right] + {\cal D}_{BB} [\hat{\rho}(t)] + {\cal D}_{T}[\hat{\rho}(t)] \label{mas}~,
\end{equation}
where ${\cal D}_{BB}$ and ${\cal D}_T$ are Bloch-Redfield dissipators for the coupling to photons and phonons, respectively. 

The photon dissipator is based on the assumption that the molecular absorbers are positioned close together relative to relevant optical wavelengths ($\sim2\pi c/\omega_A$), so that all dipoles interact with the same shared optical field via the following collective interaction Hamiltonian:
\begin{equation}
\hat{H}_{I,\text{opt}}=\sum_{i} \vec{\mu}_i\hat{\sigma}_i^x \otimes \sum_k f_k (\hat{a}_k+\hat{a}_k^\dagger) ~,
\label{eqn:INTopt}
\end{equation}
where $f_k$ and $\hat{a}_k^{(\dagger)}$ are, respectively, the coupling strength and annihilation (creation) operator for the optical mode $k$. We then couple local phonon baths to each 2LS with repeated spin-boson interaction Hamiltonians:
\begin{equation}
\hat{H}_{I,\text{vib}}=\sum_{i} \hat{\sigma}_i^z \otimes \sum_q g_{i,q} (\hat{b}_{i,q}+\hat{b}_{i,q}^\dagger) ~,
\label{eqn:INTvib}
\end{equation}
where $g_{i,q}$ and $\hat{b}_{i,q}^{(\dagger)}$ are, respectively, the coupling strength and annihilation (creation) operator for the vibrational mode $q$ for the bath linked to site $i$.

The optical and vibrational Bloch-Redfield dissipators both take the form~\cite{petru}
\begin{multline}
\mathcal{D}_{\alpha}=
\sum_{n,m}C_n C_m\Big(A_m(\omega_m)\rho_S(t)A^\dagger_n(\omega_n)\Gamma_{nm}(\omega_m) + A_n(\omega_n)\rho_S(t)A^\dagger_m(\omega_m) \Gamma^\dagger_{nm}(\omega_m) \\
- \rho_S(t)A^\dagger_m(\omega_m)A_n(\omega_n)\Gamma^\dagger_{nm}(\omega_m) -A^\dagger_n(\omega_n)A_m(\omega_m)\rho_S(t) \Gamma_{nm}(\omega_m)\Big) ~,
\label{eqn:BRdiss}
\end{multline}
where the $n,m$ summation is taken over all pairwise combinations of elements in each of the system interaction matrices from~Eq.~\eqref{eqn:INTopt} and~Eq.~\eqref{eqn:INTvib}, i.e.~the set of $\hat{\sigma}_i^x$
and $\hat{\sigma}_i^z$, respectively. The weighting terms, $C_{n}$, are determined by the transformation of these interaction matrices to the system eigenbasis (for example, the dipole contributions for transitions between different eigenstates in the optical case)~\cite{petru}. Only the real part of the spectral correlation tensor is kept, leading to a solely dissipative contribution to the dynamics. This means that Lamb shifts and any other Hamiltonian renormalisation effects are assumed to have already been accounted for when setting up $\hat{H}_S$ [Eq.~(1) of the main text]. Further, we assume a flat optical spectral density, yielding optical environment correlation functions
\begin{equation}
\Gamma_{nm}(\omega)=\frac{1}{2}\kappa_{\text{opt}}\omega^3\big(1+n(\omega)\big) ~,
\label{BR-phot-rate}
\end{equation}
where the prefactor $\kappa_{\text{opt}}$ is determined by the lifetime of an isolated 2LS, and $n(\omega)$ is the Bose-Einstein occupancy of modes, given by
\begin{equation}
n(\omega)=\frac{1}{e^{\beta\hbar\omega}-1} ~,
\label{eqn:thermPOP}
\end{equation}
with $\beta = 1 / k_B T_{BB}$, where $T_{BB}$ is the photon black-body bath temperature  and $k_B$ is Boltzmann's constant. For photon processes the transition frequencies will be of the order of single molecule excitation frequency $\omega_A$, and the factor $\omega^3$ arises from the density of modes~\cite{petru}.

For simplicity and in the absence of detailed information about the nature of the vibrational environment we also keep the phonon spectral density flat and equal at each site, this results in vibrational environment correlation terms of the form
\begin{equation}
\Gamma_{nm}(\omega)=\frac{1}{2}\kappa_{\text{vib}}\omega\big(1+n(\omega)\big) ~,
\label{BR-phon-rate}
\end{equation}
where the prefactor $\kappa_{\text{vib}}$ will be fixed by imposing a characteristic phonon spontaneous emission relaxation timescale of 1 picosecond. For instance for the dimer parameters we consider in the main text, $\Omega=2000$~cm$^{-1}=0.25$~eV, we set  $\kappa_\text{vib}=1.3\times 10^{-3}$, unitless to obtain a relaxation of the order of 1~ps between the $\ket{H}$ and $\ket{L}$ states of the dimer.  The Bose-Einstein occupation number $n(\omega)$ is here taken at room temperature (300~K). Phonon transition frequencies will be of order of the  coupling $\Omega$.
In Fig.~\ref{dip}b we show that the timescale of evolution for the populations of, and dephasing between, states connected by phonons, is indeed picoseconds in a dimer system identical to what was considered in the main text and coupled only to phonon baths with an initial state given by
\begin{equation}
\begin{pmatrix}
0 & 0 & 0 & 0 \\
0 & 0.5 & 0.5 & 0 \\
0 & 0.5 & 0.5 & 0 \\
0 & 0 & 0 & 0
\end{pmatrix}~.
\label{eqn:initial}
\end{equation}

\section{From Bloch-Redfield to Lindblad master equation}
\label{sec-BRLind}
The coupling to the photon bath is typically much weaker than the level spacing within the system. This allows the secular approximation to be applied to the Bloch-Redfield dissipator ${\cal D}_{BB}$, simplifying it to the Lindblad form~\cite{petru}. In the eigenbasis, with transitions labelled by $k$, this yields
\begin{align}
  \label{BBlind}
  {\cal D}_{BB} [\hat{\rho}] = &\sum_{k} \gamma_k(\omega) \left( 1 + n_T(\omega) \right) \left[ \hat{\sigma}_k^- \hat{\rho} \hat{\sigma}_k^+  -\frac{1}{2} \left\{ \hat{\sigma}_k^+ \hat{\sigma}_k^-, \hat{\rho} \right\} \right] \nonumber \\
  &+\sum_{k} \gamma_k(\omega) n_T(\omega)  \left[ \hat{\sigma}_k^+ \hat{\rho} \hat{\sigma}_k^-  -\frac{1}{2} \left\{ \hat{\sigma}_k^- \hat{\sigma}_k^+, \hat{\rho} \right\} \right]~,
\end{align}
where the optical rates are
\begin{equation}
    \label{opt-rate}
    \gamma_k(\omega) = \frac{\mu_k^2 \omega^3}{3 \pi \epsilon_0 \hbar c^3}~
\end{equation}
with the transition dipoles $\mu_k^2$ and  universal constants $\epsilon_0$ (vacuum permittivity), $\hbar$ (reduced Planck constant), and $c$ (speed of light).  We will now focus on a dimer system with  optical transition frequencies $\omega_L=\omega_A-|\Omega|/\hbar$, $\omega_H=\omega_A+|\Omega|/\hbar$ and with the eigenstates $\ket{G},\ket{L},\ket{H}$ and $\ket{F}$ defined in Eq.~\eqref{eig}. The thermal population of black-body photons at frequency $\omega$ is defined using~Eq.~\eqref{eqn:thermPOP}.
The curly brackets in~Eq.~\eqref{BBlind} indicate the anticommutator, and the raising/lowering operators $\hat{\sigma}_k^\pm$ in it act on the eigenstates as shown in the following scheme
\begin{align}
    &\ket{G}
    \begin{array}{cc}
         \xrightarrow{\quad \hat{\sigma}_L^+\quad}  \\
         \xleftarrow[\quad \hat{\sigma}_L^-\quad]{} 
    \end{array}
    \ket{L}
    \begin{array}{cc}
         \xrightarrow{\quad \hat{\sigma}_L^+\quad}  \\
         \xleftarrow[\quad \hat{\sigma}_L^-\quad]{} 
    \end{array}
    \ket{F} \nonumber \\
    &\ket{G}
    \begin{array}{cc}
         \xrightarrow{\quad \hat{\sigma}_H^+\quad}  \\
         \xleftarrow[\quad \hat{\sigma}_H^-\quad]{} 
    \end{array}
    \ket{H}
    \begin{array}{cc}
         \xrightarrow{\quad \hat{\sigma}_H^+\quad}  \\
         \xleftarrow[\quad \hat{\sigma}_H^-\quad]{} 
    \end{array}
    \ket{F} \, .
\end{align}
To make Eq.~\eqref{BBlind} more explicit, we first compute the transition rates between the populations of each eigenstate. We do that by projecting the black-body dissipator onto each eigenstate Eq.~\eqref{eig}, obtaining
\begin{subequations}
\begin{align}
    \bra{G} {\cal D}_{BB} [\hat{\rho}] \ket{G} = &\gamma_L(\omega_L) \left[ 1 + n_T(\omega_L) \right] \rho_{LL} + \gamma_H(\omega_H) \left[ 1 + n_T(\omega_H) \right] \rho_{HH} \nonumber \\
    &- \gamma_L(\omega_L) n_T(\omega_L) \rho_{GG} - \gamma_H(\omega_H) n_T(\omega_H) \rho_{GG} ~,\\
    \bra{L} {\cal D}_{BB} [\hat{\rho}] \ket{L} = &\gamma_L(\omega_H) \left[ 1 + n_T(\omega_H) \right] \rho_{FF} - \gamma_L(\omega_L) \left[ 1 + n_T(\omega_L) \right] \rho_{LL} \nonumber \\
    &+ \gamma_L(\omega_L) n_T(\omega_L) \rho_{GG} - \gamma_L(\omega_H) n_T(\omega_H) \rho_{LL}~, \\
    \bra{H} {\cal D}_{BB} [\hat{\rho}] \ket{H} = &\gamma_H(\omega_L) \left[ 1 + n_T(\omega_L) \right] \rho_{FF} - \gamma_H(\omega_H) \left[ 1 + n_T(\omega_H) \right] \rho_{HH} \nonumber \\
    &+ \gamma_H(\omega_H) n_T(\omega_H) \rho_{GG} - \gamma_H(\omega_L) n_T(\omega_L) \rho_{HH}~, \\
    \bra{F} {\cal D}_{BB} [\hat{\rho}] \ket{F} = &\gamma_L(\omega_H) n_T(\omega_H) \rho_{LL} + \gamma_H(\omega_L) n_T(\omega_L) \rho_{HH} \nonumber \\
    &-\gamma_L(\omega_H) \left[ 1 + n_T(\omega_H) \right] \rho_{FF} - \gamma_H(\omega_L) \left[ 1 + n_T(\omega_L) \right] \rho_{FF} ~.
\end{align}
\end{subequations}
As one can see, the dynamics of the populations are coupled only to the populations of the other states ($\rho_{kk}$) and not to the coherences ($\rho_{kk'}$ with $k\neq k'$). This allows to write a set of rate equations (i.e.~a Pauli master equation) for photon absorption and spontaneous emission for the populations of the four eigenstates.

Similarly, the dissipator describing the coupling of the dimer to a phonon bath can also be approximated in the Lindblad form and it produces dynamics where the populations are decoupled from the coherences. Specifically, for a dimer system, we have
\begin{subequations}
\begin{align}
    \bra{G} {\cal D}_T [\hat{\rho}] \ket{G} &= 0~, \\
    \bra{L} {\cal D}_T [\hat{\rho}] \ket{L} &= \Gamma_\Omega \left( 1 + n_\Omega \right) \rho_{HH} - \Gamma_\Omega n_\Omega \rho_{LL}~, \\
    \bra{H} {\cal D}_T [\hat{\rho}] \ket{H} &= -\Gamma_\Omega \left( 1 + n_\Omega \right) \rho_{HH} + \Gamma_\Omega n_\Omega \rho_{LL}~, \\
    \bra{F} {\cal D}_T [\hat{\rho}] \ket{F} &= 0~,
\end{align}
\end{subequations}
where the thermal relaxation rate $\Gamma_\Omega=2|\Omega|\kappa_{\rm vib}$ is proportional to the bath spectral density at the energy $2|\Omega|$ (with $\Omega$ being the coupling between the molecules) and the thermal population is defined with~Eq.~\eqref{eqn:thermPOP}, with $T=300$~K being the phonon temperature.

Combining the sunlight and phonon contributions, we have the rate equation for a dimer system
\begin{subequations}
\label{maslind}
\begin{align}
    \frac{d\rho_{GG}}{dt} = &\gamma_L(\omega_L) \left[ 1 + n_T(\omega_L) \right] \rho_{LL} + \gamma_H(\omega_H) \left[ 1 + n_T(\omega_H) \right] \rho_{HH} - \gamma_L(\omega_L) n_T(\omega_L) \rho_{GG} - \gamma_H(\omega_H) n_T(\omega_H) \rho_{GG}~, \\
    \frac{d\rho_{LL}}{dt} = &\gamma_L(\omega_H) \left[ 1 + n_T(\omega_H) \right] \rho_{FF} - \gamma_L(\omega_L) \left[ 1 + n_T(\omega_L) \right] \rho_{LL} + \gamma_L(\omega_L) n_T(\omega_L) \rho_{GG} - \gamma_L(\omega_H) n_T(\omega_H) \rho_{LL} \nonumber \\
    &+\Gamma_\Omega \left[ 1 + n_\Omega \right] \rho_{HH} - \Gamma_\Omega n_\Omega \rho_{LL}~, \\
    \frac{d\rho_{HH}}{dt} = &\gamma_H(\omega_L) \left[ 1 + n_T(\omega_L) \right] \rho_{FF} - \gamma_H(\omega_H) \left[ 1 + n_T(\omega_H) \right] \rho_{HH} + \gamma_H(\omega_H) n_T(\omega_H) \rho_{GG} - \gamma_H(\omega_L) n_T(\omega_L) \rho_{HH} \nonumber \\
    &-\Gamma_\Omega \left[ 1 + n_\Omega \right] \rho_{HH} + \Gamma_\Omega n_\Omega \rho_{LL} ~,\\
    \frac{d\rho_{FF}}{dt} = &-\gamma_L(\omega_H) \left[ 1 + n_T(\omega_H) \right] \rho_{FF} - \gamma_H(\omega_L) \left[ 1 + n_T(\omega_L) \right] \rho_{FF} + \gamma_L(\omega_H) n_T(\omega_H) \rho_{LL} + \gamma_H(\omega_L) n_T(\omega_L) \rho_{HH}~,
\end{align}
\end{subequations}
based on treating the influence of the environments on the dimer dynamics to second order and the Born-Markov approximation (both inherent in the Bloch-Redfield formalism), and the additional secular simplification of each dissipator to its Lindblad form.

We can compare the Bloch-Redfield approach (Eq.~\eqref{eqn:BRdiss}) -- solved numerically -- against the rate equations arising from the four-level Lindblad dissipators as presented above. For the parameter ranges we consider, we find the two results are consistent to machine precision. This is do be expected due to the large energy splitting we consider between the two single-excitation eigenstates (but we would expect the results to begin  diverging when considering near-degenerate terms from much smaller or vanishing coupling between 2LSs).
Note that the validity of the Lindblad master equation approximation for the populations of a molecular aggregate is the basis of our laser equations in the main text.

\section{Population inversion for a dimer}
\label{sec-popinv}

\begin{figure}[!hbtp]
    \centering
    \includegraphics[width=1\columnwidth]{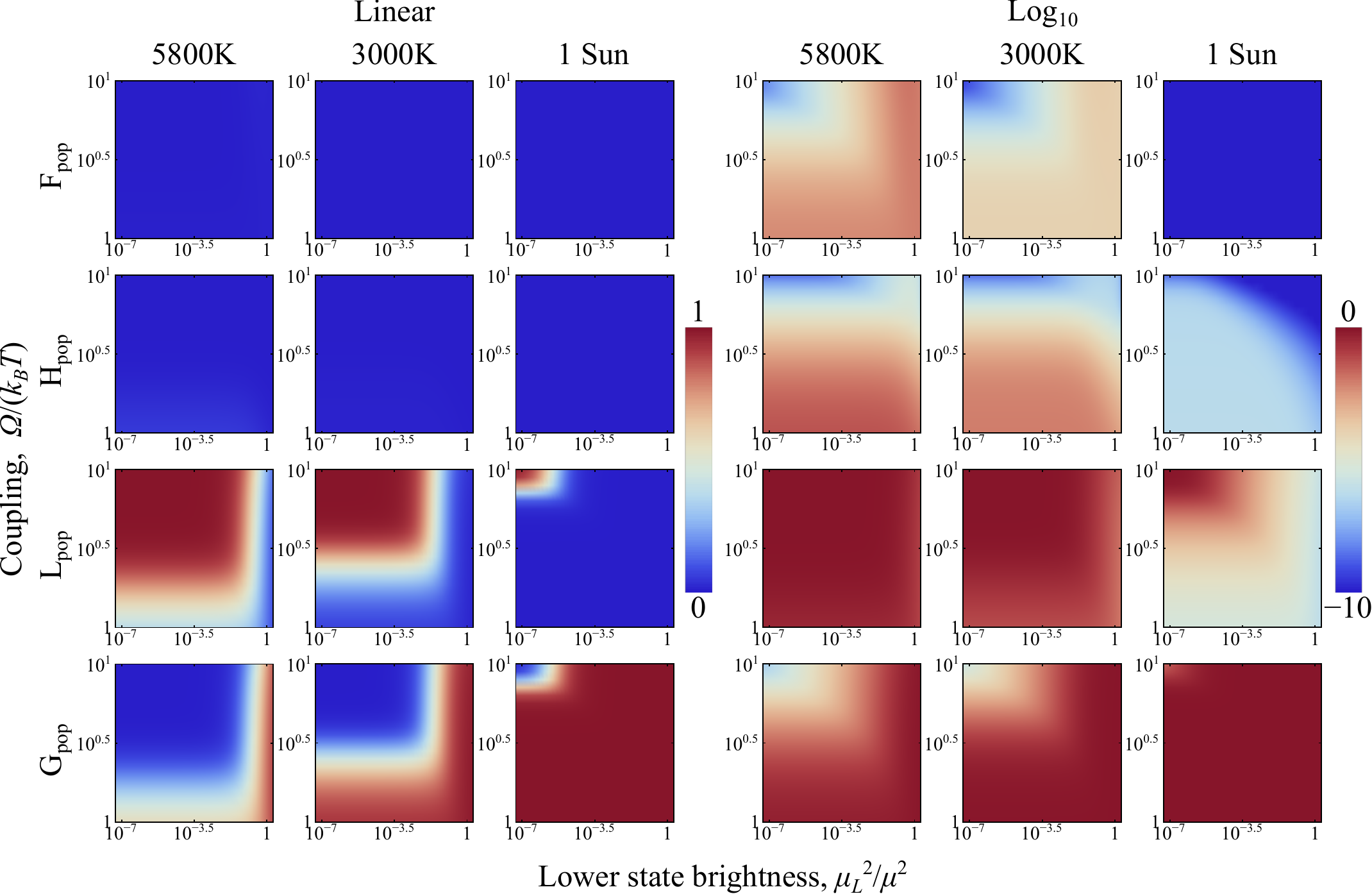}
    \caption{Total population of all four states of the dimer system at the steady-state, solving with the Bloch-Redfield approach~Eq.~\eqref{mas}. A linear scale on the color bar (left panels) is used to show population inversion in the lower states, while a logarithmic scale (right panels) shows the order of magnitude of the population in the higher energy states. We use the following dimer parameters: the single molecule excitation energy is $\hbar\omega_A=1.17$~eV and the spontaneous optical lifetime for the absorbers was fixed at $1/\gamma_0=36.8$~ns. Phonon bath at 300~K, and phonon decay rate defined to match a picosecond spontaneous decay rate for $\theta=0.07$~rad, where $\theta$ determines the relative orientation of the TDMs of the molecules in the dimer, see Fig.~\ref{dip}a. No cavity is attached. Comparison of cases with photon bath at 5800~K, 3000~K, and natural 
   `1 sun' (natural sunlight illumination). 
   }
    \label{brightnesscoupRedfieldv2}
\end{figure}

Here we analyze the parameter region for obtaining population inversion for a dimer under black-body pumping using the full Bloch-Redfield master equation. Moreover we  check the validity of
the analytical expression given in Eq.~(9) in the main text, for the
 population difference at the steady-state.
 
In Fig.~\ref{brightnesscoupRedfieldv2} we plot the populations of the dimer eigenstates at the steady-state, by solving the Master Equation~Eq.~\eqref{mas} varying the parameters $\Omega$ an $\mu_L^2$ in three cases. We remind that $\mu_L$ is the TDM of the $\ket{L}$ state of the dimer. We set photon bath temperatures at $T_{BB}=5800$~K, $T_{BB}=3000$~K, and also consider a `1 sun' case, where $T_{BB}=5800$~K, but all photon thermal population terms are reduced by the factor $f_S$, see Eq.~(12) in the main text, to account for the solid angle of the Sun as observed from Earth. We plot the populations of all four states using both linear and logarithmic scales.
The results presented here show that population inversion is possible for large enough coupling $\Omega$ in sufficiently dark dimers (as measured by $\mu_L^2 / \mu^2$). Moreover,
we see that the doubly excited state $\ket{F}$ can always be neglected, justifying the approximations used in the main text to limit our considerations to the single excitation manifold. 

\begin{figure}[!htbp]
    \centering
    \includegraphics[width=1\columnwidth]{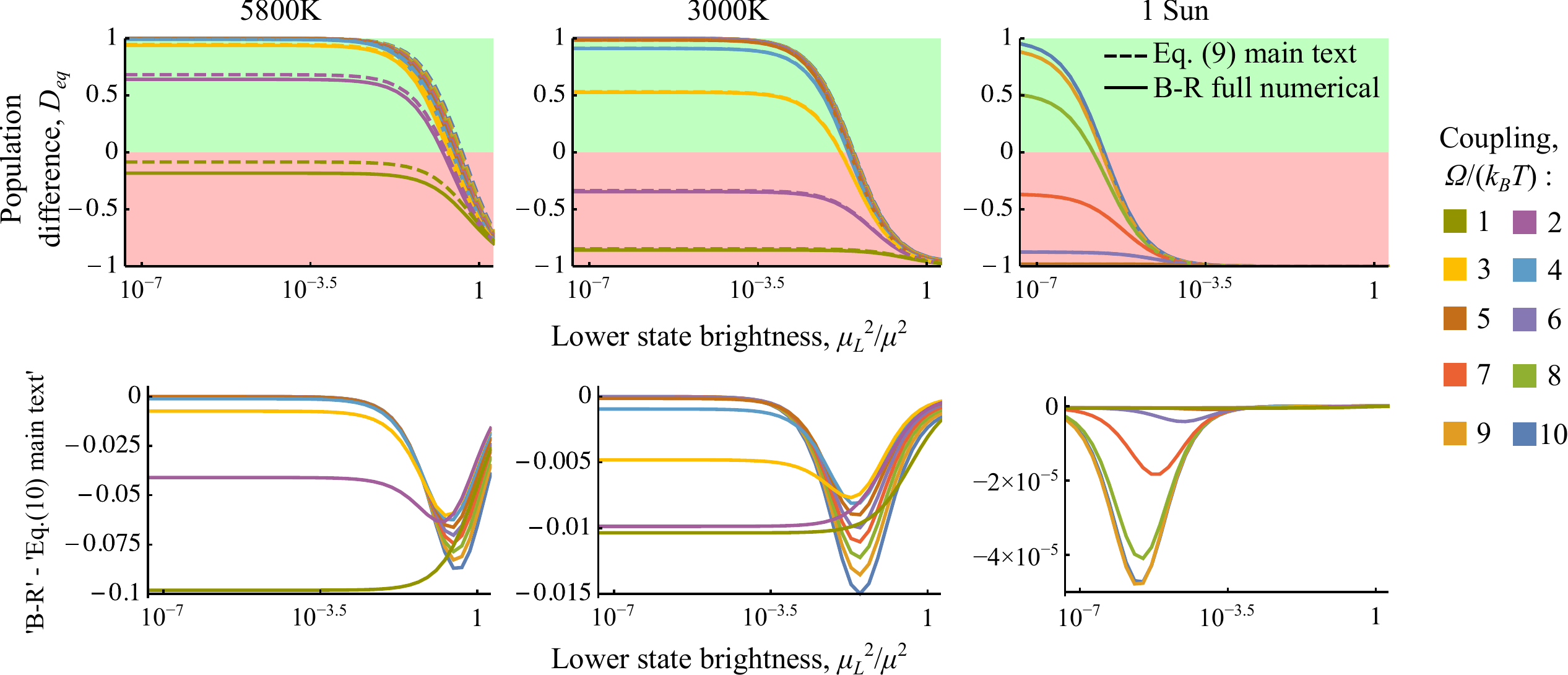}
    \caption{Comparison between the approximate steady-state solution given by Eq.~(9) in the main text  and the exact solution of Eq.~\eqref{mas} for different values of the coupling, $\Omega/(k_BT)$ (with $T=300$~K), and lower state brightness. Shown below: the differences between the results of the two methods. Dimer parameters: $\hbar\omega_A=1.17$~eV and single-molecule spontaneous optical lifetime fixed at $1/\gamma_0=36.8$~ns. Phonon bath at 300~K, and phonon decay rate defined to match a picosecond thermal relaxation time. No cavity is attached. Comparison of cases with photon bath at 5800~K, 3000~K, and `1 sun'.}
    \label{DimerBrightCoupPopDiffSuppFINv4}
\end{figure}

We now compare the results of the full  Bloch-Redfield model Eq.~\eqref{mas} against Eq.~(9) in the main text. In Fig.~\ref{DimerBrightCoupPopDiffSuppFINv4} we consider the same photon temperatures that were used in Fig.~\ref{brightnesscoupRedfieldv2} and plot the the population difference between the single excitation manifold and the ground state. The results of the Bloch-Redfield model (full curves) are compared with the analytical expression given in Eq.~(9) of the main text (dashed curves). Moreover their differences are shown in the lower panels for the three different black-body temperatures. While there are some discrepancies between the two approaches there is generally very good agreement. Note that as the temperature of the photon bath is reduced, larger  couplings, as well as weaker lower state brightness values are required to achieve population inversion.

\section{Lasing equations for molecular aggregates}
\label{sec-las}

In this section we expand on the derivation the lasing equations that is given in the main text. 

For convenience and ease of readability, we here duplicate most of the key equations, beginning with the Hamiltonian for a generic molecular aggregate $N$ molecules, written with the usual Pauli operators as
\begin{equation}
  \label{ham}
  \hat{H}_S = \sum_{j=1}^N \frac{\hbar\omega_A}{2} \hat{\sigma}_j^z + \sum_{i,j} \Omega_{i,j} \left( \hat{\sigma}_i^+\hat{\sigma}_j^- + \hat{\sigma}_j^+\hat{\sigma}_i^- \right)~,
\end{equation}
where $\Omega_{i,j}$ is the inter-molecular coupling, which here we describe through a dipole-dipole interaction~\cite{macro,phil,mukamel} (see also the sections about Bio-inspired aggregates), apart from very close-by molecules for which the interaction  is not determined by the transition dipole of the molecule alone. 
Eq.~\eqref{ham} represents a molecular aggregate where each molecule is described as a two-level system. As mentioned in the main text (and for the case of a dimer validated in Fig.~\ref{brightnesscoupRedfieldv2}), we can limit our analysis to  the ground state $\ket{G}$, which represent a state in which all molecules are in their ground state, and the single excitation manifold made of $N$ states $\ket{j}$ which represent a state in which the $j$-th molecule is excited while all the others are in the ground state. 
 The interaction of the molecular aggregate with black-body radiation and phonons is then governed by the master equation~Eq.~\eqref{mas}.
As established above, Eq.~\eqref{mas} can be  safely secularised and simplified  to Lindblad form~\cite{petru}.
Thus the coupling with the phonon bath and the black-body photon bath can be described by rate equations for the populations.  

In order to have lasing, we consider an ensemble of molecular aggregates randomly distributed inside the cavity volume with density $n_A$.
We consider a cavity mode at frequency $\omega_c$ filled with a classical field. The coupling of the molecular aggregate single excitation  eigenstate $\ket{k}$ to the resonant cavity mode is described by incoherent transition rates  for large enough dephasing  as discussed the main text. Below we report the rates induced by the cavity field for convenience,
\begin{equation}
\label{Bk}
    B_kn = n \frac{|\mu_k|^2 \omega_c}{3V \hbar \epsilon_o}
\frac{\Gamma_{\phi}}{\Gamma_{\phi}^2+(\Delta_k/\hbar)^2}~,
\end{equation}
where $\Delta_k=(E_k-\hbar\omega_c)$ is the energy detuning between the single excitation state $k$ and the mode of the cavity, $\mu_k$ is the transition dipole moment of the single excitation state $\ket{k}$, $\Gamma_{\phi}$ is the dephasing rate and $n$ is the number of photons in the cavity mode. The factor $3$ in the denominator comes from the average over the aggregate orientations. For a derivation of Eq.~\eqref{Bk}, see Eq.~\eqref{Tkk} and the analysis in the same section. 

Let us  first focus on the population of the ground state $P_G$. Due to the optical pumping from the black-body radiation there will be a  transition rate from the ground state to the single excitation states given by $R_k=n_T(\omega_k) \gamma_k(\omega_k)$, where $\gamma_k(\omega_k)$ is given by Eq.~\eqref{opt-rate} and is the spontaneous decay rate of the $k$-th eigenstate. 
Moreover, the cavity mode will drive transitions from the ground state to the states $\ket{k}$, given by the rate $B_kn$. Thus, for the population of the ground state we can write
\begin{align*}
    \frac{dP_G}{dt}= &-\sum_k R_k P_G + \sum_k [R_k +\gamma_k(\omega_k)]P_k  -\sum_k B_k n P_G + \sum_k B_k n P_k~.
\end{align*}
The total population in the single excitation manifold can be written as $P_e=\sum_k P_k$ and since we assume the other manifolds are not populated we can write $P_G+P_e=1$. 
Assuming thermal relaxation in the single excitation manifold to be the fastest time scale,  the populations in the single excitation manifold are always at thermal equilibrium so that we have
$$
P_k=P_e p_k \quad \mbox{with} \quad p_k= \frac{e^{-E_k/k_BT}}{\sum_j e^{-E_j/k_BT}}~,
$$
where $E_k$ is the energy of the excitonic state  $\ket{k}$. We consider a room temperature vibration environment by fixing $T=300$~K. Under these assumptions the population of the ground state  obeys the following equation,
\begin{align*}
\frac{dP_G}{dt}= - P_G \sum_k R_k  + P_e \sum_k [R_k+\gamma_k(\omega_k)] p_k - n P_G \sum_k B_k  + n P_e\sum_k B_k p_k~.
\end{align*}
We can now define
$$
R_u= \sum_k R_k \quad \mbox{and} \quad R_d= \sum_k [R_k+\gamma_k(\omega_k)] p_k~,
$$
and we further introduce:
$$
B_{tot}= \sum_k B_k \quad \mbox{and} \quad \langle B \rangle= \sum_k B_k p_k~,
$$
where $p_k$ is the Boltzmann occupation probability at thermal equilibrium of the state $\ket{k}$. 

With these definitions, and defining the density of aggregates in the excited states as $N_e=n_A P_e$ and the density of aggregates in the ground state as $N_G=n_A P_G$,  the laser equations coupling the population in the single excitation manifold of the molecular aggregates and the number of photons in the cavity  become
\begin{subequations}
\label{Leq1}
\begin{align}
    \frac{d N_e}{dt} =  & -R_d  N_e + R_u N_G - \langle B \rangle n N_e + B_{tot}n N_G ~,\\
    \frac{d n}{dt} = & V \langle B \rangle n N_e- VB_{tot} n N_G -n \kappa~,
\end{align}
\end{subequations}
where $VN_e$ and $VN_G$ are the total numbers of molecular aggregates in the excited and ground manifold respectively. Finally by defining the population difference per unit volume between the population in the single-excitation manifold and the population in the ground state  as $D=N_e-N_0$ we obtain the laser equations reported in the main text, see Eq.~(5).

\section{Coupling dimer to a cavity: modification of the Hamiltonian}
\label{sec-cav}

In this section we describe the coupling of the dimer to a cavity mode at frequency $\omega_c=\omega_L$, where $\omega_L$ is the transition frequency between the $\ket{G}$ and $\ket{L}$ states of the dimer.

To couple a cavity resonant with the energy of $\ket{L}$ to the dimer we begin with the laboratory frame dimer Hamiltonian in the site basis, $\hat{H}_S$ [Eq.~(1) in the main text]. In this section we set $\hbar=1$ for simplicity. We describe the interaction between the dimer and the cavity field by adding the time-dependent matrix elements $\vec{\mu}_j\cdot\vec{E}(t)$ to $\hat{H}_S$. Such matrix elements describe the interaction energy of each single-molecule TDM ($\vec{\mu}_j$, $j=1,2$) with the time-dependent cavity electric field $\vec{E}(t)=\vec{E}_0\cos(\omega_L t)$, which has a fixed polarization (determined by the direction of $\vec{E}_0=E_0\hat{\epsilon}$) and oscillates at frequency $\omega_c=\omega_L$. The resulting Hamiltonian is:
\begin{equation}
\hat{H}=
\begin{pmatrix}
2\omega_A & \vec{\mu}_1\cdot\vec{E}_0\cos(\omega_{L}t) & \vec{\mu}_2\cdot\vec{E}_0\cos(\omega_{L}t) & 0 \\
\vec{\mu}_1\cdot\vec{E}_0\cos(\omega_{L}t) & \omega_{A} & \Omega & \vec{\mu}_1\cdot\vec{E}_0\cos(\omega_{L}t) \\
\vec{\mu}_2\cdot\vec{E}_0\cos(\omega_{L}t) & \Omega & \omega_{A} & \vec{\mu}_2\cdot\vec{E}_0\cos(\omega_{L}t) \\
0 & \vec{\mu}_1\cdot\vec{E}_0\cos(\omega_{L}t) & \vec{\mu}_2\cdot\vec{E}_0\cos(\omega_{L}t) & 0
\label{labframefieldham}
\end{pmatrix}
~,
\end{equation}
written in the site basis
\begin{equation}
  \big\{ \ket{e_1}\ket{e_2}, \, \ket{e_1}\ket{g_2}, \, \ket{g_1}\ket{e_2}, \, \ket{g_1}\ket{g_2} \big\}~.
\end{equation}

The $\vec{\mu}_1\cdot\vec{E}_0$ and $\vec{\mu}_2\cdot\vec{E}_0$ terms account for the alignment of the single-molecule dipole moments with the orientation of the cavity containing the field, and are, respectively, now relabelled to $E_1$ and $E_2$. We then choose the unitary transformation matrix
\begin{equation}
\label{unitary}
\hat{U}=
\begin{pmatrix}
e^{i2 \omega_{L}t} & 0 & 0 & 0 \\
0 & e^{i \omega_{L}t} & 0 & 0 \\
0 & 0 & e^{i \omega_{L}t} & 0 \\
0 & 0 & 0 & 1
\end{pmatrix}
~,
\end{equation}
to move to a frame rotating at the frequency of the lasing transition with the transformed (prime) Hamiltonian given by
\begin{equation}
\hat{H}'=\hat{U}\hat{H}\hat{U}^\dagger+i\hat{U}\frac{d\hat{U}^\dagger}{dt}~.
\end{equation}
The transformation gives
\begin{equation}
\hat{H}'=
\begin{pmatrix}
2\omega_A-2\omega_L & \frac{E_1}{2}(1+e^{i2\omega_{L}t}) & \frac{E_2}{2}(1+e^{i2\omega_{L}t}) & 0 \\
\frac{E_1}{2}(1+e^{-i2\omega_{L}t}) & \omega_A-\omega_L & \Omega & \frac{E_1}{2}(1+e^{i2\omega_{L}t}) \\
\frac{E_2}{2}(1+e^{-i2\omega_{L}t}) & \Omega & \omega_A-\omega_L & \frac{E_2}{2}(1+e^{i2\omega_{L}t}) \\
0 & \frac{E_1}{2}(1+e^{-i2\omega_{L}t}) & \frac{E_2}{2}(1+e^{-i2\omega_{L}t}) & 0
\end{pmatrix}
~.
\end{equation}
We then drop the fast oscillating terms $e^{i2\omega_Lt}$ (oscillating at twice the lasing frequency) to leave
\begin{equation}
\hat{H}'\approx
\begin{pmatrix}
2\omega_A-2\omega_L & \frac{E_1}{2} & \frac{E_2}{2} & 0 \\
\frac{E_1}{2} & \omega_A-\omega_L & \Omega & \frac{E_1}{2} \\
\frac{E_2}{2} & \Omega & \omega_A-\omega_L & \frac{E_2}{2} \\
0 & \frac{E_1}{2} & \frac{E_2}{2} & 0
\end{pmatrix}
~.
\end{equation}
The next step is to move to the eigenbasis of the dimer. We temporarily drop the field terms to have
\begin{equation}
\hat{H}_S'=
\begin{pmatrix}
2\omega_A-2\omega_L & 0 & 0 & 0 \\
0 & \omega_A-\omega_L & \Omega & 0 \\
0 & \Omega & \omega_A-\omega_L & 0 \\
0 & 0 & 0 & 0
\end{pmatrix}
~.
\end{equation}
Using $\omega_L=\omega_A-|\Omega|$, we have that $\hat{H}_S'$ diagonalises (tilde) to
\begin{align}
\hat{\tilde{H}}_S'=
\begin{pmatrix}
2\omega_A-2\omega_L & 0 & 0 & 0 \\
0 & \omega_A-\omega_L+|\Omega| & 0 & 0 \\
0 & 0 & \omega_A-\omega_L-|\Omega| & 0 \\
0 & 0 & 0 & 0
\end{pmatrix}
=
\begin{pmatrix}
2|\Omega| & 0 & 0 & 0 \\
0 & 2|\Omega| & 0 & 0 \\
0 & 0 & 0 & 0 \\
0 & 0 & 0 & 0
\end{pmatrix}
~.
\label{HRWAnf}
\end{align}
In the rotating frame, the two lower energy states $\ket{L}$, and $\ket{G}$ appear degenerate since they are linked by a transition which is driven by the resonant lasing field in the cavity.

Finally, the interaction matrix elements with the cavity field (in the rotating frame),
\begin{equation}
\begin{pmatrix}
0 & \frac{E_1}{2} & \frac{E_2}{2} & 0 \\
\frac{E_1}{2} & 0 & 0 & \frac{E_1}{2} \\
\frac{E_2}{2} & 0 & 0 & \frac{E_2}{2} \\
0 & \frac{E_1}{2} & \frac{E_2}{2} & 0
\end{pmatrix}
~,
\end{equation}
need to be moved to the newfound dimer diagonal basis, becoming
\begin{equation}
\label{HintOmHL}
\frac{1}{\sqrt{8}}
\begin{pmatrix}
0 & E_1+E_2  & \frac{\Omega}{|\Omega|}(-E_1+E_2) & 0 \\
E_1+E_2 & 0 & 0 & E_1+E_2  \\
\frac{\Omega}{|\Omega|}(-E_1+E_2) & 0 & 0 & \frac{\Omega}{|\Omega|}(-E_1+E_2) \\
0 & E_1+E_2  & \frac{\Omega}{|\Omega|}(-E_1+E_2) & 0
\end{pmatrix}
~.
\end{equation}
In general, we obtain the full field-coupled Hamiltonian by adding Eq.~\eqref{HintOmHL} to Eq.~\eqref{HRWAnf}, that can be written as
\begin{equation}
\hat{H}_{S-C}:=\hat{\tilde{H}}'=
\begin{pmatrix}
2|\Omega| & g_H E_0 & g_L E_0 & 0 \\
g_H E_0 & 2|\Omega| & 0 & g_H E_0  \\
g_L E_0 & 0 & 0 & g_L E_0 \\
0 & g_H E_0  & g_L E_0 & 0
\end{pmatrix}
~,
\label{eqn:dimWITHf}
\end{equation}
where $g_{L,H}$ are coupling constants and $E_0=|\vec{E}_0|$ is the magnitude of the cavity electric field.
This Hamiltonian will  be substituted into Eq.~\eqref{mas}, to find the evolution of the dimer and field. 

We will be mostly interested in the case where the lasing mode of the cavity is matched to the frequency  of the $\ket{L}\leftrightarrow\ket{G}$ transition and the orientation of the molecular aggregate is random w.r.t the cavity field polarization.  In this case we may set $g_H E_0= \mu_H E_0 /(2 \sqrt{3})$ and $g_L E_0= \mu_L E_0 /(2 \sqrt{3})$.
In the case where the lasing mode of the cavity is matched to the frequency and polarisation of the $\ket{L}\leftrightarrow\ket{G}$ transition, in  we may set $g_H E_0=0$ and $g_L E_0=\mu_L E_0/2$ through requiring $\vec{\mu}_1\cdot\vec{E}=-\vec{\mu}_2\cdot\vec{E}$.

\section{Including dephasing and the coupling to the cavity in the Bloch-Redfield master equation}
\label{sec-deph}

 In Eq.~\eqref{mas}, dephasing  between the ground state and higher excitation manifolds is not properly included (rather, such a Bloch-Redfield approach only captures vibrationally driven dephasing within each manifold). Therefore, in the absence of introducing additional dephasing Lindblad operators, dephasing between different excitation subspaces only occurs on optical (nanosecond) timescales. However, much faster electronic dephasing will be present in any realistic situation, typically on a (sub)-picosecond timescale. 
 For this reason, we supplement our  master equation Eq.~\eqref{mas} with additional pure dephasing Lindblad operators as described below. We shall see that this dephasing plays a crucial role in that it gives rise to effectively incoherent energy exchange with the resonantly cavity, provided it is faster than the Rabi frequency. 

We illustrate the inclusion of pure dephasing using the example of a dimer system with the usual eigenbasis
\begin{equation}
  \big\{ \ket{F}, \, \ket{H}, \, \ket{L}, \, \ket{G} \big\}~.
\end{equation}

Specifically, we introduce phenomenological dephasing processes between each pair of eigenstates, apart from the $\ket{H}-\ket{L}$ pair. Indeed,  we do not include a phenomenological operator linking the two single-excitation states as dephasing (as well as relaxation) between these states is already included by the phonon dissipator, as is evidenced in the decay of coherence seen in Fig.~\ref{dip}b.  We constructed a range of dephasing Lindblad dissipators 
\begin{equation}
{\cal D}_{\phi}[\hat{\rho}]=\frac{2}{7}\Gamma_\phi\sum_k \hat{L}_k\hat{\rho}\hat{L}^\dag_k-\frac{1}{2}\{ \hat{L}^\dag_k\hat{L}_k,\hat{\rho} \} 
\label{Dphi}
\end{equation}
using the following $\hat{L}_k$ operators:
\begin{align}
\begin{pmatrix}
1 & 0 & 0 & 0 \\
0 & -1 & 0 & 0 \\
0 & 0 & 0 & 0 \\
0 & 0 & 0 & 0
\end{pmatrix}~,~
\begin{pmatrix}
1 & 0 & 0 & 0 \\
0 & 0 & 0 & 0 \\
0 & 0 & -1 & 0 \\
0 & 0 & 0 & 0
\end{pmatrix}~,~ 
\begin{pmatrix}
1 & 0 & 0 & 0 \\
0 & 0 & 0 & 0 \\
0 & 0 & 0 & 0 \\
0 & 0 & 0 & -1
\end{pmatrix}~,~ 
\begin{pmatrix}
0 & 0 & 0 & 0 \\
0 & 1 & 0 & 0 \\
0 & 0 & 0 & 0 \\
0 & 0 & 0 & -1
\end{pmatrix}~,~
\begin{pmatrix}
0 & 0 & 0 & 0 \\
0 & 0 & 0 & 0 \\
0 & 0 & 1 & 0 \\
0 & 0 & 0 & -1
\end{pmatrix}~.
\label{deph-term}
\end{align}

\begin{figure}[!htbp]
    \centering
    \begin{tabular}{l}
        {\Huge \bf a} \\
        \includegraphics[width=0.8\columnwidth]{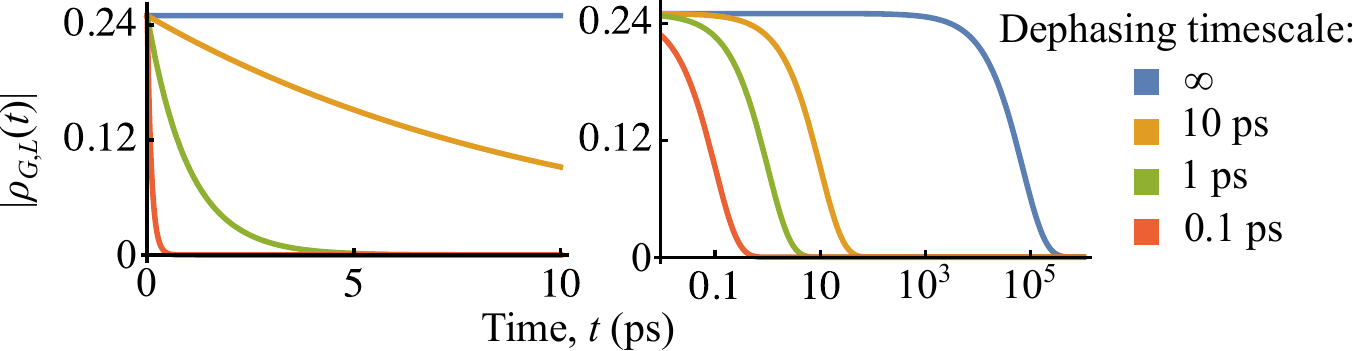}
    \end{tabular}
    \begin{tabular}{l}
        {\Huge \bf b} \\
        \includegraphics[scale=0.6]{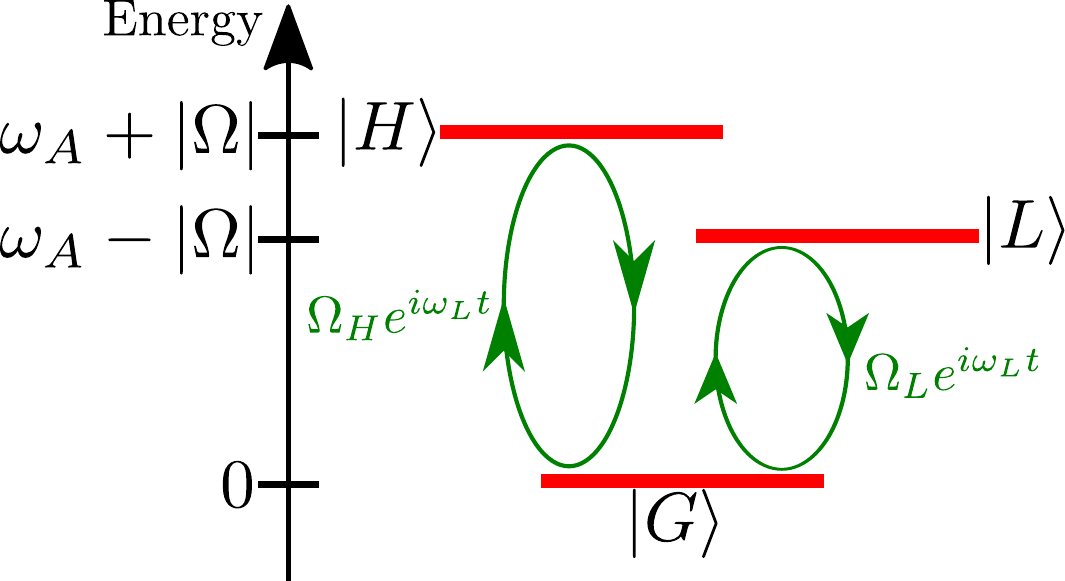}
    \end{tabular}
    \caption{{\bf (a)} Plots of the decay of the $\rho_{G,L}(t)$ coherence term in a dimer connected to a 3000~K photon bath for solar proccesses and local phonon baths but no cavity. Phenomenological dephasing operators are added at varying rates and this produces the expected decay on picosecond timescales. The parameters match those used in Fig.~\ref{dip}b. {\bf (b)} Dimer levels excluding the double-excitation $\ket{F}$ state, with a coherent coupling (curves) induced by the cavity field. Here $\hbar=1$ and the cavity frequency is $\omega_L=\omega_A-|\Omega|$. The Rabi frequencies are $\Omega_L=\mu_L E_0/(2\sqrt{3})$ and $\Omega_H=\mu_H E_0/(2\sqrt{3})$, where $\mu_{L,H}$ are the magnitudes of the TDMs, $\sqrt{3}$ comes from an average over the dimer orientations and $E_0$ is the amplitude of the cavity field.}
    \label{dimerphendephasing}
\end{figure}

We plot the evolution of our coherence term of interest, $\rho_{G,L}(t)$, for various dephasing rates in Fig.~\ref{dimerphendephasing}a, using the Bloch-Redfield equation given in~Eq.~\eqref{mas} with the addition of ${\cal D}_{\phi}[\hat{\rho}]$ and without the coupling with the photon bath. With the inclusion of these phenomenological dissipators we observe decay in the coherence term at the anticipated rapid timescales. The factor of $2/7$ in Eq.~\eqref{Dphi} ensures that the observed decoherence of $\rho_{G,L}(t)$ decays as $e^{-\Gamma_\phi t}$. 

Finally we can write the Bloch-Redfield master equation which includes both the dephasing terms and the Hamiltonian of the system-cavity coupling, Eq.~\eqref{eqn:dimWITHf}.  Adding the new dissipative terms to Eq.~\eqref{mas} we obtain
\begin{equation}
    \frac{d\hat{\rho}(t)}{dt} = -\frac{i}{\hbar} \left[ \hat{H}_{S-C}, \hat{\rho}(t) \right] + {\cal D}_{BB} [\hat{\rho}(t)] + {\cal D}_{T} [\hat{\rho}(t)] + {\cal D}_{\phi} [\hat{\rho}(t)]~. \label{mas2}
\end{equation}

\section{Full 4-level Bloch-Redfield: coherent laser equations}
\label{sec-4levBR}

In the following, we write a closed system of laser equations for the dimer coupled to the cavity field, in the frame of a semi-classical description.
As far as the dimer is concerned, its dynamics is described by the master equation Eq.~\eqref{mas2}. Note that the electric field magnitude $E_0=E_0(t)$ is a time-dependent quantity, because it exchanges energy with the dimers and loses energy out of the cavity. For this reason 
Eq.~\eqref{mas2} must be extended by an equation determining the evolution of the electric field.

The evolution of the classical field in the cavity  derives from Maxwell Equations~\cite{scullybook},
\begin{equation}
\label{Maxwell}
    \nabla^2 \vec{E}(t) - \frac{1}{c^2}\frac{\partial^2}{\partial t^2} \vec{E}(t) = -\frac{1}{\epsilon_0 c^2} \frac{\partial^2}{\partial t^2} \vec{P}(t)~,
\end{equation}
where $\vec{E}(t)=E_0(t)\,\hat{\epsilon}\cos(\omega_Lt)$ is the oscillating electric field in the cavity with polarization $\hat{\epsilon}$
and $\vec{P}(t)=P_0(t)\hat{\epsilon}\cos(\omega_Lt)$ is the oscillating polarization per unit volume, 
where $E_0(t)$ and $P_0(t)$ are their slowly-varying envelopes.
Since they vary on timescales that are much slower than $\sim 2\pi / \omega_L$, we can apply the slowly-varying envelope approximation (SVEA)~\cite{scullybook}, so that Eq.~\eqref{Maxwell} becomes
\begin{equation}
\label{SVEA}
    \frac{dE_0}{dt} = \frac{i\omega_L}{2\epsilon_0}P_0 \, .
\end{equation}
 We consider an ensemble of dimers with density $n_A$ in a volume $V$, so that the polarization component parallel to the field is $P_0=2n_A\langle\left(\hat{\vec{\mu}} \cdot \hat{\epsilon}\right)\rangle$. Here the average TDM is the expectation value of the dipole operator averaged over the dimer orientations, $\langle\left(\hat{\vec{\mu}} \cdot \hat{\epsilon}\right)\rangle = \langle{\rm Tr}[\left(\hat{\vec{\mu}} \cdot \hat{\epsilon}\right) \hat{\rho}]\rangle _{\rm or}$, and the TDM operator in the dimer eigenbasis is
\begin{equation}
    \hat{\vec{\mu}}=
    \begin{pmatrix}
        0 & \vec{\mu}_H & \vec{\mu}_L & 0 \\
        \vec{\mu}_H & 0 & 0 & \vec{\mu}_H \\
        \vec{\mu}_L & 0 & 0 & \vec{\mu}_L \\
        0 & \vec{\mu}_H & \vec{\mu}_L & 0
    \end{pmatrix}
    ~.
\end{equation}
Taking the trace ${\rm Tr}[\left(\hat{\vec{\mu}} \cdot \hat{\epsilon}\right) \hat{\rho}]$ we have
\begin{align}
    \label{pol}
    P_0= 2n_A \langle\left[ \left(\vec{\mu}_L\cdot\hat{\epsilon}\right) \left( \rho_{LG} + \rho_{FL} \right) + \left(\vec{\mu}_H\cdot\hat{\epsilon}\right) \left( \rho_{HG} + \rho_{FH} \right) \right]\rangle_{\rm or} \, ,
\end{align}
where the subscript ``or'' accounts for the average over different dimer orientations. Replacing Eq.~\eqref{pol} into Eq.~\eqref{SVEA} we have
\begin{align}
    \label{field1}
    \frac{dE_0}{dt} = \frac{i\omega_Ln_A}{\epsilon_0} \langle\left[ \left(\vec{\mu}_L\cdot\hat{\epsilon}\right) \left( \rho_{LG} + \rho_{FL} \right) + \left(\vec{\mu}_H\cdot\hat{\epsilon}\right) \left( \rho_{HG} + \rho_{FH} \right) \right]\rangle_{\rm or} \, .
\end{align}
Now we proceed by performing the orientational averaging.
It can be shown [see Eq.~\eqref{cohapp} in the following section] that the coherence terms are proportional to the corresponding TDM, for example $(\rho_{LG}+\rho_{FL})\propto (\vec{\mu}_L\cdot\hat{\epsilon})$. For this reason, the average over dipole orientations is proportional to the squared amplitude of the TDM along $\hat{\epsilon}$, namely $\langle(\vec{\mu}_L\cdot\hat{\epsilon})(\rho_{LG}+\rho_{FL})\rangle_{\rm or}\propto \langle(\vec{\mu}_L\cdot\hat{\epsilon})^2\rangle_{\rm or}$, and similarly for the $(\vec{\mu}_H\cdot\hat{\epsilon})$ term. When the dimers are randomly oriented,  $\langle(\vec{\mu}_L\cdot\hat{\epsilon})^2\rangle_{\rm or}=\mu_L^2/3$. Finally, we add the term $-\kappa E_0/2$ accounting for the losses from the cavity, so that we have
\begin{equation}
\label{field2}
    \frac{dE_0}{dt}= -\frac{\kappa E_0}{2} -i \alpha n_A \big[g_H(\rho_{F,H}+\rho_{H,G}) + g_L(\rho_{F,L}+\rho_{L,G})\big]~,
\end{equation}
where $\kappa$ is the field loss rate, $\alpha=2\omega_L/\epsilon_0$ and $g_{L,H}=\mu_{L,H}/(2\sqrt{3})$.
Note that in the case of an ordered ensemble of dimers, 
with $\vec{\mu}_L$ TDMs parallel with the cavity electric field, it should be $g_L=\mu_L/2$ and $g_H=0$.
Thus the final laser equations are given by the two coupled equations Eq.~\eqref{mas2} and Eq.~\eqref{field2}. Before showing the results of these full coherent laser equations, we introduce in the following sections further approximations leading to the working set of rate equations.

\section{Results with 4-level Bloch-Redfield static field model}
\label{sec-BRstat}

While the most complete version of our model considers a cavity field that is coupled dynamically to the molecular aggregates, see  Fig.~2b in the main text, considering the coupling of an ensemble of molecular aggregates with a non-evolving static field allows us to focus on a few more subtle behavioural aspects.

In the main text, the laser equations [see Eqs.~(6) and Eqs.~(7)], have been derived under the assumption that the population of the double excited state $\ket{F}$ of the dimer can be neglected  and that the dephasing rate  between $\ket{L}$ and $\ket{G}$ is larger than the Rabi frequency induced by the coupling of the $\ket{L} \leftrightarrow \ket{G}$ transition with the cavity mode. Nevertheless, when a strong intensity field is present in a cavity, this could also induce transitions between $\ket{L}$ and $\ket{F}$, despite the fact that this transition is detuned from the cavity frequency by $2\Omega$ and hence suppressed.  In particular, the suppression will cease to be effective once the cavity field is sufficiently intense and the Rabi frequency approaches the detuning. 

Moreover, if the polarisation of the cavity lasing mode is not quite parallel with the dipole moment of the  $\ket{L} \leftrightarrow \ket{G}$ transition, it can additionally induce an off-resonant transition between $\ket{G}$ and $\ket{H}$ and a resonant transition between $\ket{H}$ and $\ket{F}$. Nevertheless we do not consider the case of disordered ensemble of molecular aggregates here. 

\begin{figure}[!htbp]
    \centering
    \includegraphics[width=0.75\linewidth]{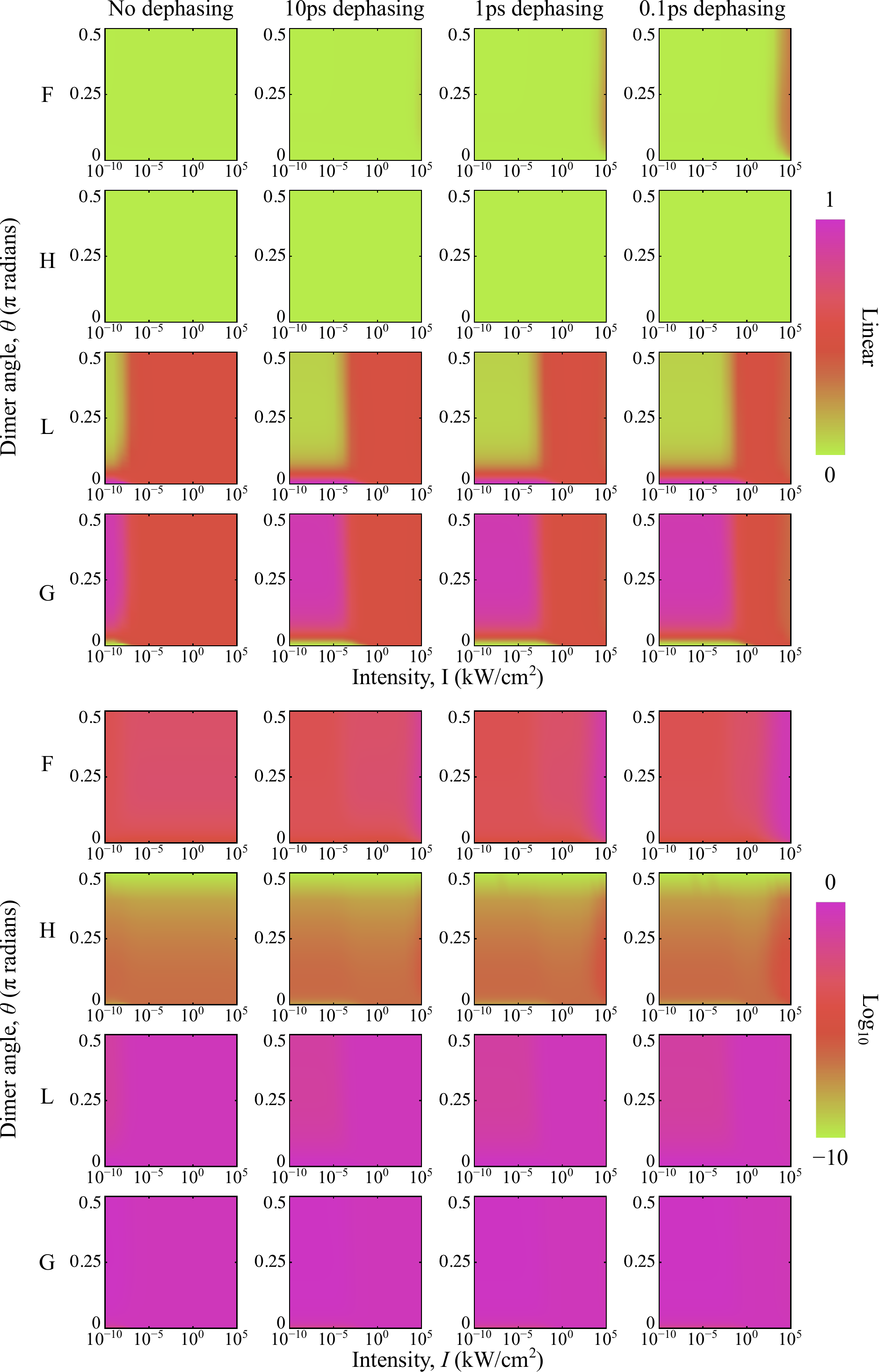}
    \caption{Parameter scan of steady states using the static field model. Results are displayed in linear and logarithmic scales. The intensity of the coupled static field, $I$, and the dimer angle $\theta$ are varied across the parameter scan. The range of dephasing rates for phenomenological dephasing dissipators matches those used in other models. We use parameters from the main text: $\hbar\omega_A=1.17$~eV and the spontaneous optical lifetime for the absorbers was fixed at $1/\gamma_0=36.8$~ns. Photon bath at 3000~K, phonon bath at 300~K, and phonon decay rate defined to match a picosecond spontaneous decay rate for $\theta=0.07$~rad.}
    \label{fig:StaticField}
\end{figure}

In order to understand when it is possible to neglect the $\ket{F}$ state even in presence of a cavity field,  we consider the coupling of the dimer molecule to an ideally aligned cavity [i.e. we set $g_H=0$ and $g_L=\mu_L/2$  in Eq.~\eqref{eqn:dimWITHf}]. To relate the Rabi frequency of the field to the intensity we use $(\hbar=1)$
\begin{equation}
\Omega_L = 2\frac{\mu}{\sqrt{\epsilon_0c}} \sin \theta \sqrt{I(\infty)} ~,
\end{equation}
where (using our assumed default dipole parameters)
\begin{equation}
\frac{\mu}{\sqrt{\epsilon_0c}} = 4.106\times 10^{-9} \frac{\rm eV}{(\text{W}/\text{m}^2)^{1/2}}~.
\end{equation}
In Fig.~\ref{fig:StaticField} we show the steady-state populations of the dimer eigenstates as a function of the dimer angle, $\theta$, and intensity of the cavity field, $I$. 
The results show, while the $\ket{H}$ state is always negligible (for the intensity range considered), the fully excited state  $\ket{F}$ can mostly be ignored, and only for very large intensities (larger than we use in our results) do we approach a regime where the transition linking $\ket{L}$ to $\ket{F}$ becomes important. 
We also note that the larger the dephasing is, the larger the needed intensity in the cavity to excite the dimer becomes.

We also note that there is a minimum intensity level, below which behaviour is dominated by the phonon and photon dissipators rather than the coupling to the cavity. Once the cavity coupling takes over population is evenly distributed between the states coupled by lasing transitions, namely $\ket{G}$, $\ket{L}$ and occasionally $\ket{F}$.

\section{Electric field with strong dephasing: incoherent rates}
\label{sec-rates}

In the main text we derive the lasing equations assuming that dephasing is much stronger than the coupling to the cavity. Here we show in detail how a master equation reduces to a rate equation under that assumption. 

At first we derive an expression for the transition rate between the ground state and single-excitation manifold states induced by the cavity field. 
For simplicity, we neglect the $\ket{F}$ state but our results are nonetheless valid more generally. Let us introduce an Hamiltonian coupling induced by the cavity, as given  in Eq.~\eqref{eqn:dimWITHf} and shown pictorially in Fig.~\ref{dimerphendephasing}b ($\hbar=1$). We relabel the coupling strength (Rabi frequency) as $\Omega_{L,H}:=g_{L,H}E_0$. The cavity field is resonant with the $\ket{L}-\ket{G}$ transition at frequency $\omega_L=\omega_A-|\Omega|$, and it is therefore detuned from the frequency of the $\ket{H}-\ket{G}$ transition.

Following the same approach described in the previous sections [see~Eq.~\eqref{unitary} and following], we write the Hamiltonian in the rotating frame as
\begin{equation}
    \hat{H}_{S-C} =
    \begin{pmatrix}
        2|\Omega| & 0 & \Omega_H \\
        0 & 0 & \Omega_L \\
        \Omega_H & \Omega_L & 0
    \end{pmatrix}~,
\end{equation}
represented in the basis $\mathcal{B} = \{ \ket{H}, \ket{L}, \ket{G} \}$. For the current purpose, we may  consider a simplified master equation without thermal relaxation and coupling to the photon bath,
\begin{equation}
    \label{mas3lev}
    \frac{d\hat{\rho}}{dt} = -i \left( \hat{H}_{S-C}\hat{\rho} - \hat{\rho}\hat{H}_{S-C} \right) + \mathcal{D}_{\phi} [\hat{\rho}]~,
\end{equation}
where the density matrix and the dephasing operator are, respectively,
\begin{equation}
    \hat{\rho} =
    \begin{pmatrix}
        \rho_{HH} & \rho_{HL} & \rho_{HG} \\
        \rho_{LH} & \rho_{LL} & \rho_{LG} \\
        \rho_{GH} & \rho_{GL} & \rho_{GG}
    \end{pmatrix}~,
    \qquad \qquad
    \mathcal{D}_{\phi}[\hat{\rho}] =
    \begin{pmatrix}
        0 & -\Gamma_{HL}\rho_{HL} & -\Gamma_{GH}\rho_{HG} \\
        -\Gamma_{HL}\rho_{LH} & 0 & -\Gamma_{GL}\rho_{LG} \\
        -\Gamma_{GH}\rho_{GH} & -\Gamma_{GL}\rho_{GL} & 0
    \end{pmatrix}
    \, .
\end{equation}
The dephasing rates $\Gamma_{HL}$, $\Gamma_{GH}$ and $\Gamma_{GL}$ describe the decay of coherence due to energy fluctuations and interaction with the (vibrational) environment. ~Eq.~\eqref{mas3lev} reads explicitly
\begin{subequations}
    \label{mas3levex}
    \begin{align}
        \frac{d\rho_{GL}}{dt} &= -i\Omega_L \left( \rho_{LL} - \rho_{GG} \right) -i \Omega_H \rho_{HL} - \Gamma_{GL} \rho_{GL}~, \\
        \frac{d\rho_{HL}}{dt} &= -i\Omega_H \left( \rho_{GL} - \rho_{HG} \right) -2i|\Omega| \rho_{HL} - \Gamma_{HL} \rho_{HL}~, \\
        \frac{d\rho_{GH}}{dt} &= -i\Omega_H \left( \rho_{HH} - \rho_{GG} \right) -i \Omega_L \rho_{LH} +2i|\Omega| \rho_{GH} - \Gamma_{GH} \rho_{GH}~, \\
        \frac{d\rho_{LL}}{dt} &= -i\Omega_L \left( \rho_{GL} - \rho_{LG} \right)~, \\
        \frac{d\rho_{HH}}{dt} &= -i\Omega_H \left( \rho_{GH} - \rho_{HG} \right) \, ,
    \end{align}
\end{subequations}
where $\Omega_{L,H}=\mu_{L,H} E_0/(2\sqrt{3})$ with $\mu_{L,H}$ being the magnitudes of the TDMs, and $\sqrt{3}$ arising from orientational averaging.
When decoherence is very fast, namely
\begin{equation}
    \label{ratecond}
    \min \{ \Gamma_{HL}, \Gamma_{GH}, \Gamma_{GL} \} \gg \max \{ \Omega_H, \Omega_L \} \, ,
\end{equation}
we can assume that coherences ($\rho_{k\neq k'}$) will adiabatically follow populations $\rho_{k k}$. Thus, setting the first three derivatives to zero, we have
\begin{subequations}
    \label{cohapp}
    \begin{align}
        \rho_{GL} &\approx -i\frac{\Omega_L}{\Gamma_{GL}} \left( \rho_{LL} - \rho_{GG} \right) -i \frac{\Omega_H}{\Gamma_{GL}} \rho_{HL} ~,\\
        \rho_{HL} &\approx -i\frac{\Omega_H}{\Gamma_{HL}+2i|\Omega|} \left( \rho_{GL} - \rho_{HG} \right)~, \\
        \rho_{GH} &\approx -i\frac{\Omega_H}{\Gamma_{GH}-2i|\Omega|} \left( \rho_{HH} - \rho_{GG} \right) -i \frac{\Omega_L}{\Gamma_{GH}-2i|\Omega|} \rho_{LH} \, .
    \end{align}
\end{subequations}
On the right-hand sides of~Eq.~\eqref{cohapp}, 
the terms involving coherences are of the order $|\rho_{k\neq k'}|\sim \Omega_{H,L} / \Gamma_{k k'} \ll 1$, whereas populations have larger values, $\rho_{kk} \sim 1$. Therefore, we can safely drop the coherence terms from those right-hand sides and substitute the resulting simplified relationship back into the population equations~Eq.~\eqref{mas3levex}. This yields
\begin{subequations}
    \begin{align}
        \frac{d\rho_{LL}}{dt} &\approx -2\frac{\Omega_L^2}{\Gamma_{GL}} \left( \rho_{LL} - \rho_{GG} \right)~, \\
        \frac{d\rho_{HH}}{dt} &\approx -2\frac{\Omega_H^2}{\Gamma_{GH}} \frac{1}{1+(2|\Omega|/\Gamma_{GH})^2} \left( \rho_{HH} - \rho_{GG} \right) \, .
    \end{align}
\end{subequations}
Consequently, the full master equation for the time evolution of the entire density matrix is very well-approximated by just two rate equations for the populations
\begin{subequations}
    \begin{align}
        \frac{d\rho_{LL}}{dt} &\approx -T_{GL} \left( \rho_{LL} - \rho_{GG} \right)~, \\
        \frac{d\rho_{HH}}{dt} &\approx -T_{GH} \left( \rho_{HH} - \rho_{GG} \right)~,
    \end{align}
\end{subequations}
with the resonant and off-resonant transition rates given by, respectively,
\begin{equation}
    T_{GL} = 2\frac{\Omega_L^2}{\Gamma_{GL}}~, \qquad \qquad T_{GH} = 2\frac{\Omega_H^2}{\Gamma_{GH}} \frac{1}{1+(2|\Omega|/\Gamma_{GH})^2} \, .
\end{equation}
Note that these expressions are more generally valid and applicable between weakly coupled levels (both resonant and detuned) in the presence of fast dephasing, as is shown in Ref.~\cite{levyang}. Indeed, in general we can write
\begin{equation}
\label{Tkk}
     T_{kk'} = 2\frac{\Omega_{kk'}^2}{\Gamma_{kk'}} \frac{1}{1+(\Delta_{kk'}/\Gamma_{kk'})^2} \, ,
\end{equation}
where $T_{kk'}$ is the incoherent transition rate between the states $\ket{k}$ and $\ket{k'}$, $\Omega_{kk'}$ is the coherent coupling between the states, $\Gamma_{kk'}$ is the dephasing rate between them and $\Delta_{kk'}=\omega_{kk'}-\omega_c$ is the detuning between the $\ket{k}-\ket{k'}$ transition frequency and the cavity frequency $\omega_c$. Here $\omega_{kk'}=|E_k-E_{k'}|/\hbar$ and $E_{k,k'}$ is the energy of the $\ket{k},\ket{k'}$-th state. 

Note that including photon and phonon (Bloch-Redfield) dissipators to the master equation (which can be done additively within second order perturbation expansions) only affects the precise values of the dephasing terms $\Gamma_{kk'}$ and will not alter the form and validity of Eq.~\eqref{Tkk}.

\section{Electric field with strong dephasing: incoherent laser equation}
\label{sec-rateeq}

In the following, we use the incoherent rates of~Eq.~\eqref{Tkk} to derive a set of rate equations for the dimer levels driven by the cavity under strong dephasing, including also the effect of photons and phonons. We start from the approximate Lindblad master equation~Eq.~\eqref{maslind} comprising a set of rate equations for photon absorption, spontaneous emission and phonon-induced thermal relaxation. We include the effect of the cavity by adding semi-classical transition rates~Eq.~\eqref{Tkk} between the levels $\ket{G}-\ket{L}$, $\ket{H}-\ket{F}$ (resonant transitions) and $\ket{L}-\ket{F}$, $\ket{G}-\ket{H}$ (non-resonant transitions, with a detuning frequency $|2\Omega|/\hbar$), assuming for simplicity that the dephasing rate has the same value $\Gamma_{kk'}=\Gamma_\phi$ for all transitions. Moreover, we derive a rate equation for the intensity of the cavity field $I=\epsilon_0|E_0|^2/2$: we start from Eq.~\eqref{field2} and we replace the coherence terms with the approximate expressions that we derived in Eq.~\eqref{cohapp} dropping higher-order terms proportional to the coherences, under the assumption of strong dephasing. The resulting system of coupled equations reads
\begin{subequations}
\label{maslindcav}
\begin{align}
    \frac{d\rho_{GG}}{dt} = &\gamma_L(\omega_L) \left[ 1 + n_T(\omega_L) \right] \rho_{LL} + \gamma_H(\omega_H) \left[ 1 + n_T(\omega_H) \right] \rho_{HH} - \gamma_L(\omega_L) n_T(\omega_L) \rho_{GG} - \gamma_H(\omega_H) n_T(\omega_H) \rho_{GG} \nonumber \\
    &+ T_{GL}(I) \left[ \rho_{LL} - \rho_{GG} \right] + T_{GH}(I) \left[ \rho_{HH} - \rho_{GG} \right]~, \\
    \frac{d\rho_{LL}}{dt} = &\gamma_L(\omega_H) \left[ 1 + n_T(\omega_H) \right] \rho_{FF} - \gamma_L(\omega_L) \left[ 1 + n_T(\omega_L) \right] \rho_{LL} + \gamma_L(\omega_L) n_T(\omega_L) \rho_{GG} - \gamma_L(\omega_H) n_T(\omega_H) \rho_{LL} \nonumber \\
    &+\Gamma_\Omega \left[ 1 + n_\Omega \right] \rho_{HH} - \Gamma_\Omega n_\Omega \rho_{LL} + T_{GL}(I) \left[ \rho_{GG} - \rho_{LL} \right] + T_{LF}(I) \left[ \rho_{FF} - \rho_{LL} \right]~, \\
    \frac{d\rho_{HH}}{dt} = &\gamma_H(\omega_L) \left[ 1 + n_T(\omega_L) \right] \rho_{FF} - \gamma_H(\omega_H) \left[ 1 + n_T(\omega_H) \right] \rho_{HH} + \gamma_H(\omega_H) n_T(\omega_H) \rho_{GG} - \gamma_H(\omega_L) n_T(\omega_L) \rho_{HH} \nonumber \\
    &-\Gamma_\Omega \left[ 1 + n_\Omega \right] \rho_{HH} + \Gamma_\Omega n_\Omega \rho_{LL} +T_{GH}(I) \left[ \rho_{GG} - \rho_{HH} \right] + T_{HF}(I) \left[ \rho_{FF} - \rho_{HH} \right]~, \\
    \frac{d\rho_{FF}}{dt} = &-\gamma_L(\omega_H) \left[ 1 + n_T(\omega_H) \right] \rho_{FF} - \gamma_H(\omega_L) \left[ 1 + n_T(\omega_L) \right] \rho_{FF} + \gamma_L(\omega_H) n_T(\omega_H) \rho_{LL} + \gamma_H(\omega_L) n_T(\omega_L) \rho_{HH} \nonumber \\
    &+T_{LF}(I) \left[ \rho_{LL} - \rho_{FF} \right] + T_{HF}(I) \left[ \rho_{HH} - \rho_{FF} \right]~, \\
    \frac{dI}{dt} = &b(I)\left\{\frac{\mu_L^2}{\mu^2}\left[ \rho_{LL}-\rho_{GG} + \frac{\rho_{FF}-\rho_{LL}}{1+(2|\Omega|/\hbar\Gamma_\phi)^2} \right]+\frac{\mu_H^2}{\mu^2}\left[ \rho_{FF}-\rho_{HH} + \frac{\rho_{HH}-\rho_{GG}}{1+(2|\Omega|/\hbar\Gamma_\phi)^2} \right]\right\} -\kappa I~,
\end{align}
\end{subequations}
where we have introduced the transition rates
\begin{gather}
    T_{GL}(I) = \frac{\mu_L^2 I}{3\hbar^2\epsilon_0c\Gamma_\phi}\, , \quad T_{LF}(I) = \frac{\mu_L^2 I}{3\hbar^2\epsilon_0c\Gamma_\phi[1+(2\Omega/\hbar\Gamma_\phi)^2]} \, , \quad T_{HF}(I) = \frac{\mu_H^2 I}{3\hbar^2\epsilon_0c\Gamma_\phi}\, , \nonumber \\
    T_{GH}(I) = \frac{\mu_H^2 I}{3\hbar^2\epsilon_0c\Gamma_\phi[1+(2\Omega/\hbar\Gamma_\phi)^2]} \, , \quad b(I) = \frac{n_A\mu^2\omega_L I}{3\hbar\epsilon_0\Gamma_\phi}~,
    \label{Trates}
\end{gather}
which all depend on the field intensity $I$. The approximations leading to Eq.~\eqref{maslindcav} are allowed only if the dephasing rate is much faster than the coherent oscillations induced by the laser field in the cavity. We now proceed to write the condition of validity for the rate equations~Eq.~\eqref{ratecond} in terms of the field intensity. Approximating $\max\{ \Omega_H, \Omega_L\} \approx \Omega_H \approx \mu E_0/(\sqrt{6}\hbar)$ for $\mu_L^2 \ll \mu^2$, we define the parameter
\begin{equation}
    \label{etaImax}
    \eta = \frac{\max \{ \Omega_L, \Omega_H\}}{\Gamma_\phi} = \sqrt{\frac{I}{\tilde{I}}} \qquad {\rm where} \qquad \tilde{I} \approx \frac{3c\epsilon_0\hbar^2\Gamma_\phi^2}{\mu^2} \, .
\end{equation}
According to the condition~Eq.~\eqref{ratecond}, the rate equation description is valid when $\eta \ll 1$, that is when $I\ll \tilde{I}$. For the parameters used in the main text, $\mu=10.157$~D and $\Gamma_\phi=1/(10$~ps$)$, we have $\tilde{I}=80$~kW/cm$^2$ which is larger than the stationary values of the intensities considered in the main text.

\section{Comparing coherent and incoherent laser equations}
\label{sec-BRrate}

\begin{figure}[!htbp]
    \centering
    {\Large {\bf (a)} $\Gamma_{\phi}=1/(10~{\rm ps})$, $n_A=5\times10^{-4}$~mmol/L, $\tilde{I}=80$~kW/cm$^2$}
    \begin{tabular}{cc}
        \includegraphics[width=0.275\columnwidth]{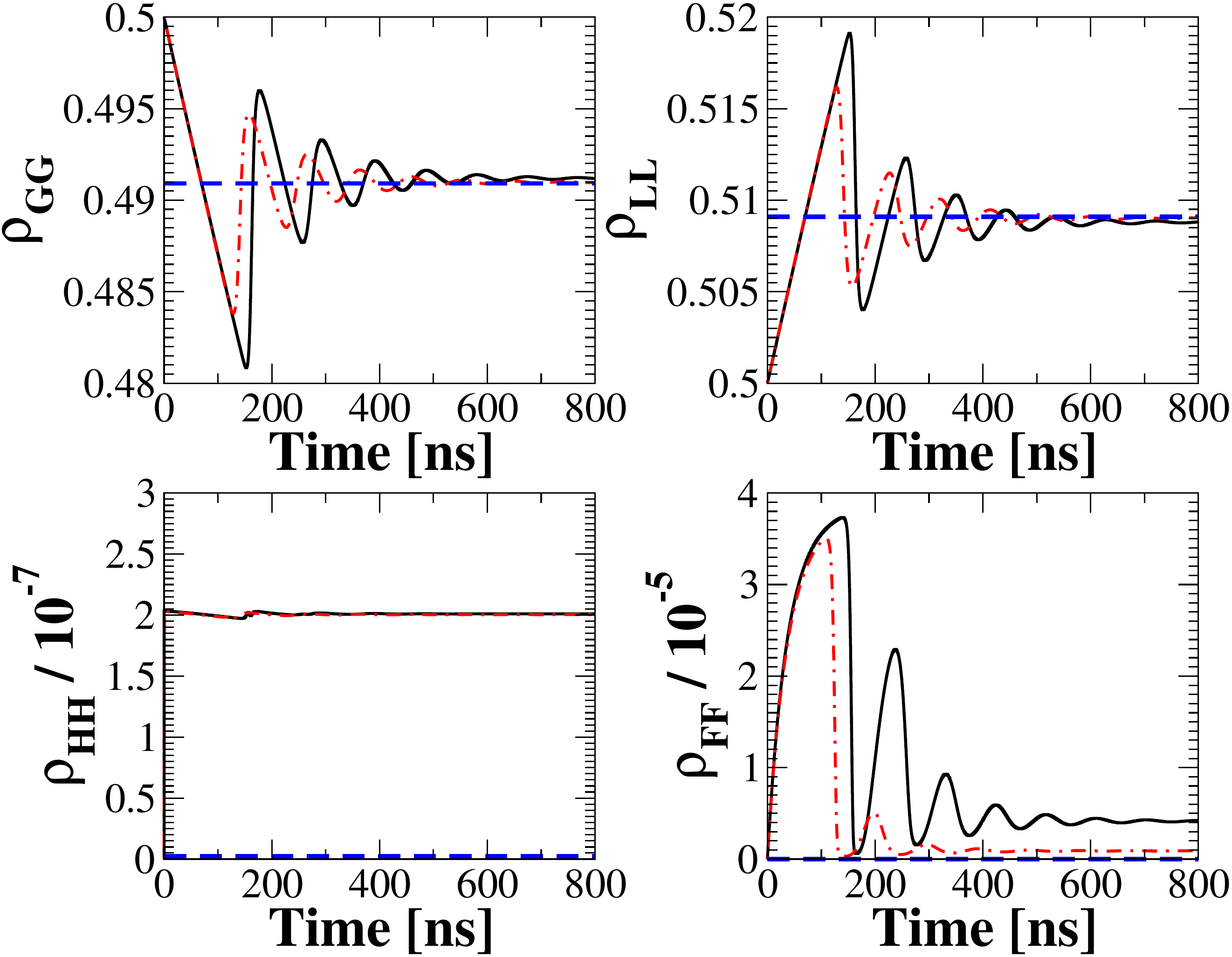} &
    \includegraphics[width=0.31\columnwidth]{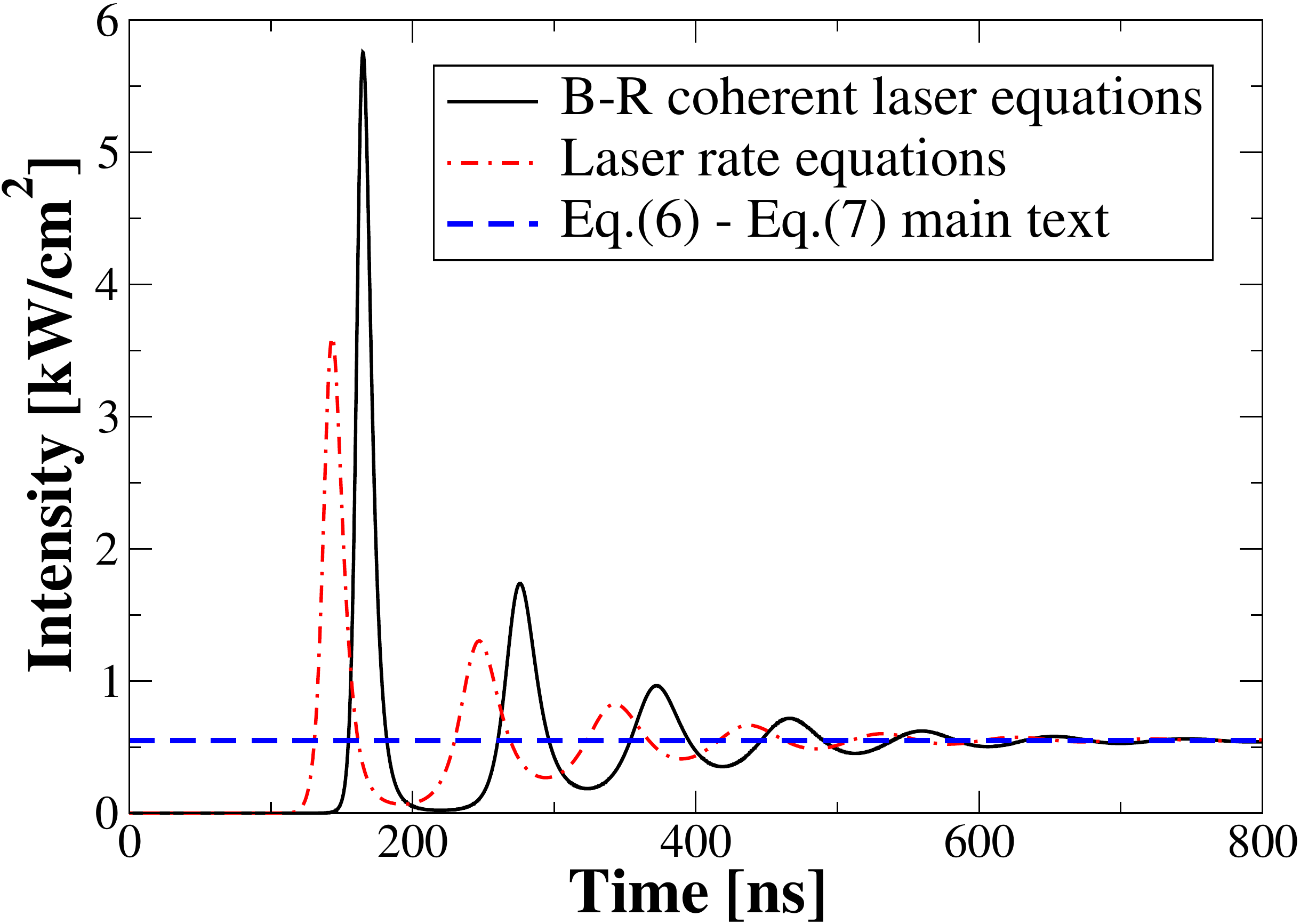} \\
    & \\
    & \\
    \end{tabular}

    {\Large {\bf (b)} $\Gamma_{\phi}=1/(1~{\rm ps})$, $n_A=5\times10^{-3}$~mmol/L, $\tilde{I}=8$~MW/cm$^2$}
    \begin{tabular}{cc}
        \includegraphics[width=0.275\columnwidth]{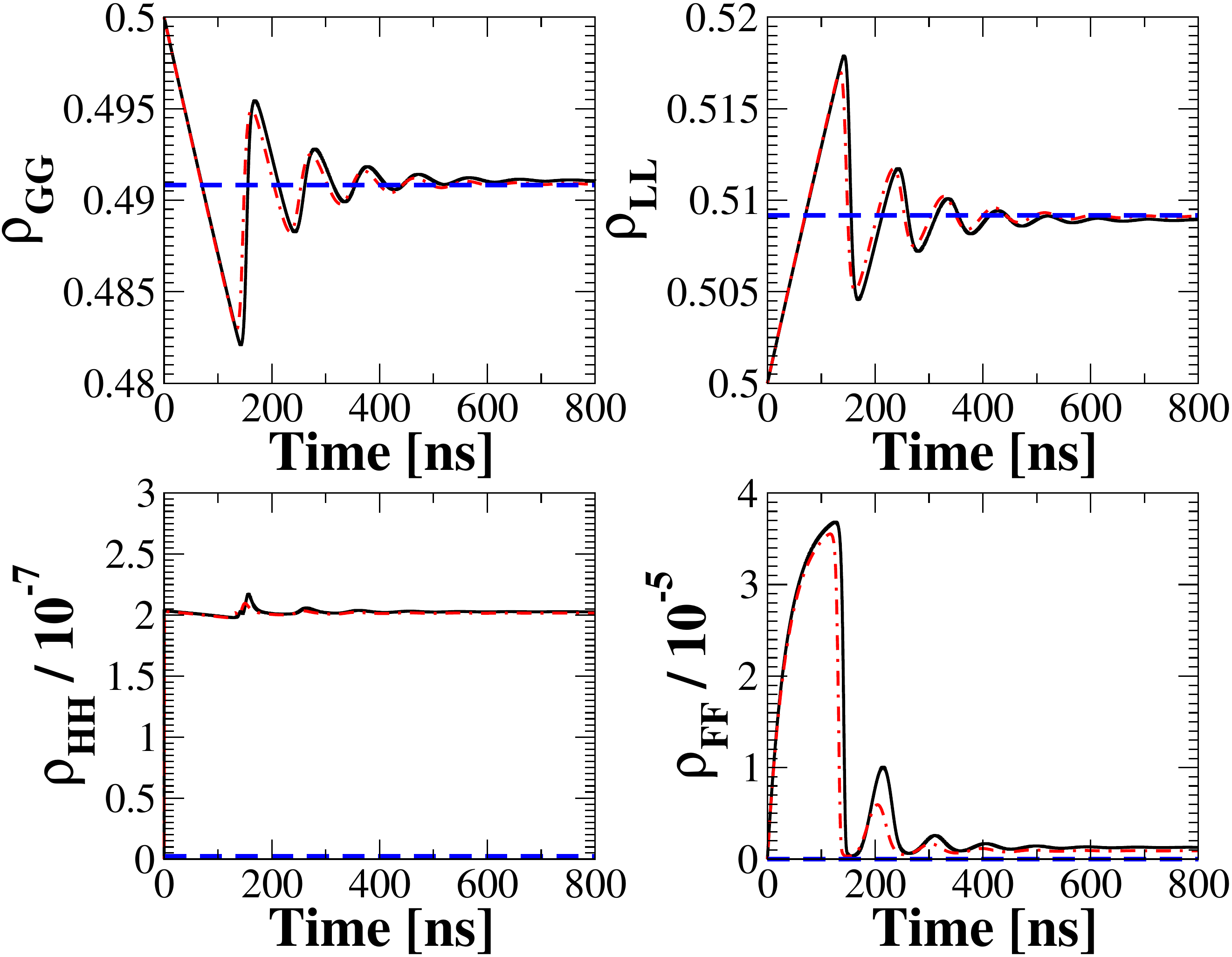} &
    \includegraphics[width=0.31\columnwidth]{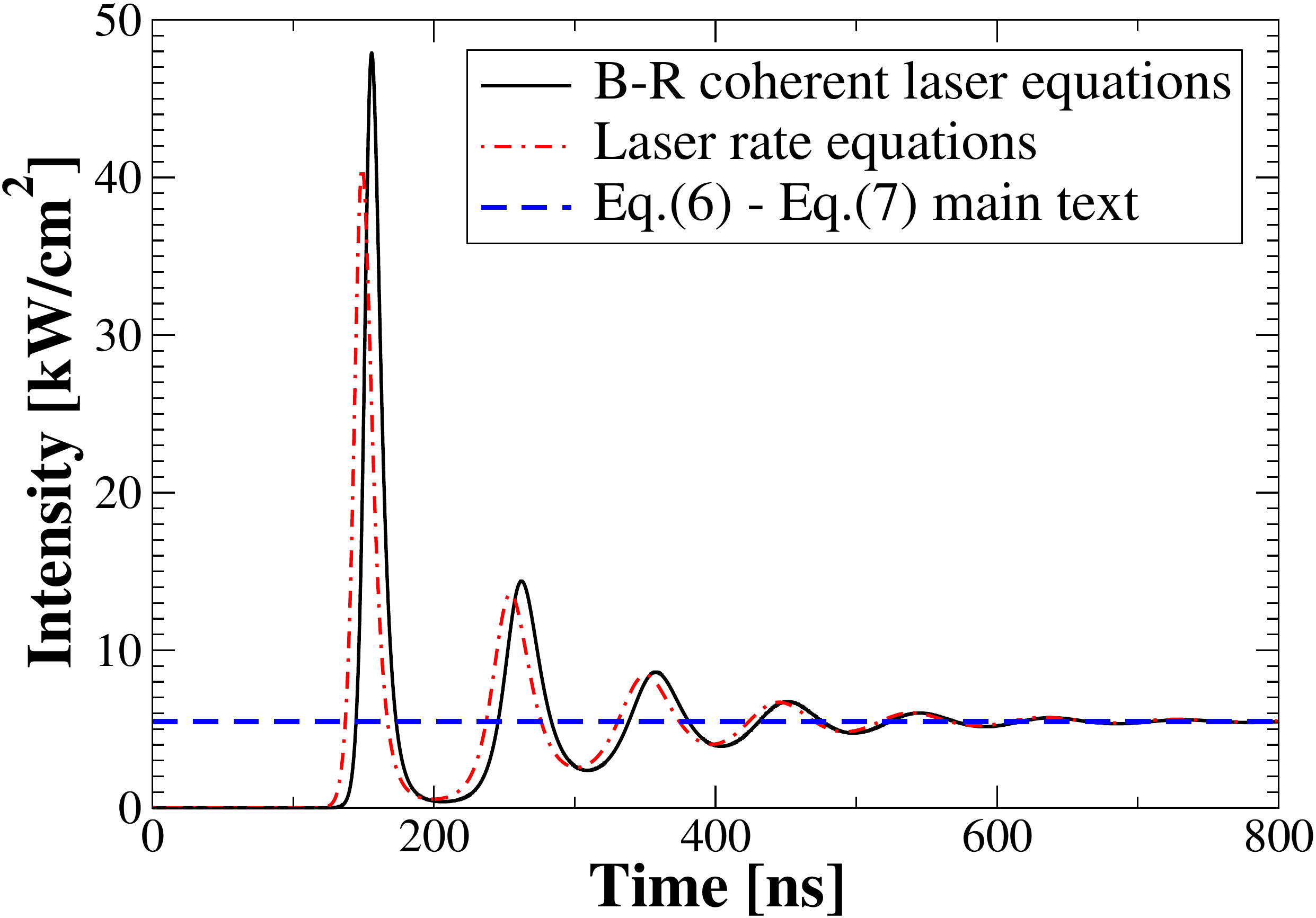} \\
    & \\
    & \\
    \end{tabular}
    
    {\Large {\bf (c)} $\Gamma_{\phi}=1/(100~{\rm fs})$, $n_A=5\times10^{-2}$~mmol/L, $\tilde{I}=800$~MW/cm$^2$}
    \begin{tabular}{cc}
        \includegraphics[width=0.275\columnwidth]{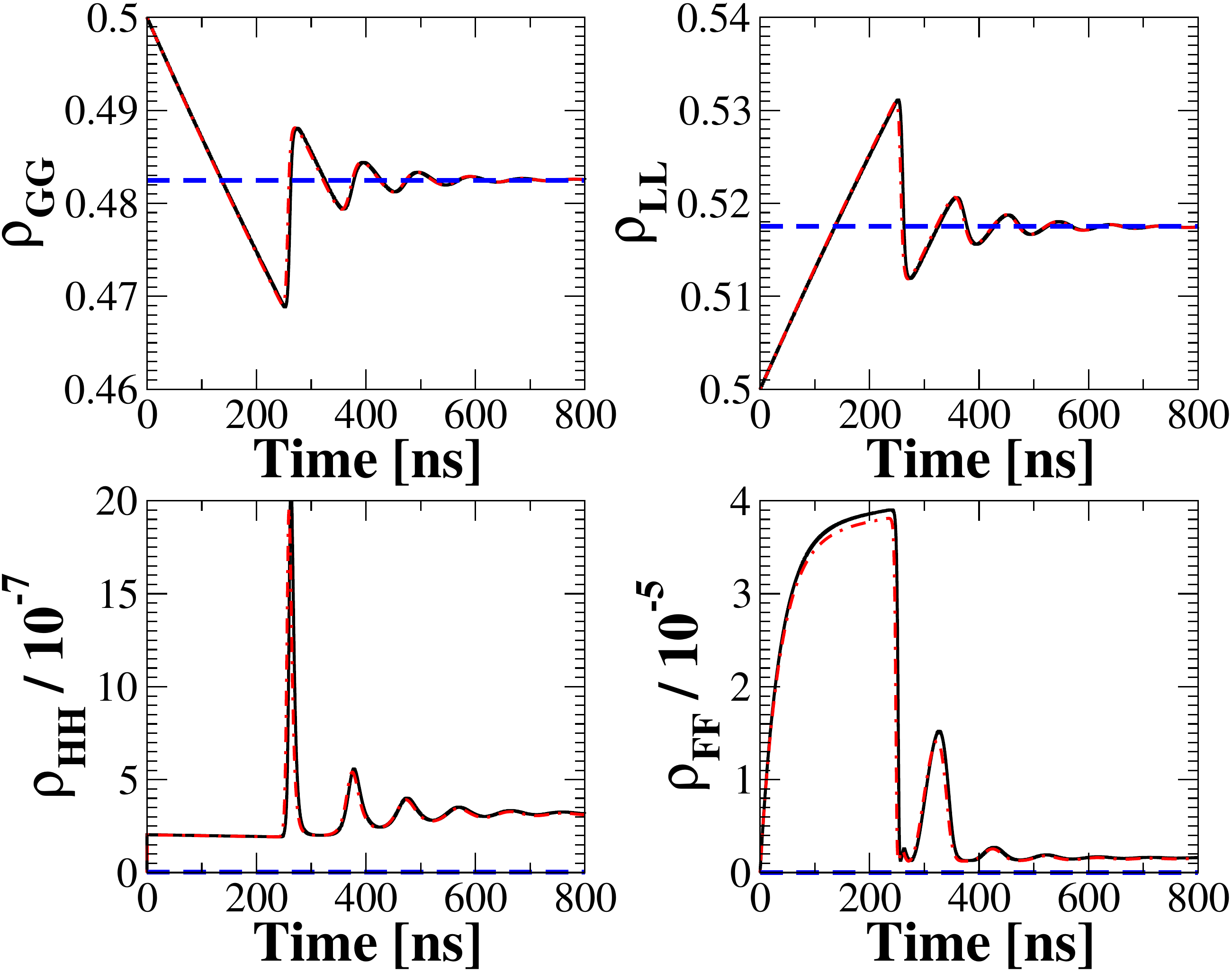} &
    \includegraphics[width=0.31\columnwidth]{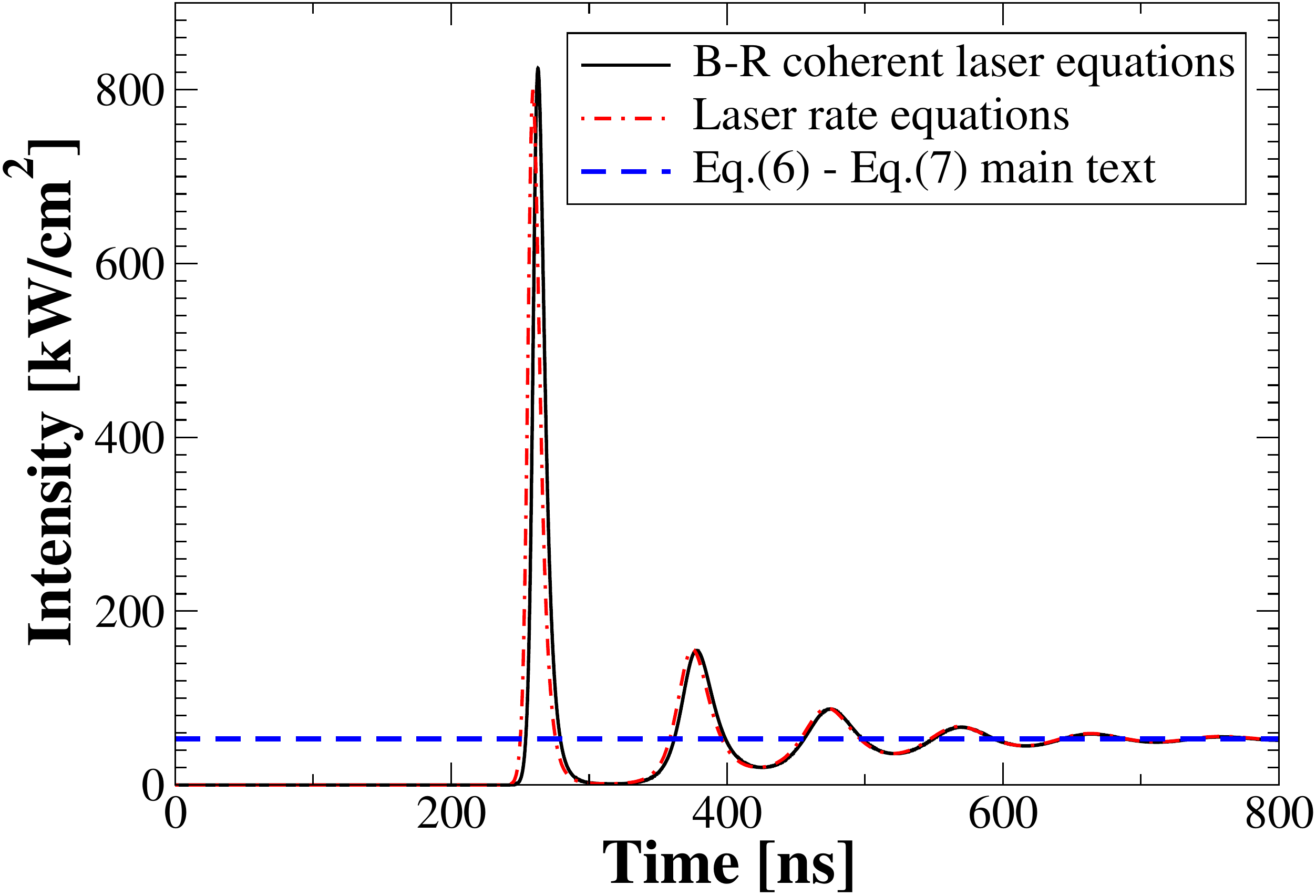} \\
    & \\
    & \\
    \end{tabular}
    
    {\Large {\bf (d)} no dephasing, $n_A=5\times10^{-4}$~mmol/L, $\tilde{I}=5.7$~mW/cm$^2$}
    \begin{tabular}{cc}
        \includegraphics[width=0.275\columnwidth]{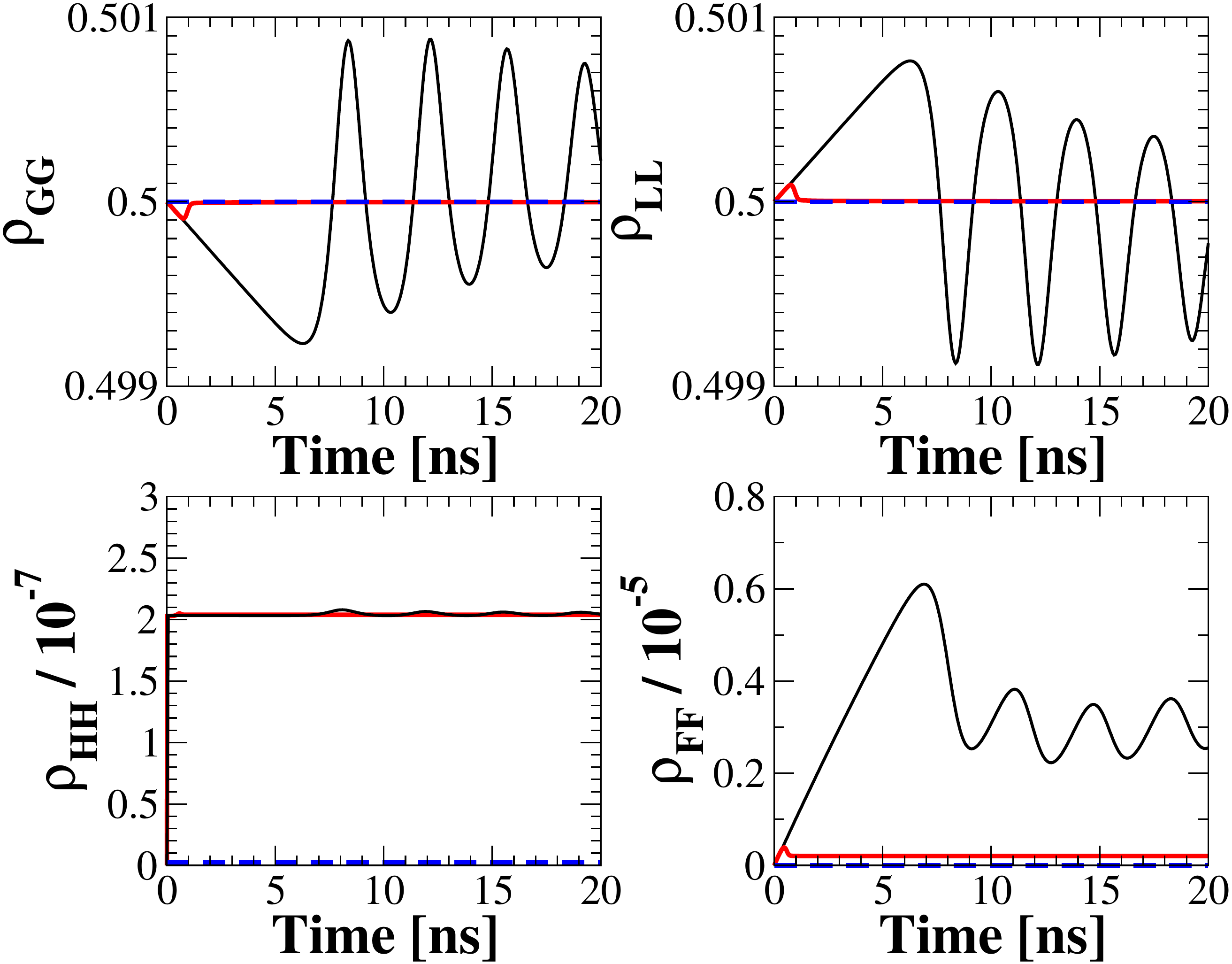} &
    \includegraphics[width=0.31\columnwidth]{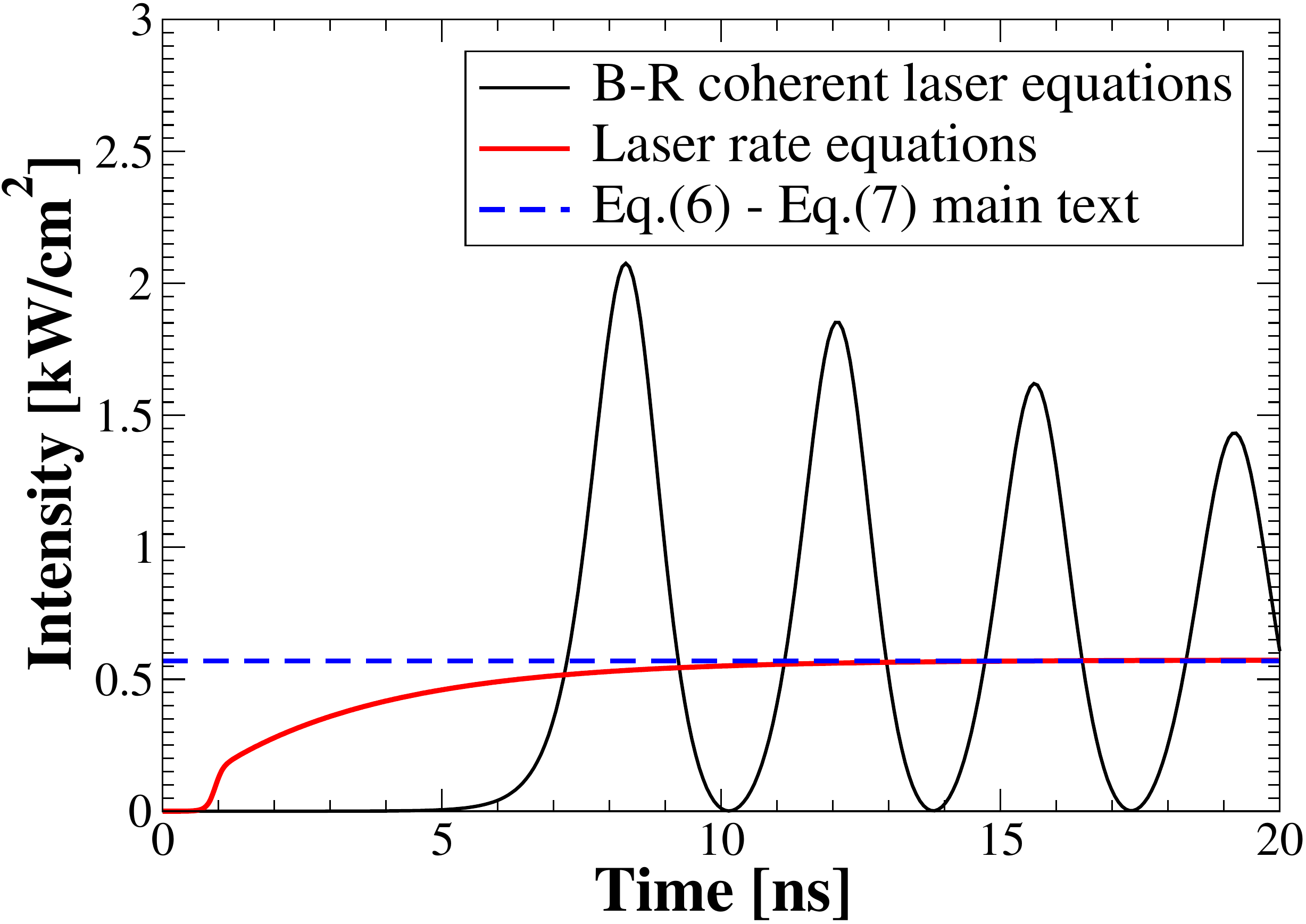}
    \end{tabular}
    \caption{Level populations and laser intensity obtained with the Bloch-Redfield (B-R) coherent model~Eq.~\eqref{mas2}-Eq.~\eqref{field2} and with the laser rate equations~Eq.~\eqref{maslindcav}. The blue dashed lines represent the steady-state solution of the laser equations in the main text. $\tilde{I}$ is the maximal intensity for which the rate equations are expected to work, see~Eq.~\eqref{etaImax}. In panel (d), for the ``laser rate equation'' and ``main text'' cases we set the dephasing rate from the spontaneous emission timescale ($\Gamma_{\phi}=1/(36.78~{\rm ns})$). Parameters: $\mu=10.157$~D, $\hbar\omega_A=1.17$~eV, $\Omega=2000$~cm$^{-1}$, $T_{BB}=3000$~K, $\kappa/(2\pi)=50$~MHz, $\langle \chi \rangle=0.005$.}
    \label{dyncav}
\end{figure}

In this section we compare the full Bloch-Redfield coherent laser equations~Eq.~\eqref{mas2} and Eq.~\eqref{field2} with the incoherent laser equation Eq.~\eqref{maslindcav}. Both will be contrasted against the (stationary) laser equation presented in the main text, see Eqs.~(6,7). Note that the full Bloch-Redfield coherent laser equations are the most general (and also most difficult to solve), while the incoherent laser equations assumed strong dephasing so that energy exchange with the cavity can be considered incoherent. Finally, the laser equation presented in the main text not only assumed strong dephasing as in the incoherent laser equation, but they also  assume fast thermal relaxation which has not been assumed in deriving Eq.~\eqref{maslindcav}. 

In order to validate the laser equation presented in the main text we have computed the evolution of the  intensity as a function of time starting with the following initial conditions for the dimer density matrix
\begin{equation}
\rho(t=0)=
\begin{pmatrix}
0 & 0 & 0 & 0 \\
0 & 0 & 0 & 0 \\
0 & 0 & 0.5 & 0 \\
0 & 0 & 0 & 0.5
\end{pmatrix}~,
\end{equation}
where only the $\ket{G},\ket{L}$ states are initially equally populated and as initial value of the field in the cavity we set $E_0(t=0)=868$~Vm$^{-1}$. 
Only the case of disordered dimers is considered. 
A first comparison between the fully coherent model and the laser equation is presented in Fig.~2b  in the main text for the case of a black body pumping of $T=3000$~K. In the following, we shall also limit our consideration to the same black-body temperature. 

In Fig.~\ref{dyncav} we plot the populations of all the four levels and the laser intensity against time. In panel (a) the parameters ensure a stationary laser intensity of 0.5~kW/cm$^2$ at $T_{BB}=3000$~K and $\tilde{I}=80$~kW/cm$^2$, which implies $\eta=0.08$, so that the rate equation approximation Eq.~\eqref{maslindcav} is valid. By comparing the rate equations Eq.~\eqref{maslindcav} results (red dash-dotted lines) with the full Bloch-Redfield coherent laser equations  results (black continuous lines), one can see that they are very similar. Moreover,  the stationary intensity reached by both approaches  is also equal to the main text laser equation, see horizontal dashed line.
The laser equations derived in the main text remain valid also for shorter dephasing times, as we show in panels (b,c), where higher dephasing values are used (1~ps and 0.1~ps), so that $\tilde{I}=8$~MW/cm$^2$ (panel b) and $\tilde{I}=800$~MW/cm$^2$ (panel c). Note that in panels (b,c) we also use higher densities than panel (a), in order to reach the lasing threshold. With these parameters we obtain a stationary laser intensity of 5~kW/cm$^2$ (panel b), implying $\eta=0.03$, and intensity of 50~kW/cm$^2$ (panel c), implying $\eta=0.008$ so that the rate equations (red dash-dotted lines) remain valid in these two cases. Finally, in panel (d) we consider the case with ``no dephasing'', in the sense that no extra dephasing is included in the system, and the main source of decoherence is due to the coupling to the photons. Therefore, we compare  the full Bloch-Redfield coherent laser equations results with the rate equations Eq.~\eqref{maslindcav}, where we choose $\Gamma_\phi=\gamma_0$, that is we assume that the dephasing rate is given only by  spontaneous decay. From panel (d) one can see that the coherent model has a very different dynamics from the rate equations. The coherent model shows non-damped oscillations (we verified that there is no damping up to 650~ns), while the rate equations quickly reach their steady-state values in few nanoseconds. For the parameters in panel (d) we obtain an average stationary laser intensity of 0.5~kW/cm$^2$ with $\tilde{I}=5.7$~mW/cm$^2$, implying $\eta=296$, so that the rate equations are not expected to be valid. 

The results presented in this section thus confirm the validity of our laser equations, see Eq.~(6,7)  in the main text, under the assumption of  strong dephasing and quick thermal relaxation. Both these assumptions are realistic in molecular aggregates at room temperature. 

Finally we would like to note that
the timescale required for reaching the stationary regime strongly depends on the black-body temperature. For instance, for $T_{BB}=3000$~K equilibrium is reached in $\sim 100$~ns, while for natural sunlight timescales up to $10$~ms are needed to reach the equilibrium lasing intensity. As this would require heavy numerical simulations, we are not showing this case here. 
In general, the equilibration timescale is given by $\max \left\{ [(R_u+R_d)(n_A/2n_A^{th})]^{-1}, 1/\kappa \right\}$, with the threshold density $n_A^{th}$ given implicitly by Eq.~(8) in the main text. 

Moreover, within the considered range of parameters the double-excitation $\ket{F}$ state never acquires any significant population. Therefore, when $\eta \ll 1$ and the 4-level rate-equation description is valid, we can also safely ignore the $\ket{F}$ state and assume thermal relaxation within the single-excitation manifold, obtaining the main-text theory. In other words, the $\eta \ll 1$ condition is enough to justify the approximations made in our theory in the main text.

Finally we would like to comment on the effect of faster dephasing as considered here. The main effect of faster dephasing is to lower the incoherent driving of the cavity, thus inducing a larger critical threshold density for lasing. As for the level of darkness required to have lasing, Eq.~(11) in the main text establishes an upper bound for the brightness, in order to have lasing. The only dependence on dephasing in Eq.~(11) of the main text is in the rates $\langle B \rangle,B_{tot}$. In any case, when the aggregate has only one transition that is resonant with the cavity (and all others are well detuned with energy differences larger than $\hbar\Gamma_\phi$), then $\langle B \rangle \approx B_{tot}$ and the upper bound for the dimer brightness is dephasing-independent. In all of our calculations we indeed focused to such regime. 

On the other hand, the threshold density Eq.~(8) in the main text always depends on $\Gamma_\phi$, since the right-hand side of Eq.~(8) in the main text increases proportionally to $\Gamma_\phi$ [see Eq.~(4) in the main text neglecting the contribution of the off-resonant states].

\section{Bio-inspired aggregate: Purple Bacteria}
\label{sec-purp}

In the main text we apply our lasing theory to a bio-mimetic molecular aggregate inspired by the antenna complex of purple bacterium \emph{Rhodobacter Sphaeroides}. Here we report some supplemental details about the structure of the aggregate and we discuss the validity of the lasing equations in such aggregates.

\begin{figure}[!htbp]
    \centering
    \includegraphics[width=\linewidth]{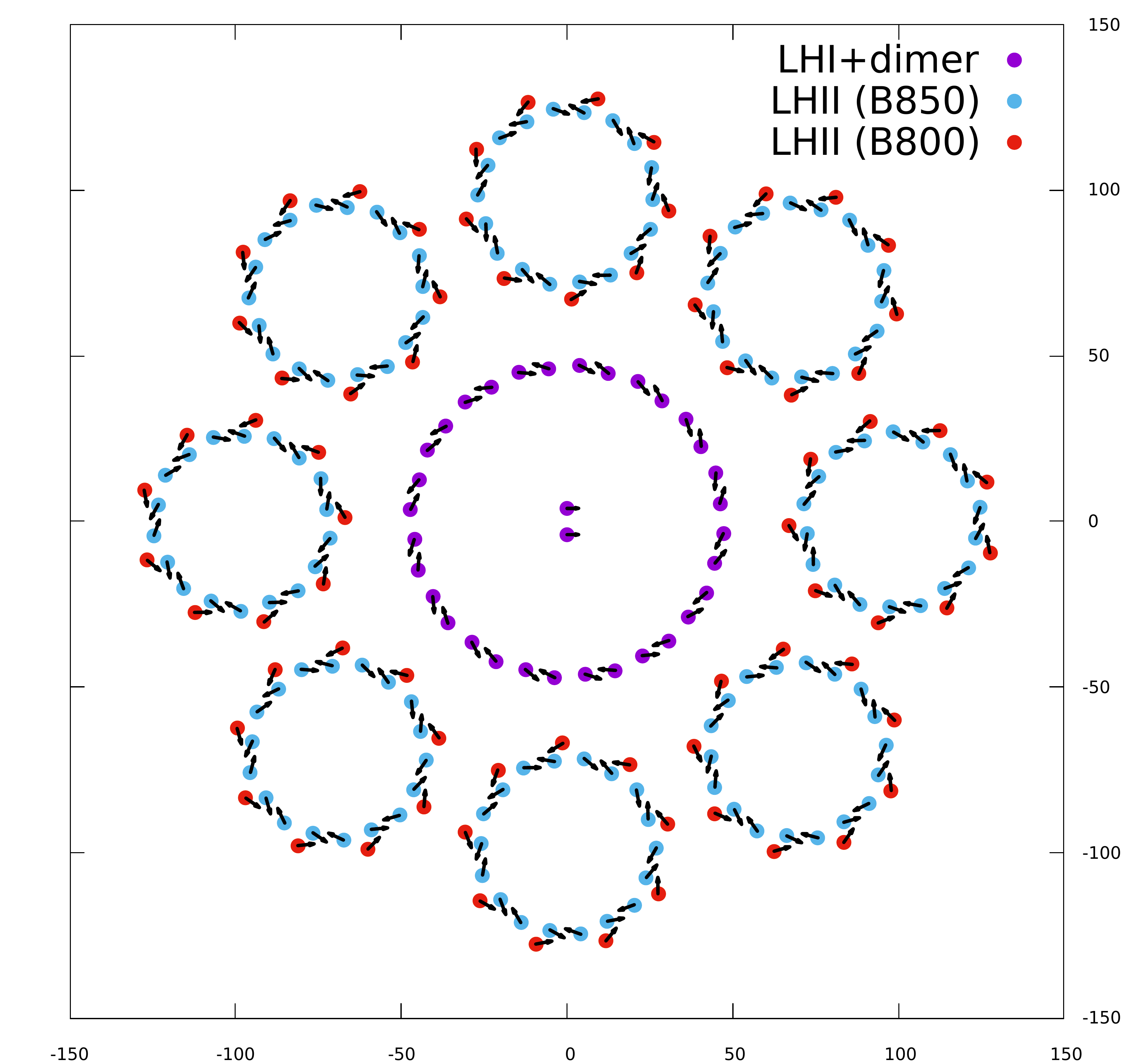}
    \caption{Positions (circles) and transition dipole orientations (arrows) for all the molecules in the aggregate (see also Table~\ref{tab-posdip}). View from top. The axes are in \AA~units. Magnified version of Fig.~3e of the main text.}
    \label{fig-posdip}
\end{figure}

Purple bacteria antenna complexes are formed by bacteriochlorophyll-a (BChl) molecules, aggregated in different kinds of ring-like structures~\cite{schulten3,kassal2}. The most common  BChl aggregates are those called light-harvesting complex II (LHII). These are either 9-fold~\cite{kassal2} or 8-fold~\cite{schulten:1,schulten:2} structures, formed by two stacked rings. In this manuscript we consider the 9-fold structure, which is subdivided into a lower ring made of 18 BChl molecules and an upper ring made of nine molecules. The lower ring is called B850 since it has a fluorescence peak at 850~nm, while the upper ring is called B800 because it emits at 800~nm. The LHII rings are distributed on bacterial membranes, surrounding the larger light-harvesting I (LHI) complexes. LHI can have a ring structure~\cite{schulten3} made of 32 molecules or an S-like shape~\cite{kassal2} made of 56 molecules. In our analysis we consider the ring-like structure for LHI, also called B875 for its emission peak at $875$~nm.  Around the LHI complex we place 8 equally spaced LHII aggregates (see Fig.~\ref{fig-posdip}) mimicking natural antennae. The minimal distance between neighboring LHII aggregates is 22.4~\AA\ as in Ref.~\cite{kassal2}, while the minimal distance between LHII and LHI is 25.3~\AA. 
In Fig.~\ref{fig-posdip} we show a top-view of the aggregate (a magnification of Fig.~3e of the main text), while in Table~\ref{tab-posdip} we report the positions of the molecules and the unit vectors for their transition dipole moment.
Following Ref.~\cite{schulten:1,schulten:2}, we describe the excitation transfer between the molecules of the aggregate using the single-excitation Hamiltonian Eq.~(1) in the main text, with the positions and dipole directions from Table~\ref{tab-posdip} and using the parameters reported in Table~I in the main text.

\begin{table}[!htbp]
    \centering
    \begin{tabular}{cccccc}
    \toprule
    $x$ [\AA] & $y$ [\AA] & $z$ [\AA] & $\mu_x/\mu$ & $\mu_y/\mu$ & $\mu_z/\mu$ \\
    \midrule
    \multicolumn{2}{c}{\bf LHII (B850)} & & & & \\
    \midrule
    123.289	 & 	-5.157	 & 	-0.06	 & 	0.48	 & 	0.865	 & 	0.147	 \\ 
    124.618	 & 	4.179	 & 	0.067	 & 	-0.33	 & 	-0.934	 & 	0.136	 \\ 
    120.678	 & 	12.253	 & 	-0.047	 & 	-0.182	 & 	0.972	 & 	0.145	 \\ 
    115.707	 & 	20.304	 & 	0.043	 & 	0.34	 & 	-0.93	 & 	0.143	 \\ 
    107.485	 & 	23.988	 & 	-0.056	 & 	-0.76	 & 	0.633	 & 	0.146	 \\ 
    98.534	 & 	26.991	 & 	0.055	 & 	0.862	 & 	-0.489	 & 	0.135	 \\ 
    89.822	 & 	24.479	 & 	-0.06	 & 	-0.989	 & 	-0.017	 & 	0.147	 \\ 
    81.072	 & 	20.962	 & 	0.067	 & 	0.974	 & 	0.182	 & 	0.136	 \\ 
    76.05	 & 	13.513	 & 	-0.047	 & 	-0.751	 & 	-0.644	 & 	0.145	 \\ 
    71.563	 & 	5.183	 & 	0.043	 & 	0.635	 & 	0.759	 & 	0.143	 \\ 
    72.483	 & 	-3.78	 & 	-0.056	 & 	-0.168	 & 	-0.975	 & 	0.146	 \\ 
    74.358	 & 	-13.033	 & 	0.055	 & 	-0.007	 & 	0.991	 & 	0.135	 \\ 
    80.889	 & 	-19.322	 & 	-0.06	 & 	0.509	 & 	-0.848	 & 	0.147	 \\ 
    88.31	 & 	-25.141	 & 	0.067	 & 	-0.644	 & 	0.753	 & 	0.136	 \\ 
    97.272	 & 	-25.766	 & 	-0.047	 & 	0.933	 & 	-0.328	 & 	0.145	 \\ 
    106.73	 & 	-25.487	 & 	0.043	 & 	-0.975	 & 	0.171	 & 	0.143	 \\ 
    114.032	 & 	-20.208	 & 	-0.056	 & 	0.928	 & 	0.342	 & 	0.146	 \\ 
    121.108	 & 	-13.958	 & 	0.055	 & 	-0.855	 & 	-0.502	 & 	0.135	 \\ 
    \midrule
    \multicolumn{2}{c}{\bf LHII (B800)} & & & & \\
    \midrule
    126.628	 & 	11.769	 & 	16.843	 & 	-0.771	 & 	0.619	 & 	0.149	 \\ 
    112.396	 & 	27.476	 & 	16.751	 & 	-0.987	 & 	-0.013	 & 	0.162	 \\ 
    91.382	 & 	30.349	 & 	16.774	 & 	-0.754	 & 	-0.643	 & 	0.131	 \\ 
    73.494	 & 	18.908	 & 	16.843	 & 	-0.151	 & 	-0.977	 & 	0.149	 \\ 
    67.007	 & 	-1.271	 & 	16.751	 & 	0.504	 & 	-0.848	 & 	0.162	 \\ 
    75.026	 & 	-20.906	 & 	16.774	 & 	0.934	 & 	-0.332	 & 	0.131	 \\ 
    93.878	 & 	-30.677	 & 	16.843	 & 	0.922	 & 	0.358	 & 	0.149	 \\ 
    114.597	 & 	-26.205	 & 	16.751	 & 	0.482	 & 	0.861	 & 	0.162	 \\ 
    127.592	 & 	-9.443	 & 	16.774	 & 	-0.18	 & 	0.975	 & 	0.131	 \\
    \bottomrule
    \end{tabular}
    \quad
    \begin{tabular}{cccccc}
    \toprule
    $x$ [\AA] & $y$ [\AA] & $z$ [\AA] & $\mu_x/\mu$ & $\mu_y/\mu$ & $\mu_z/\mu$ \\
    \midrule
    {\bf LHI} & & & & & \\
    \midrule
    44.706	 & 	-12.591	 & 	-0.099	 & 	0.634	 & 	0.76	 & 	0.147	 \\ 
    47.167	 & 	-3.677	 & 	0.099	 & 	-0.452	 & 	-0.886	 & 	0.098	 \\ 
    46.122	 & 	5.475	 & 	-0.099	 & 	0.295	 & 	0.944	 & 	0.147	 \\ 
    44.984	 & 	14.653	 & 	0.099	 & 	-0.079	 & 	-0.992	 & 	0.098	 \\ 
    40.515	 & 	22.709	 & 	-0.099	 & 	-0.089	 & 	0.985	 & 	0.147	 \\ 
    35.952	 & 	30.752	 & 	0.099	 & 	0.307	 & 	-0.947	 & 	0.098	 \\ 
    28.741	 & 	36.485	 & 	-0.099	 & 	-0.459	 & 	0.876	 & 	0.147	 \\ 
    21.448	 & 	42.169	 & 	0.099	 & 	0.646	 & 	-0.757	 & 	0.098	 \\ 
    12.591	 & 	44.706	 & 	-0.099	 & 	-0.76	 & 	0.634	 & 	0.147	 \\ 
    3.677	 & 	47.167	 & 	0.099	 & 	0.886	 & 	-0.452	 & 	0.098	 \\ 
    -5.475	 & 	46.122	 & 	-0.099	 & 	-0.944	 & 	0.295	 & 	0.147	 \\ 
    -14.653	 & 	44.984	 & 	0.099	 & 	0.992	 & 	-0.079	 & 	0.098	 \\ 
    -22.709	 & 	40.515	 & 	-0.099	 & 	-0.985	 & 	-0.089	 & 	0.147	 \\ 
    -30.752	 & 	35.952	 & 	0.099	 & 	0.947	 & 	0.307	 & 	0.098	 \\ 
    -36.485	 & 	28.741	 & 	-0.099	 & 	-0.876	 & 	-0.459	 & 	0.147	 \\ 
    -42.169	 & 	21.448	 & 	0.099	 & 	0.757	 & 	0.646	 & 	0.098	 \\ 
    -44.706	 & 	12.591	 & 	-0.099	 & 	-0.634	 & 	-0.76	 & 	0.147	 \\ 
    -47.167	 & 	3.677	 & 	0.099	 & 	0.452	 & 	0.886	 & 	0.098	 \\ 
    -46.122	 & 	-5.475	 & 	-0.099	 & 	-0.295	 & 	-0.944	 & 	0.147	 \\ 
    -44.984	 & 	-14.653	 & 	0.099	 & 	0.079	 & 	0.992	 & 	0.098	 \\ 
    -40.515	 & 	-22.709	 & 	-0.099	 & 	0.089	 & 	-0.985	 & 	0.147	 \\ 
    -35.952	 & 	-30.752	 & 	0.099	 & 	-0.307	 & 	0.947	 & 	0.098	 \\ 
    -28.741	 & 	-36.485	 & 	-0.099	 & 	0.459	 & 	-0.876	 & 	0.147	 \\ 
    -21.448	 & 	-42.169	 & 	0.099	 & 	-0.646	 & 	0.757	 & 	0.098	 \\ 
    -12.591	 & 	-44.706	 & 	-0.099	 & 	0.76	 & 	-0.634	 & 	0.147	 \\ 
    -3.677	 & 	-47.167	 & 	0.099	 & 	-0.886	 & 	0.452	 & 	0.098	 \\ 
    5.475	 & 	-46.122	 & 	-0.099	 & 	0.944	 & 	-0.295	 & 	0.147	 \\ 
    14.653	 & 	-44.984	 & 	0.099	 & 	-0.992	 & 	0.079	 & 	0.098	 \\ 
    22.709	 & 	-40.515	 & 	-0.099	 & 	0.985	 & 	0.089	 & 	0.147	 \\ 
    30.752	 & 	-35.952	 & 	0.099	 & 	-0.947	 & 	-0.307	 & 	0.098	 \\ 
    36.485	 & 	-28.741	 & 	-0.099	 & 	0.876	 & 	0.459	 & 	0.147	 \\ 
    42.169	 & 	-21.448	 & 	0.099	 & 	-0.757	 & 	-0.646	 & 	0.098	 \\ 
    \midrule
    {\bf Dimer} & & & & & \\
    \midrule
    0.0	 & 	-4.0	 & 	0.0	 & 	$\cos \theta$	 & 	0.0	 & 	$\sin \theta$ \\ 
    0.0	 & 	4.0	 & 	0.0	 & 	$\cos \theta$	 & 	0.0	 & 	$-\sin \theta$ \\
    \bottomrule
    \end{tabular}
    \caption{Positions and normalized transition dipole vectors for the LHI complex, the dimer and one of the eight LHII complexes forming the aggregate in Fig.~\ref{fig-posdip}.}
    \label{tab-posdip}
\end{table}

By diagonalizing the Hamiltonian, Eq.~(1) of the main text, for the whole bio-inspired aggregate, we obtain the energies and TDMs for all eigenstates of the system. This information allows us to use our laser equations~(5,6,7) in the main text, under the assumption of quasi-instantaneous thermal relaxation and incoherent driving induced by the cavity field. 
We would like to stress that, while incoherent driving induced by the cavity field is well justified,  the fast thermal relaxation assumption might fail in very large systems. Nevertheless here we show that this assumption remains valid for the large aggregates considered in this section. Indeed, due to the cooperative coupling between the superradiant states of the LHII-LHI complex and LHI-dimer system, thermalization is estimated to occur in tens of~picoseconds~\cite{schulten:4}. We now proceed by estimating the remaining relevant timescales: the driving rate induced by the cavity coupling and the optical rates (spontaneous and stimulated emission) induced by the photon field. 
Specifically, from Eq.~(4) in the main text we can estimate the cavity-induced transition rate as $n(B_{tot}+\langle B \rangle)$. When the $\ket{L}$ state of the dimer is well-gapped below the rest of the single excitation manifold, its Boltzmann occupation is close to 1 and we have $B_{tot} \approx \langle B \rangle \approx B_L$. Owing to Eq.~(4) in the main text we also have $n(B_{tot}+\langle B \rangle) \approx 2nB_L = \Omega_L^2/\Gamma_\phi = \mu_L^2 I / (\hbar^2 c \epsilon_0 \Gamma_\phi)$. For the parameters used here ($\mu=10.157$~D, $\Gamma_\phi=1/(10$~ps)) and the maximal intensity that we consider ($I=1$~kW/cm$^2$), we obtain $n(B_{tot}+\langle B \rangle) \approx 7.78(\mu_L/\mu)^2$~ns$^{-1}$. In the extreme case $\mu_L=\mu$ we have $n(B_{tot}+\langle B \rangle) \approx 1/(126$~ps$)$. In most cases $\mu_L \ll \mu$ and the intensity is less than 1~kW/cm$^2$, so laser-induced transitions occur on a timescale slower than $126$~ps.  Moreover, as explained in the main text, 
 thermal relaxation is certainly faster than optical pumping and decay rates, which  range from a few nanoseconds (large aggregates, high black-body temperature) down to milliseconds (small aggregates, natural sunlight). Another relevant timescale is the realistic extraction rate $\kappa$ from the cavity of about three nanoseconds which we considered.
 Finally, note that under natural sunlight pumping (where the maximal laser intensity considered is $10$~W/cm$^{2}$) these other timescales 
 are considerably longer and thus Eqs.~(5,6,7) of the main text retain their  validity for larger aggregates.     
 
 These estimates justify the use of the laser equations of the main text for the bio-inspired aggregate considered here. To further validate this and to show how to extend our analysis to even larger aggregates, where thermal relaxation might not be the fastest timescale, here we analyze a set of incoherent laser equations, similar to Eq.~\eqref{maslindcav}, where instantaneous thermalization is not assumed. 
 
For this purpose we focus on the LHI-Dimer aggregate, composed of a total of 34 molecules, 32 for the LHI and two for the dimer. In this case we will consider the eigenstates of the LHI aggregate and the dimer aggregate separately. Let us define $P_k$ as the population of one of these 34 states: $k=1,\dots,32$ for the LHI eigenstates and $k=33,34$ for the dimer eigenstates ($\ket{L}$ and $\ket{H}$). This separation is motivated from the fact that thermalization on the LHI or dimer aggregate alone will be faster (few picoseconds) than the thermalization over the whole LHI-Dimer aggregate. This is because the coupling within the aggregates (of the order of several hundreds of cm$^{-1}$) is much stronger than the coupling between the aggregates (few cm$^{-1}$). 
Moreover, let us define $P_G$ as the probability to be in the ground state and $n$ the number of photons in the cavity, so that 
the incoherent laser equations read
\begin{subequations}
\label{ILE}
    \begin{align}
        \frac{dP_G}{dt}&=-\sum_k R_k P_G + \sum_k [R_k +\gamma_k(\omega_k)]P_k  -\sum_k B_k n P_G + \sum_k B_k n P_k~, \\
        \frac{dP_k}{dt}&= R_k P_G - [R_k +\gamma_k(\omega_k)]P_k  + B_k n P_G -  B_k n P_k +\sum_j \left(T_{k,j}P_j -T_{j,k} P_k \right) ~, \\
        \frac{d n}{dt} &= V \sum_k B_k nP_k - V\sum_k B_k n P_G -n \kappa~,
    \end{align}
\end{subequations}
where the definitions of $R_k,\gamma_k(\omega_k), B_k$ and $\kappa$ can be found in the previous sections of the supplementary material. The rates $T_{j,k}$ between the eigenstates of each aggregate are taken from Eq.~\eqref{BR-phon-rate},
 $$
 T_{j,k} = \frac{\kappa_{\rm vib}\omega_{j,k}}{1-e^{-\hbar\omega_{j,k}/(k_BT)}} \qquad \mbox{with} \qquad \omega_{j,k}=\frac{E_k-E_j}{\hbar}
 $$
where $\kappa_{\rm vib}=1.3\times 10^{-3}$ has the same value that we used when studying the dimer in the previous sections. 
The coupling between the LHI and the dimer is taken into account by considering only the coupling of one superradiant state of the LHI with the $\ket{H}$ dimer state. Indeed, these two states are resonant and the coupling between all other states is either much smaller or largely off-resonant. Starting from~\cite{schulten16ps}, the thermal transfer rate between the LHI ring and the special pair of the RC system has been estimated to be $\approx 1/(16~{\rm ps})$. The thermal transfer rate between two aggregates is defined as  $K=\sum_{n,m} = p_n T_{n,m}$, where $n$ labels the eigenstates of the first aggregate and $m$ labels the eigenstates of the second aggregate, $p_n$ is the thermal population of the $n$-th eigenstate and $T_{n,m}$ is the transfer rate between the $n$-th eigenstate of the first aggregate and the $m-$th eigenstate of the second aggregate. For the case of the LHI-dimer complex, only the coupling between one superradiant state of the ring and the $\ket{H}$ dimer state is relevant, so that $K=p_2 T_{2,34}$. Given that $p_2 \approx 0.212$ at room temperature,  we consider only the coupling $T_{2,34}=T_{34,2}=1/(p_2 \times 16~{\rm ps})$~\cite{schulten16ps} and set all the other couplings between the eigenstates of LHI and the dimer to zero. The value of the coupling between the superradiant state of the LHI aggregate and the   $\ket{H}$ dimer state has been chosen to be identical to that estimated for natural structures between the LHI aggregate and the special pair~\cite{schulten:4}. The laser intensity can now be obtained from the number of photons $n$ using Eq.~(7) in the main text. 

\begin{figure}[!htbp]
    \centering
    \includegraphics[width=0.5\linewidth]{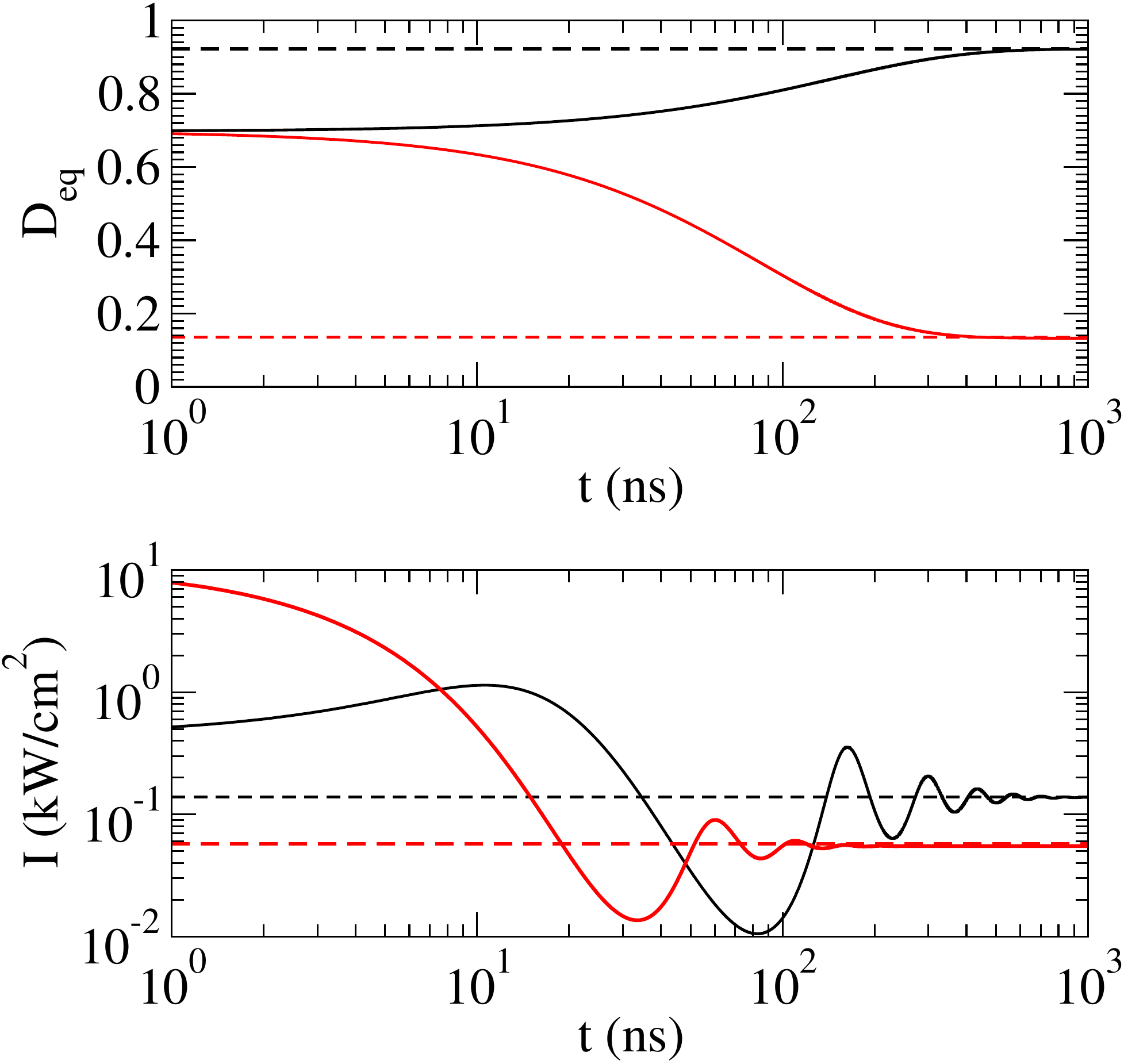}
    \caption{Comparison between theoretical prediction of Eqs.~(5,6,7)  in the main text and incoherent rate laser equations Eq.~\eqref{ILE} for the LHI-dimer aggregate (Fig.~3b in the main text). Initial conditions for both panels: $P_G(t=0)=0.15$, $P_{34}(t=0)=0.85$ (initial population of the $\ket{H}$ dimer state), $P_{k\ne 34}(t=0)=0$ and $n(t=0)/V=10^{11}$~cm$^{-3}$.
    Upper panel: population difference $D_{eq}$ {\it vs.} time: the horizontal dashed lines correspond to Eq.~(9) in the main text, whereas the solid lines have been obtained from Eq.~\eqref{ILE} in absence of the coupling to the cavity. Lower panel: laser intensity {\it vs} time: 
    the horizontal dashed lines correspond to Eq.~(7) in the main text whereas the solid lines to Eq.~\eqref{ILE}.
    In both panels we consider black body pumping at $T_{BB}=3000$~K, and an aggregate density of $n_A=10^{-5}$~mmol/L. Two different values of the dimer brightness are shown: $\langle \chi \rangle=9.7 \times 10^{-3}$ (black curves) and $\langle \chi \rangle=0.185$ (red curves). The other dimer and LHI parameters are  discussed in the main body. }
    \label{LHI-Dimer-I}
\end{figure}

In Fig.~\ref{LHI-Dimer-I} we compare the numerical solutions obtained from Eq.~\eqref{ILE} with the theoretical prediction for stationary values obtained from Eqs.~(5,6,7) of the main text. As one can see, both the stationary population difference (upper panel) and the stationary laser intensity (lower panel) obtained with Eq.~\eqref{ILE} are in excellent agreement with our theoretical predictions obtained assuming instantaneous thermal relaxation over the whole aggregate. This figure shows the case of black-body pumping at $3000$~K. For natural sunlight irradiation, since the pumping and the coupling to the cavity is much weaker, we expect that our assumption of fast thermal relaxation will be  valid even for much larger aggregates than the one considered in Fig.~\ref{LHI-Dimer-I}.

Whilst the rate equation model presented above will naturally not capture some of the more subtle details of the excitation dynamics of an LHI ring, it nonetheless resolves the thermalization process dynamically and is thus more realistic than assuming instant thermalization. Moreover, finer details in the dynamics will not affect the lasing predictions and efficiency, provided thermalization does indeed occur sufficiently quickly, as is captured by the above rate equations. 

\section{Threshold density at 3000~K for purple bacteria bio-inspired aggregates}
\label{sec-3000purp}

\begin{figure}[!htbp]
    \centering
    \includegraphics[width=0.6\columnwidth]{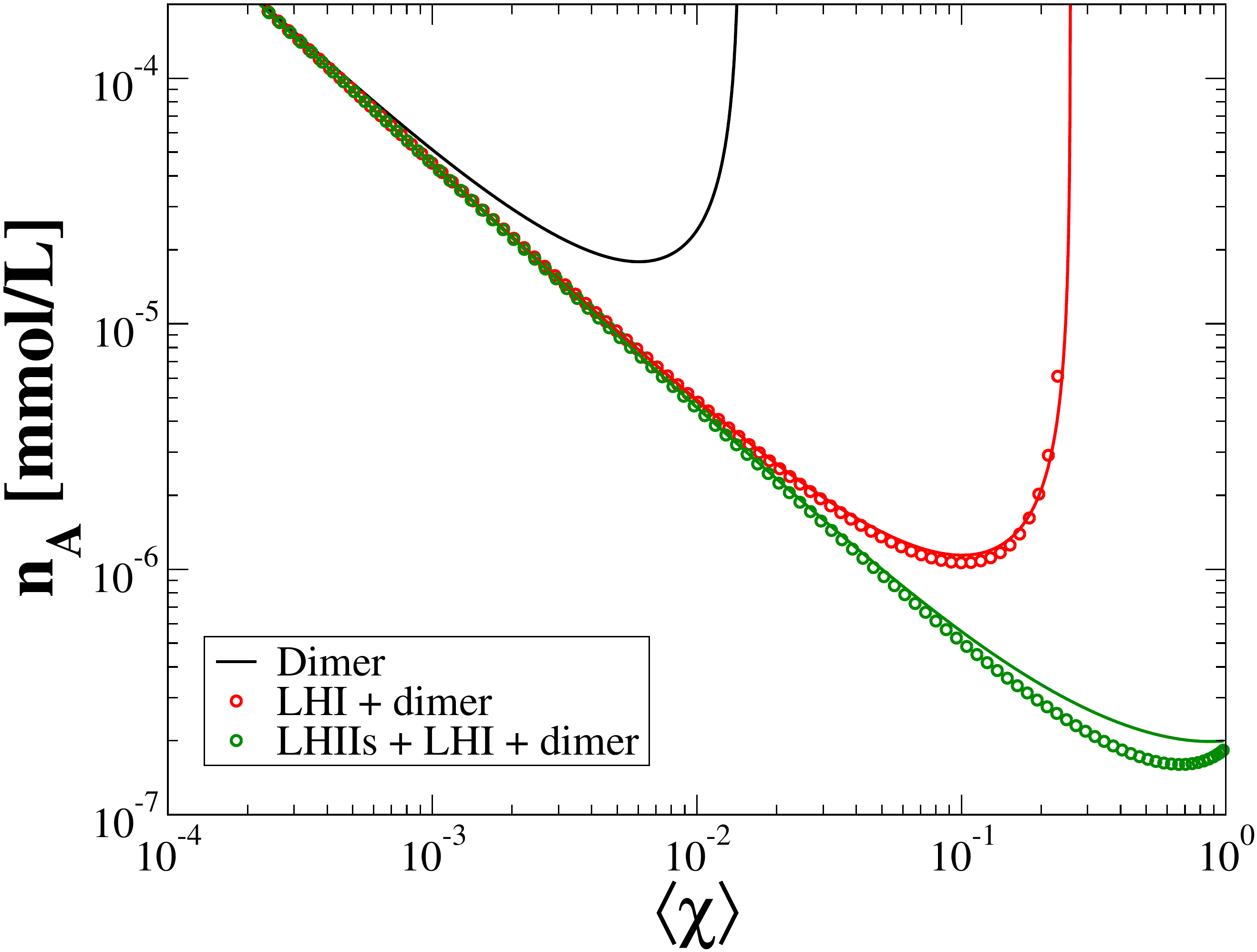}
    \caption{Threshold density for dimer, LHI+dimer and LHIIs+LHI+dimer aggregates. The red line is a dimer with $R_u$ rescaled by $0.85(34/2)$ where 34 is the number of molecules in the LHI + dimer aggregate. The green line is a dimer with $R_u$ rescaled by $0.45(250/2)$, where 250 is the number of  molecules in the LHIIs + LHI + dimer aggregate. The parameters are the same as in Fig.~3h of the main text, apart from the black-body temperature, here $T_{BB}=3000$~K.}
    \label{agg-comp}
\end{figure}

In Fig.~3h of the main text we compare the threshold density under natural sunlight pumping for a dimer and two aggregates (LHI + dimer and LHIIs + LHI + dimer). We show that the threshold density of the aggregates is mapped to the threshold density of a dimer whose $R_u$ pumping factor is multiplied by $N/2$, where $N$ is the total number of molecules in the whole aggregate.  Here, we show the same effect under pumping from a 3000~K black-body source. In Fig.~\ref{agg-comp} the threshold density is plotted for a dimer and for the two aggregates. The maximal $\langle \chi \rangle$ required for lasing increases with the aggregate size, and the threshold density decreases with increasing $N$. Similarly to the natural sunlight case (shown in the main text), we can reproduce the aggregate threshold density using a dimer with an enhanced $R_u$ factor. However, differently to the natural sunlight case, now the scaling factor is smaller than $N/2$. Specifically, we reproduce the ``LHI + dimer'' case using a dimer with $0.85(N/2)R_u$ and the ``LHIIs + LHI + dimer'' case using a dimer with $0.45(N/2)R_u$. The scaling here is not as simple as the natural sunlight case because of energy-dependent terms contained within $R_u$. Specifically, $R_u=\sum_k \gamma_k(\omega_k) n_T(\omega_k)$ is the sum of the dipole strengths of the eigenstates (contained in $\gamma_k$) multiplied by the black-body spectrum of the source. In the natural sunlight case ($T_{BB}=5800$~K) the black-body spectrum is basically flat over the energy range of the aggregate, so that $R_u$ scales as $N$, the sum of all the dipole strengths in the aggregate. In the $T_{BB}=3000$~K case, instead, the black-body spectrum decreases with the energy in the aggregate range. So, the additional dipole strength that is provided by the LHIIs in the aggregate is proportionally less effective and needs to be multiplied by a correction factor accounting for the reduced spectral intensity. This explains why the enhancement is smaller than $N/2$ at this temperature.

\bibliography{supp}